\documentclass[5p]{elsarticle}

\usepackage[T1]{fontenc}
\usepackage[shortcuts]{extdash}
\usepackage{color}
\usepackage{amsmath}
\usepackage{amssymb}
\usepackage{longtable}
\usepackage{hyperref}

\journal{Astroparticle Physics}

\DeclareMathOperator{\sinc}{sinc}

\newcommand{\Across}{A_{\times}}
\newcommand{\Amp}{\mathcal{A}}
\newcommand{\Aplus}{A_{+}}
\newcommand{\DSmax}{X^{\mathrm{max}}}
\newcommand{\DSthr}{X^{*}}
\newcommand{\Delay}[1]{\Delta_{\mathrm{#1}}}
\newcommand{\Erot}{E_{\mathrm{rot}}}
\newcommand{\Fcross}{F_{\times}}
\newcommand{\Fplus}{F_{+}}
\newcommand{\Fstat}{$\mathcal{F}$-statistic}
\newcommand{\Ixx}{I_{xx}}
\newcommand{\Iyy}{I_{yy}}
\newcommand{\Izz}{I_{zz}}
\newcommand{\Lgw}{\dot{E}_{\mathrm{GW}}}
\newcommand{\MVHMM}{\MV_{\mathrm{HMM}}}
\newcommand{\MVany}{\MV_{\cdots}}
\newcommand{\MVbin}{\MV\ubin}
\newcommand{\MVfdd}{\MV\ufdd}
\newcommand{\MVfd}{\MV\ufd}
\newcommand{\MVfreq}{\MV\ufreq}
\newcommand{\MVrange}[1]{\Big|#1\Big]}
\newcommand{\MVsky}{\MV\usky}
\newcommand{\MV}{\mathcal{B}}
\newcommand{\NS}{_{\mathrm{NS}}}
\newcommand{\OmegaO}{\Omega_{o}}
\newcommand{\OmegaS}{\Omega_{s}}
\newcommand{\PhiC}{\Phi^{C}}
\newcommand{\SSB}{_{\mathrm{SSB}}}
\newcommand{\Sh}{S_{h}}
\newcommand{\Tcohmax}{T_{\mathrm{coh,max}}}
\newcommand{\Tcoh}{T_{\mathrm{coh}}}
\newcommand{\Tspan}{T}
\newcommand{\WVol}{\mathcal{V}}
\newcommand{\dErot}{\dot{E}_{\mathrm{rot}}}
\newcommand{\depth}{\mathcal{D}}
\newcommand{\dist}{D}
\newcommand{\fddrot}{\ddot{f}\rot}
\newcommand{\fdd}{\ddot{f}}
\newcommand{\fdrmode}{\dot{f}_{\alpha}}
\newcommand{\fdrot}{\dot{f}\rot}
\newcommand{\fd}{\dot{f}}
\newcommand{\frmode}{f_{\alpha}}
\newcommand{\frot}{f\rot}
\newcommand{\hcross}{h_{\times}}
\newcommand{\hplus}{h_{+}}
\newcommand{\hrmode}{h_{\alpha}}
\newcommand{\ifo}{_{\mathrm{det}}}
\newcommand{\ndot}[1]{^{(#1)}}
\newcommand{\oneF}{\mathcal{F}}
\newcommand{\order}{\mathcal{O}}
\newcommand{\paramsp}{\mathcal{P}}
\newcommand{\phiEuler}{\phi_{E}}
\newcommand{\psiEuler}{\psi_{E}}

\newcommand{\rawMVany}{\rawMV_{\cdots}}
\newcommand{\rawMVbin}{\rawMV\ubin}
\newcommand{\rawMVfdd}{\rawMV\ufdd}
\newcommand{\rawMVfd}{\rawMV\ufd}
\newcommand{\rawMVfreq}{\rawMV\ufreq}
\newcommand{\rawMVsky}{\rawMV\usky}
\newcommand{\rawMV}{B}
\newcommand{\rot}{_{\mathrm{rot}}}
\newcommand{\solar}{\odot}
\newcommand{\tasc}{t_{\mathrm{asc}}}
\newcommand{\tauO}{\tau_{o}}
\newcommand{\tauS}{\tau_{s}}
\newcommand{\twoF}{2\mathcal{F}}
\newcommand{\ubin}{_{\mathrm{bin}}}
\newcommand{\ufdd}{_{\fdd}}
\newcommand{\ufd}{_{\fd}}
\newcommand{\ufreq}{_{f}}
\newcommand{\umax}{_\mathrm{max}}
\newcommand{\uobs}{_\mathrm{obs}}
\newcommand{\usky}{_{\mathrm{sky}}}
\newcommand{\uvec}[1]{\hat{#1}}

\newcommand{\NSEARCHES}{297}
\newcommand{\NPAPERS}{80}
\newcommand{\NOHTHREEPAPERS}{17}
\newcommand{\MINYEAR}{2003}
\newcommand{\MAXYEAR}{2022}

\biboptions{sort&compress}

\begin{document}

\begin{frontmatter}

  \title{Searches for continuous gravitational waves from neutron stars: \\ A twenty-year retrospective}

  \author[cga,ozgrav]{Karl Wette}
  \ead{karl.wette@anu.edu.au}
  \affiliation[cga]{
    addressline={Centre for Gravitational Astrophysics, Australian National University},
    city={Canberra},
    state={ACT},
    postcode={2601},
    country={Australia}
  }
  \affiliation[ozgrav]{
    addressline={Australian Research Council Centre of Excellence for Gravitational Wave Discovery (OzGrav)},
    city={Hawthorn},
    state={VIC},
    postcode={3122},
    country={Australia}
  }

  \begin{abstract}%
    Seven years after the first direct detection of gravitational waves, from
    the collision of two black holes, the field of gravitational wave astronomy
    is firmly established. A first detection of continuous gravitational waves
    from rapidly-spinning neutron stars could be the field's next big
    discovery. I review the last twenty years of efforts to detect continuous
    gravitational waves using the LIGO and Virgo gravitational wave detectors. I
    summarise the model of a continuous gravitational wave signal, the
    challenges to finding such signals in noisy data, and the data analysis
    algorithms that have been developed to address those challenges. I present a
    quantitative analysis of \NSEARCHES\ continuous wave searches from \NPAPERS\
    papers, published from \MINYEAR\ to \MAXYEAR, and compare their
    sensitivities and coverage of the signal model parameter space.
  \end{abstract}

  \begin{keyword}
    gravitational waves \sep neutron stars \sep data analysis
  \end{keyword}

\end{frontmatter}

\section{Introduction}\label{sec:introduction}

\begin{figure*}[!t]%
  \centering%
  \includegraphics[width=0.95\linewidth]{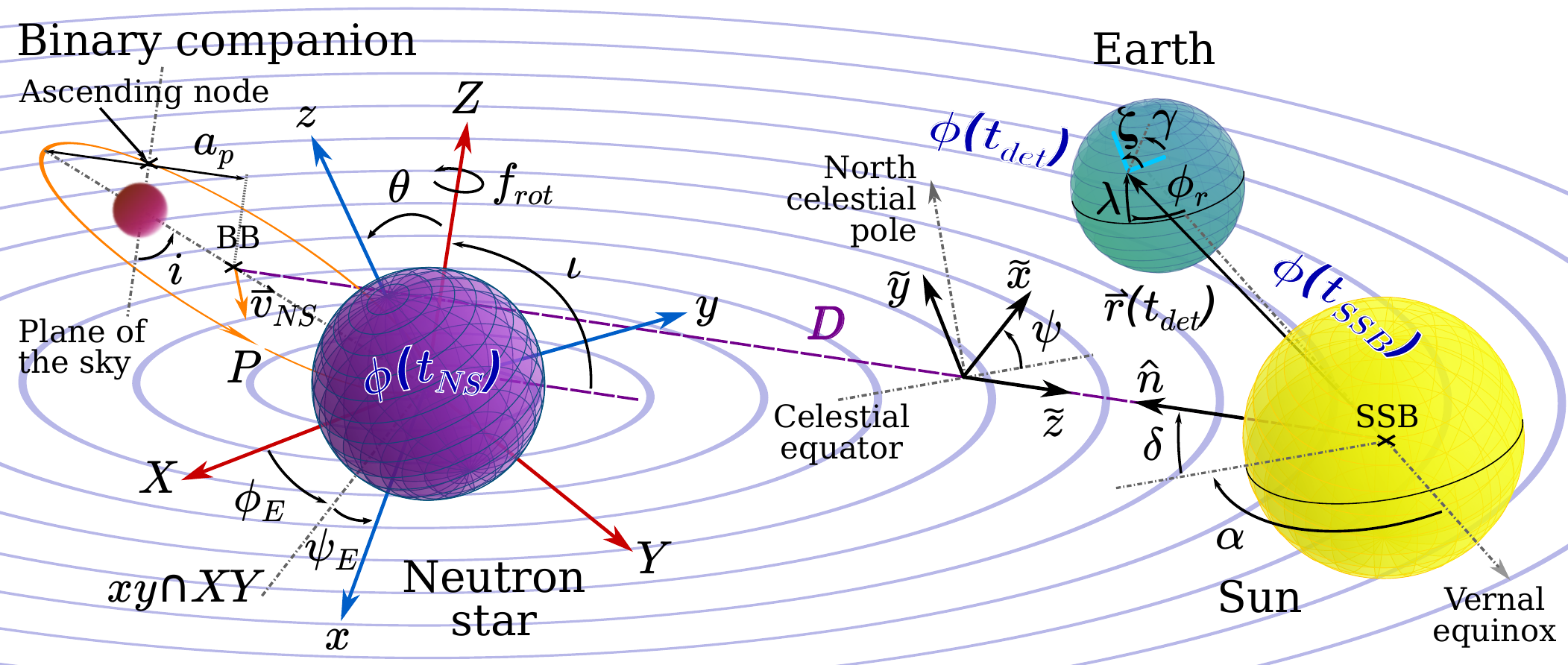}%
  \caption{\label{fig:cw-signal-model}%
    CW signal model, illustrated by a rigidly rotating neutron star with a
    possible binary companion.  Parameters are: $\frot$ is the rotation
    frequency of the star; $(\phiEuler, \theta, \psiEuler)$ orient the body frame
    $(x, y, z)$ with respect to an inertial frame $(X, Y, Z)$; $(a_p, P)$ are
    the projected semi-major axis and period of the neutron star's orbit around
    the binary companion, if present; $\vec{v}\NS$ is the linear velocity of the
    neutron star, or of the binary barycentre (BB) if a binary companion is
    present; $\uvec{n}$ is a unit vector pointing from the Solar System
    barycentre (SSB) to the neutron star or BB; $\iota$ is the angle from the
    line of sight (parallel to $\uvec{n}$) to the inertial $Z$ axis; $\dist$ is
    the distance from the SSB to the neutron star or BB;
    $(\alpha, \delta, \psi)$ orient the wave frame
    $(\tilde{x}, \tilde{y}, \tilde{z})$ in equatorial coordinates;
    $\vec{r}(t\ifo)$ denotes the position of the detector with respect to the
    SSB; $(\phi_r, \lambda, \gamma)$ orient the interferometric detector with
    respect to the Earth; and $\zeta$ is the angle between the interferometer
    arms.  The CW signal phase at the neutron star, SSB, and detector are
    labelled $\phi(t\NS)$, $\phi(t\SSB)$, and $\phi(t\ifo)$ respectively.%
  }%
\end{figure*}

\begin{quote}%
  Together with a young collaborator, I arrived at the interesting result that
  gravitational waves do not exist, though they had been assumed a certainty to
  the first approximation.  This shows that the non-linear general relativistic
  field equations can tell us more or, rather, limit us more than we have
  believed up to now.%
  \begin{flushright}%
    Einstein, in a letter to Born, 1936~\cite{Kenn2005-EnsVrPhyRv}.%
  \end{flushright}%
\end{quote}
Progress in science is rarely, if ever, a story of uninterrupted success;
rather, a journey of trial and error, initial presumptions giving way to new
discoveries. The century-long study of gravitational waves, from theoretical
conjecture in 1916~\cite{Eins1916-NhrInDrFldDGr, Eins1918-UbrGrv} to direct
detection in 2015~\cite{LIGOVirg2016-ObsGrvWvBnBHMr}, certainly follows this
template. On that difficult journey, it is understandable that even Einstein at
one time doubted their existence~\cite{EinsRose1937-GrvWvs}; yet it is just as
well that his initial predictions have stood.  Gravitational waves -- wave-like
solutions to the Einstein field equations -- are fulfilling their promise to
``tell us more'' about the Universe that could be achieved through traditional
astronomy. Analysis of the growing catalogue of detections of gravitational
waves from colliding pairs of black holes and neutron
stars~\cite{LIGOVirg2016-BnBlHMrFAdvLObsR, LIGOVirg2017-GWObsGrvWBNtSIn,
  LIGOVirg2019-GGrvTrnCtCBMObLVDFSObR, LIGOVirg2021-GCmBnClsObLVDFHTObsR, GWTC3}
by the LIGO and Virgo detectors~\cite{LIGO2015-AdvLIG, Virg2015-AdVScnIntGrWDt}
has, e.g.: confirmed the existence of stellar-mass black
holes~\cite{LIGOVirg2016-ObsGrvWvBnBHMr}; strengthened links between the merger
of binary neutron stars, short-hard gamma-ray bursts, kilonovae, and the
production of heavy elements~\cite{LIGOEtAl2017-MltObsBnNtSMr}; yielded insights
into the mass spectrum of black holes, and by inference their stellar
progenitors~\cite{LIGOVirg2019-BBHPplPrpInFSObRAdLAV,
  LIGOVirg2021-PplPrCOSLIGrvTrC}; and opened an independent avenue towards
resolving tensions in cosmology regarding measurement of the Hubble
constant~\cite{LIGOEtAl2017-GrvStSMsrHbCn, DESEtAl2019-FMsrHCnDStSUDESGLIBnrMG,
  LIGOVirg2021-GrvMsrHCnFlSObRALV}.

As well as the important discoveries and insights from observing binary black
hole and neutron star mergers, it is hoped that the gravitational wave view of
the Universe will continue to widen in the coming years. Perhaps we will observe
the gravitational aftermath of a supernova, the faint hum of gravitational waves
from rapidly-spinning neutron stars, signatures of dark matter or particles
beyond the Standard Model, or the even fainter murmur of gravitational waves
from the very early Universe. It is the second of these -- the search for
so-called \emph{continuous gravitational waves} -- that is the subject of this
review.

General relativity predicts gravitational waves from an astronomical body only
when it possesses a time-varying quadrupole moment.  As an example, if the body
is not symmetric about its rotation axis, so that the distribution of its mass
is seen to ``move'' when rotated, its \emph{mass} quadrupole moment will vary
with time. (Velocity perturbations within the star may also give rise to a
time-varying \emph{current} quadrupole moment; see
Sec.~\ref{sec:r-mode-oscillations}.)  While the orbit of neutron stars or black
holes around each other presents an obvious asymmetry, the extent to which a
single neutron may sustain a non\-/axisymmetric shape is expected to be much
smaller~\cite{Rude1969-NtrStrPlPr, HoroKada2009-BrkStNtSCrGrvW}. It is assumed
that all neutron stars possess a magnetic field~\cite{Reis2001-MgnFlNtStOvr}
which, so long as it is not aligned with the star's rotation axis, will distort
the star in a non\-/axisymmetric way~\cite{BonaGour1996-GrvWPlEmMgFIDst}. It is
likely therefore that all neutron stars radiate \emph{some} continuous
gravitational waves. It remains to be discovered, however, whether the strength
of those waves -- arising either from magnetic distortion, or through some other
mechanism -- is sufficient to be \emph{detectable} using the instruments and
data analysis techniques available to us.

In this review I look back at the last twenty years of searches for continuous
gravitational waves (CWs). I begin in Sec.~\ref{sec:cw-signal-model} with an
overview of the CW signal model. I then discuss the many challenges faced by CW
searches in Sec.~\ref{sec:chall-cont-wave}, and suggest metrics by which we may
compare their performance in Sec.~\ref{sec:cont-wave-perf-meas}. After a brief
summary of gravitational wave detectors and observations
(Sec.~\ref{sec:brief-hist-grav}), I review CW searches performed from \MINYEAR\
to \MAXYEAR\ in Sec.~\ref{sec:cont-wave-search}, and the algorithms employed by
those searches in Sec.~\ref{sec:cont-wave-alg}. I conclude with a summary, and
suggestions for further reading, in Sec.~\ref{sec:summary}.

\section{Continuous wave signal model}\label{sec:cw-signal-model}

Figure~\ref{fig:cw-signal-model} gives an overview of the CW signal model.
Gravitational waves in general relativity exist in two polarisations:
\emph{plus} and \emph{cross}. A gravitational wave is modelled as a time series
$h(t)$ in the calibrated output of a gravitational wave detector:
\begin{equation}
  \label{eq:hoft}
  h(t) = \Fplus(t) \hplus(t) + \Fcross(t) \hcross(t) \,. \\
\end{equation}
The functions $\Fplus(t)$ and $\Fcross(t)$ are the responses of the detector to
each polarisation. These depend on the polarisation basis
$(\tilde{x}, \tilde{y}, \tilde{z})$, oriented by the polarisation angle $\psi$,
and a model of the response of the detector to a gravitational wave. An
interferometric detector may be described~\cite{JaraEtAl1998-DAnGrvSgSpNSSDtc}
by its location on Earth -- its latitude $\lambda$, and local sidereal time
$\phi_r$ at a given reference time -- the orientation of its arms $\gamma$ and
the angle $\zeta$ between them.

A CW signal is described by the two functions
$\hplus(t)$ and $\hcross(t)$ in terms of numerous parameters
(Fig.~\ref{fig:cw-signal-model}). Conventionally the CW signal parameters are
divided into: \emph{amplitude parameters} (Sec.~\ref{sec:cw-signal-amplitude}),
which control the overall amplitude of the signal; and \emph{phase parameters}
(Sec.~\ref{sec:cw-signal-phase}), which control its phase.

\subsection{Amplitude parameters}\label{sec:cw-signal-amplitude}

\begin{table*}[!t]%
  \newcounter{oldequation}%
  \setcounter{oldequation}{\value{equation}}%
  \setcounter{equation}{0}%
  \renewcommand{\theequation}{T\thetable.\arabic{equation}}%
  \centering%
  \caption{\label{tab:cw-amp}%
    CW signal amplitude parameters, for the general case, and for the biaxial,
    triaxial aligned, and $r$-mode oscillations cases. Expressions are given
    for: the amplitudes $C_{22}, C_{21}$ and initial phases
    $\PhiC_{22}, \PhiC_{21}$ of the $l=2, m=2$ and $l=2, m=1$ harmonics,
    respectively; and for the gravitational wave luminosity $\Lgw$. The graphics
    at left illustrate the shape and orientation of the neutron star for the
    rigidly rotating star cases; for the $r$-mode case, the pattern on the
    neutron star surface is proportional to the $r$-mode velocity
    perturbations~\cite{LindEtAl1998-GrvRdInsHYnNtS}.%
  }%
  \begin{tabular*}{\textwidth}{p{0.15\textwidth} @{} p{0.8\textwidth}}%
    \hline\hline
    \multicolumn{2}{c}{Triaxial non-aligned (general case); see Sec.~\ref{sec:rigidly-rotat-stars}} \\
    \multicolumn{2}{p{0.95\textwidth}}{ { \setlength{\abovedisplayskip}{0pt} \setlength{\belowdisplayskip}{0pt}
    \begin{flalign}
      \begin{split}
        \label{eq:cw-amp-general-C}
        C_{22} &= \frac{1}{2} \Big[ h_0^2 \cos^4\theta \cos^4\psiEuler + ( h_p \sin^2\theta - h_0 \sin^2\psiEuler )^2  + 2 h_0 ( h_p \sin^2\theta + h_0 \sin^2\psiEuler ) \cos^2\theta \cos^2\psiEuler \Big]^{1/2} \,, \\
        C_{21} &= \frac{1}{2} \Big[ h_0^2 \sin^2\theta \sin^22\psiEuler + \frac{1}{4} (h_0 \cos2\psiEuler + h_0 - 2 h_p)^2 \sin^22\theta \Big]^{1/2} \,,
      \end{split}
    \end{flalign} } } \\
    \raisebox{-\height}{\includegraphics[width=\linewidth]{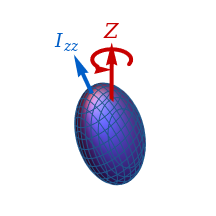}} 
    & { \setlength{\abovedisplayskip}{0pt} \setlength{\belowdisplayskip}{0pt}
      \begin{flalign}
        \begin{split}
          \label{eq:cw-amp-general-PhiC}
          \tan(2\phiEuler - \PhiC_{22}) &= \frac{h_0 \cos\theta \sin2\psiEuler}{h_0 \sin^2\psiEuler - h_p \sin^2\theta - h_0 \cos^2\theta \cos^2\psiEuler} \,, \\
          \tan(\phiEuler - \PhiC_{21}) &= \cos\theta \Big[ \cot\psiEuler - \frac{2 h_p}{h_0} \csc2\psiEuler \Big] \,,
        \end{split} \\
        \label{eq:cw-amp-general-Lgw}
        \Lgw &= \frac{\pi^2 c^3}{10 G} ( 64 C_{22}^2 + C_{21}^2 ) \dist^2 \frot^2 \,.
      \end{flalign} } \\
    \hline
    \multicolumn{2}{c}{Biaxial ($h_0 = 0$); see Sec.~\ref{sec:rigidly-rotat-stars}} \\
    \raisebox{-\height}{\includegraphics[width=\linewidth]{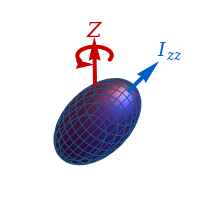}} 
    & { \setlength{\abovedisplayskip}{0pt} \setlength{\belowdisplayskip}{0pt}
      \begin{flalign}
        \label{eq:cw-amp-biax-C}
        C_{22} &= \frac{1}{2} h_p \sin^2\theta \,, & C_{21} &= \frac{1}{2} h_p \sin2\theta \,, \\
        \label{eq:cw-amp-biax-PhiC}
        \PhiC_{22} &= 2\phiEuler + \pi = 2\Phi_0 + \pi \,, & \PhiC_{21} &= \phiEuler + \frac{\pi}{2} = \Phi_0 + \pi \,, \\
        \label{eq:cw-amp-biax-Lgw}
        \Lgw &= \frac{8 \pi ^2 c^3}{5 G} \Theta \dist^2 \frot^2 h_p^2 = \frac{2048 \pi ^6 G}{5 c^5} \Theta \Izz^2 \epsilon_p^2 \frot^6 \,, & \Theta &= \frac{(17 - 15 \cos2\theta) \sin^2\theta}{32} \,.
      \end{flalign} } \\
    \hline
    \multicolumn{2}{c}{Triaxial aligned ($\theta = 0$); see Sec.~\ref{sec:rigidly-rotat-stars}} \\
    \raisebox{-\height}{\includegraphics[width=\linewidth]{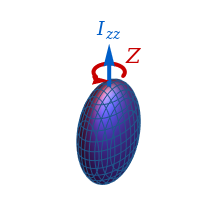}} 
    & { \setlength{\abovedisplayskip}{0pt} \setlength{\belowdisplayskip}{0pt}
      \begin{flalign}
        \label{eq:cw-amp-trialign-C}
        C_{22} &= \frac{1}{2} h_0 \,, & C_{21} &= 0 \,, \\
        \label{eq:cw-amp-trialign-PhiC}
        \PhiC_{22} &= 2 ( \phiEuler + \psiEuler ) = \phi_0 + \pi \,, \\
        \label{eq:cw-amp-trialign-Lgw}
        \Lgw &= \frac{2 \pi ^2 c^3}{5 G} \dist^2 f^2 h_0^2 = \frac{32 \pi ^6 G }{5 c^5} \Izz^2 \epsilon ^2 f^6 \,.
      \end{flalign} } \\
    \hline
    \multicolumn{2}{c}{$r$-mode; see Sec.~\ref{sec:r-mode-oscillations}} \\
    \raisebox{-\height}{\includegraphics[width=\linewidth]{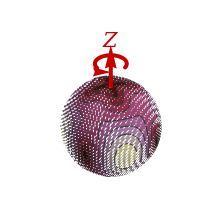}} 
    & { \setlength{\abovedisplayskip}{0pt} \setlength{\belowdisplayskip}{0pt}
      \begin{flalign}
        \label{eq:cw-amp-rmode-C}
        C_{22} &= \frac{1}{2} \hrmode \,, & C_{21} &= 0 \,, \\
        \label{eq:cw-amp-rmode-PhiC}
        \PhiC_{22} &= \phi_0 + \pi \,, \\
        \label{eq:cw-amp-rmode-Lgw}
        \Lgw &= \frac{1024 \pi^9 G}{25 c^7} \alpha ^2 \tilde{J}^2 M^2 R^6 \frmode^8 \,.
      \end{flalign} } \\
    \hline\hline
  \end{tabular*}%
  \setcounter{equation}{\value{oldequation}}%
\end{table*}

The basic model of a CW source is a rotating, non\-/axisymmetrically-deformed
neutron star which generates gravitational waves through a time-changing mass
quadrupole (Sec.~\ref{sec:rigidly-rotat-stars}). It is assumed that the
non\-/axisymmetry is \emph{rigid}, i.e.\ does not change over typical
observation periods. As a result, the CW amplitude parameters of this model may
be considered independently of the precise mechanism responsible for the
deformation. Extensions to the basic model are free precession
(Sec.~\ref{sec:free-precession}), $r$-mode oscillations
(Sec.~\ref{sec:r-mode-oscillations}), and accretion from a binary companion star
(Sec.~\ref{sec:accr-from-binary}).

\subsubsection{Rigidly rotating stars}\label{sec:rigidly-rotat-stars}

Provided that its internal velocities are non-relativistic, a rigidly-rotating
neutron star is described by its classical moment of inertia tensor
$I_{ij}$~\cite{Whee1966-SprStr, Chau1967-GrvRdtNtSt, OstrGunn1969-NtrPlsITh}. A
reference frame $(x, y, z)$ may always be found in which $I_{ij}$ is diagonal;
the diagonal elements are the principal moments $\Ixx$, $\Iyy$, and $\Izz$.
Examining whether or not the principal moments are equal to each other leads to
three equivalence classes:
\begin{enumerate}

\item $\Ixx = \Iyy = \Izz$: If all three principal moments are equal, the star
  is symmetric, and no gravitational waves are emitted.

\item $\Ixx = \Iyy \ne \Izz$: If two principal moments (conventionally $\Ixx$
  and $\Iyy$) are equal, the star is \emph{biaxial}. The degree of deformation
  along the $z$ axis is characterised by the \emph{poloidal ellipticity}
  \begin{equation}
    \label{eq:ellip-pol} \epsilon_p = \frac{ | \Ixx - \Izz | }{ \Izz }.
  \end{equation}

\item $\Ixx \ne \Iyy \ne \Izz$: If no principal moments are equal, the star is
  \emph{triaxial}. In addition to $\epsilon_p$, the degree of deformation
  perpendicular to the $z$ axis is characterised by the \emph{equatorial
    ellipticity}
  \begin{equation}
    \label{eq:ellip} \epsilon = \frac{ | \Ixx - \Iyy | }{ \Izz }.
  \end{equation}

\end{enumerate}
The body frame $(x, y, z)$ is related to an inertial frame $(X, Y, Z)$, where
the star rotates about the $Z$ axis.

The general form of the CW signal, following the notation
of~\cite{Jone2010-GrvWEmRtSprNtS, Jone2015-PrCRCntGrvWSrStSNS,
  PitkEtAl2015-FRFPrDlhSrGrvWSpNS} (cf.~\cite{BejgKrol2014-SrcGrvWvKPlTSFr}), is
\begin{align}
  \begin{split}
    \label{eq:hoft-plus}
    \hplus(t) &= -C_{22} ( 1 + \cos^2\iota ) \cos[ 2\phi\rot(t) + \PhiC_{22} ] \\
    &\quad -\frac{1}{2} C_{21} \sin\iota \cos\iota \cos[ \phi\rot(t) + \PhiC_{21} ] \,,
  \end{split} \\
  \begin{split}
    \label{eq:hoft-cross}
    \hcross(t) &= -2 C_{22} \cos\iota \sin[ 2\phi\rot(t) + \PhiC_{22} ] \\
    &\quad -\frac{1}{2} C_{21} \sin\iota \sin[ \phi\rot(t) + \PhiC_{21} ] \,.
  \end{split}
\end{align}
The inclination angle $\iota$ is measured from the line of sight
(Fig.~\ref{fig:cw-signal-model}) and its angular momentum vector (which is
assumed parallel to the $Z$ axis).  The CW signal contains two
harmonics, with amplitudes $C_{22}, C_{21}$ and initial phases
$\PhiC_{22}, \PhiC_{21}$, corresponding to the $l=2, m=2$ and $l=2, m=1$
spherical harmonic components of the mass quadrupole, respectively. The
time-dependent phase of the signal scales as $\phi\rot(t) \sim 2 \pi \frot$
where $\frot$ is the rotation frequency of the star
(Sec.~\ref{sec:cw-signal-phase}); the two harmonics are therefore at
approximately once and twice the star's rotation
frequency. Equations~\eqref{eq:hoft-plus} and~\eqref{eq:hoft-cross} give the
most general form of the signal, and may be
specialised~\cite{Chau1970-GrvRdtObRtMd, ZimmSzed1979-GrvWRtPrRBSMAppPl,
  Zimm1980-GrvWRtPrRBIIGSlCmpUF-II, BonaGour1996-GrvWPlEmMgFIDst,
  JaraEtAl1998-DAnGrvSgSpNSSDtc} to particular relationships between
$(\Ixx, \Iyy, \Izz)$ and orientations of $(x, y, z)$, as discussed in this
section.

\begin{figure}[t]%
  \centering%
  \includegraphics[width=0.95\linewidth]{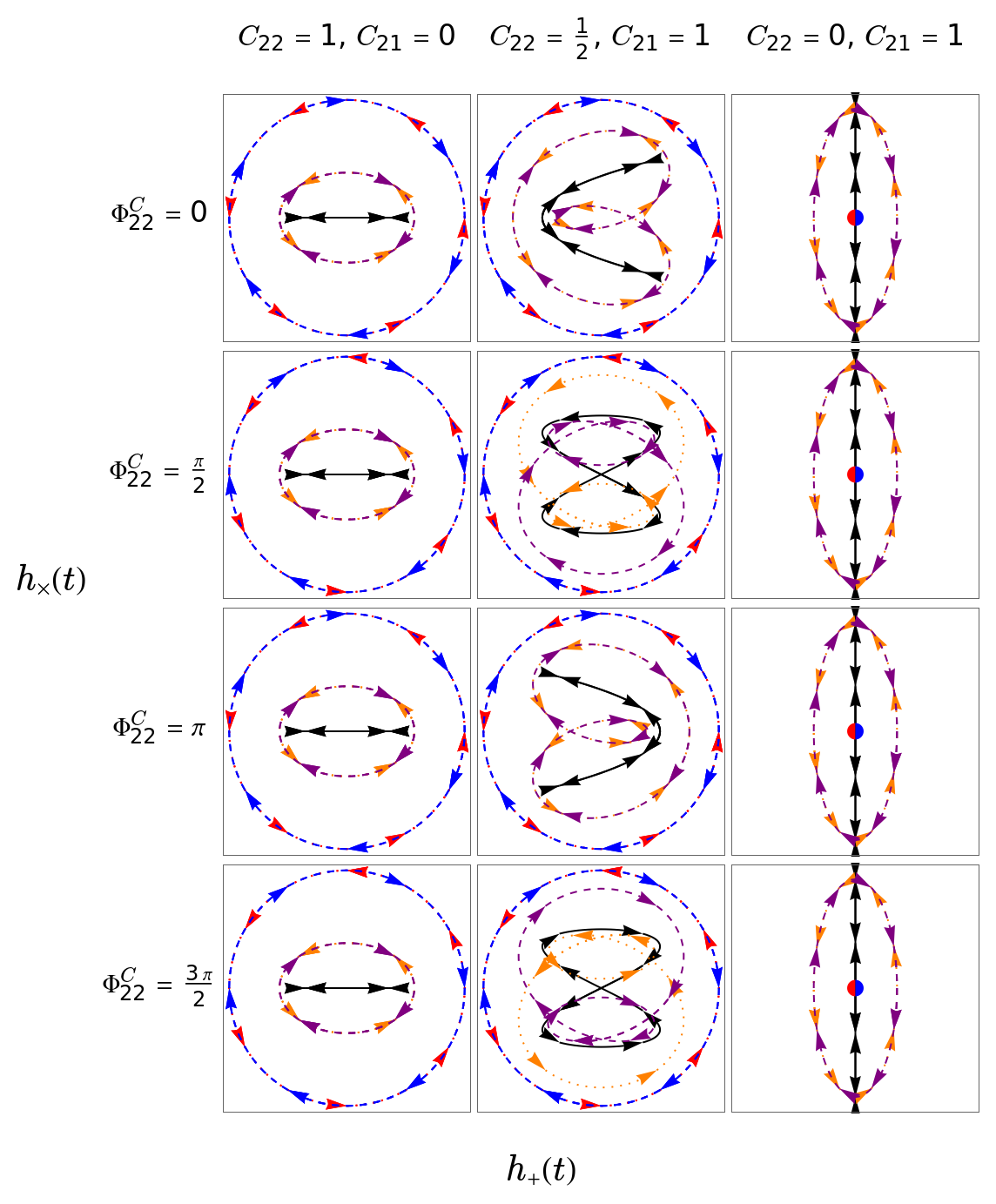}%
  \caption{\label{fig:cw-amplitudes-polarisation}%
    CW signal polarisations $\hcross(t)$ versus $\hplus(t)$, for selected values
    of $C_{22}, C_{21}$ (columns), and of $\PhiC_{22}$ (rows), with
    $\PhiC_{21} = 0$, and with $\iota = 0$ (red, dotted), $\iota = 3\pi/8$
    (orange, dotted), $\iota = \pi/2$ (black, solid), $\iota = 5\pi/8$ (purple,
    dashed), and $\iota = \pi$ (blue, dashed). Arrows indicate the direction of
    time.  Note that the scales of the plots in each column are 4 to 2 to 1
    (left to right).%
  }%
\end{figure}

Figure~\ref{fig:cw-amplitudes-polarisation} illustrates the polarisation state
of the CW signal. We see that the $C_{22}$ harmonic generates
linearly\-/polarised waves when $\iota = \pi/2$, right\-/hand (left\-/hand)
circularly\-/polarised waves when $\iota = 0$ ($\iota = \pi$), and
elliptically\-/polarised waves at other values of $\iota$. In contrast, the
$C_{21}$ harmonic disappears when $\iota \in \{0, \pi\}$, and is linearly
polarised (when $\iota = \pi/2$) at right angles to the $C_{22}$ harmonic. When
both harmonics are present, emission is still circularly polarised at
$\iota \in \{0, \pi\}$ (as the $C_{21}$ harmonic disappears here), and is
otherwise non-elliptical in nature.

Table~\ref{tab:cw-amp} lists expressions for $C_{22}$, $C_{21}$, $\PhiC_{22}$,
and $\PhiC_{21}$~\cite{Jone2015-PrCRCntGrvWSrStSNS,
  PitkEtAl2015-FRFPrDlhSrGrvWSpNS}.  I have given these expressions in terms of
the following quantities (cf.~\cite{JaraEtAl1998-DAnGrvSgSpNSSDtc,
  BejgKrol2014-SrcGrvWvKPlTSFr}):
\begin{align}
  \label{eq:h0}
  h_0 &= \frac{4 \pi ^2 G}{c^4 \dist} \Izz \epsilon f^2 \,, & f &= 2 \frot \,, \\
  \label{eq:hp}
  h_p &= \frac{16 \pi ^2 G}{c^4 \dist} \Izz \epsilon_p \frot^2 \,.
\end{align}
Equation~\eqref{eq:h0} is familiar from~\cite{JaraEtAl1998-DAnGrvSgSpNSSDtc} as
the characteristic CW amplitude in the most commonly assumed triaxial aligned
case (see below). Note that $h_0$ is occasionally re-defined in terms of
$\epsilon_p$~\cite[e.g.][]{JaraEtAl1998-DAnGrvSgSpNSSDtc,
  BejgKrol2014-SrcGrvWvKPlTSFr}; in this review I define $h_0$ solely as in
Eq.~\eqref{eq:h0}, and introduce $h_p$ as the equivalent amplitude in terms of
$\epsilon_p$. Similarly, in this review $f$ (often referred to as the
``gravitational wave frequency'' in the context of the triaxial aligned case) is
always equal to twice the rotation frequency $\frot$.

The top panel in Table~\ref{tab:cw-amp} is for the general case of a
\emph{triaxial non-aligned} star, where $\Ixx \ne \Iyy \ne \Izz$, and the body
frame $(x, y, z)$ is orientated arbitrarily with respect to the inertial frame
$(X, Y, Z)$. This orientation is specified by three Euler angles
(\cite{Jone2010-GrvWEmRtSprNtS, Jone2015-PrCRCntGrvWSrStSNS,
  PitkEtAl2015-FRFPrDlhSrGrvWSpNS}, see Fig.~\ref{fig:cw-signal-model}):
$\phiEuler$~\footnote{I use $\phiEuler$ for this angle in preference to the
  $\phi_0$ of~\cite{Jone2010-GrvWEmRtSprNtS, Jone2015-PrCRCntGrvWSrStSNS,
    PitkEtAl2015-FRFPrDlhSrGrvWSpNS} to avoid confusion, since this convention
  differs from other conventions~\cite[e.g.][]{JaraEtAl1998-DAnGrvSgSpNSSDtc,
    Prix2007-SrCnGrvWMMltFs} which also use $\phi_0$ or similar notations.}
specifies an initial rotation about $z$, $\theta$ gives the inclination of $z$
with respect to $Z$, and $\psiEuler$~\footnote{I use $\psiEuler$ for this angle
  in preference to the $\psi$ of~\cite{Jone2010-GrvWEmRtSprNtS,
    Jone2015-PrCRCntGrvWSrStSNS} and the $\lambda$
  of~\cite{PitkEtAl2015-FRFPrDlhSrGrvWSpNS}, to avoid confusion in
  Fig.~\ref{fig:cw-signal-model} with the polarisation angle and the detector
  latitude respectively.} gives the orientation of the $x$--$y$ plane about $z$
with respect to the $X$--$Y$ plane.

\begin{figure*}[!t]%
  \centering%
  \includegraphics[width=0.95\linewidth]{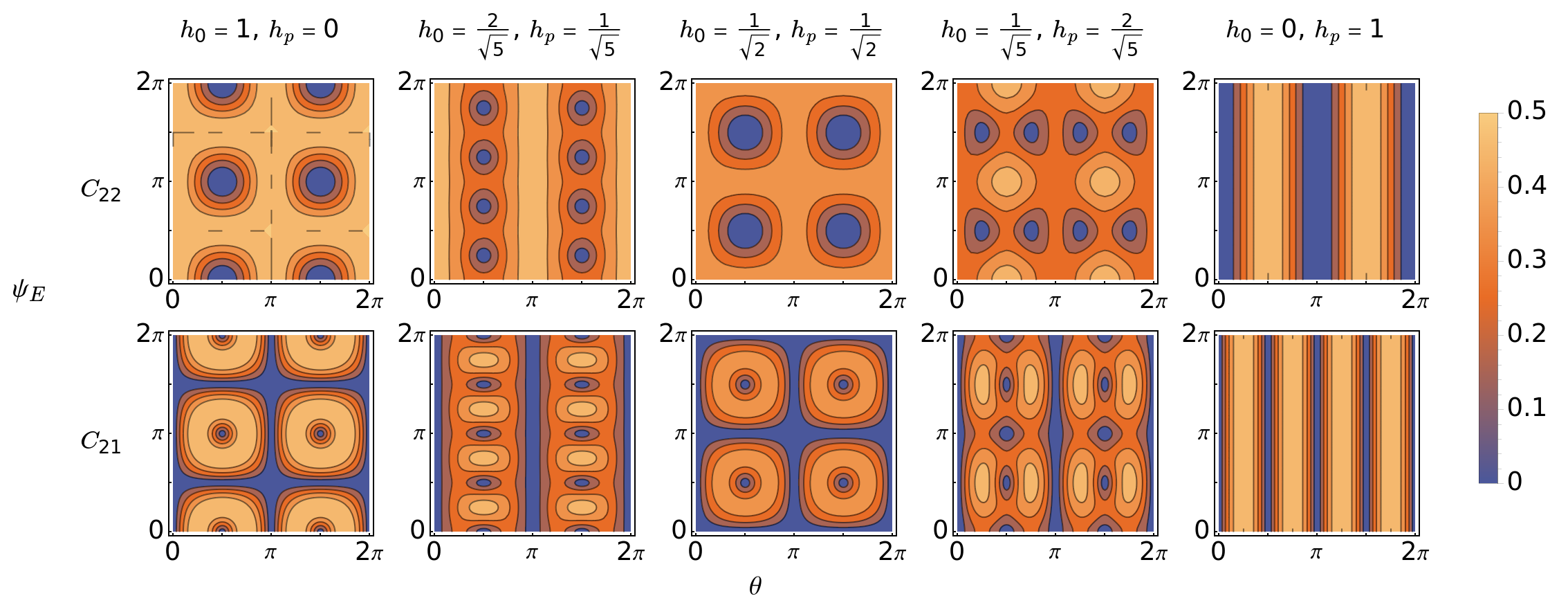}%
  \caption{\label{fig:cw-amplitudes-C22-C21}%
    CW signal amplitudes $C_{22}$ (top row), $C_{21}$ (bottom row) as functions
    of $\theta, \psiEuler$ for selected values of $h_0, h_p$ (columns).%
  }%
\end{figure*}

A triaxial non-aligned star may in general radiate CWs at both $f = 2\frot$ and
$\frot$, depending on their respective amplitudes $C_{22}$ and $C_{21}$
[Table~\ref{tab:cw-amp},
Eqs.~\eqref{eq:cw-amp-general-C}]. Figure~\ref{fig:cw-amplitudes-C22-C21} shows
$C_{22}, C_{21}$ as functions of $\theta$ and $\psiEuler$, for 5 choices of
$h_0, h_p$: $h_0$ only, $h_0 = 2 h_p$, $h_0 = h_p$, $h_p = 2 h_0$, and $h_p$
only. (The $h_0, h_p$ are normalised so that $h_0^{2} + h_p^{2} = 1$.)  We see
that $C_{22}, C_{21}$ exhibit periodicities over $\theta, \psiEuler$, with
periods~\cite{Jone2015-PrCRCntGrvWSrStSNS, PitkEtAl2015-FRFPrDlhSrGrvWSpNS}
depending on the relative contribution to the CW signal from equatorial
($h_0 \propto \epsilon$) and poloidal ($h_p \propto \epsilon_p$) deformations.
The gravitational wave luminosity from a triaxial non-aligned star is given by
Eq.~\eqref{eq:cw-amp-general-Lgw}; for equivalent amplitudes
($C_{22} \approx C_{21}$), the $f$ harmonic is much more efficient at radiating
energy than the $\frot$ harmonic~\cite{UshoEtAl2000-DfrAcNtSCGrvWEm}. (Note
that, as $C_{22}, C_{21} \propto 1/D$, $\Lgw$ is independent of distance.)

\begin{figure}[t]%
  \centering%
  \includegraphics[width=0.95\linewidth]{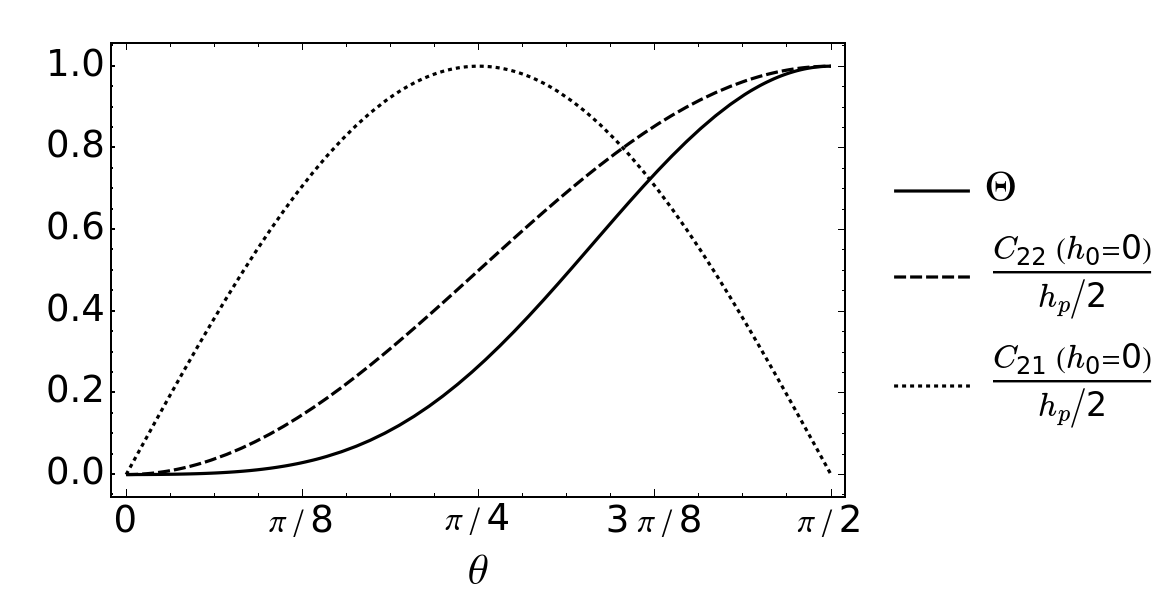}%
  \caption{\label{fig:cw-amplitudes-Theta-C21-C22-vs-theta}%
    $C_{22}$, $C_{21}$, and $\Theta$ for a biaxial star as functions of
    $\theta$.%
  }%
\end{figure}

When $h_0$ is zero, $C_{22}, C_{21}$ are independent of $\psiEuler$
(Fig.~\ref{fig:cw-amplitudes-C22-C21}, rightmost column). This gives the case of
a \emph{biaxial} star (2nd panel, Table~\ref{tab:cw-amp}), which is only
deformed poloidally. Here the amplitudes $C_{22}$ and $C_{21}$
[Eqs.~\eqref{eq:cw-amp-biax-C}] simplify to expressions involving only $h_p$ and
$\theta$. Figure~\ref{fig:cw-amplitudes-Theta-C21-C22-vs-theta} plots
$C_{22}, C_{21}$ (normalised by $h_p/2$) as a function of $\theta$. Emission at
$\frot$ is maximal at $\theta = \pi/4$, but disappears at $\theta = \pi/2$;
emission at $f$, however, grows monotonically with $\theta$. The gravitational
wave luminosity is proportional to a quantity
$\Theta(\theta)$~[Eqs.~\eqref{eq:cw-amp-biax-Lgw}], plotted in
Fig.~\ref{fig:cw-amplitudes-Theta-C21-C22-vs-theta}, and is maximal when
$\Theta = 1$ at $\theta = \pi/2$.

When one of the body frame axes (conventionally $z$) aligns with the inertial
frame $Z$ axis, $\theta = 0$ and the star rotates about a principal moment of
inertia. Such a \emph{triaxial aligned} star emits only at $f$ (3rd panel,
Table~\ref{tab:cw-amp}); $C_{21} = 0$, and $C_{22}$ is given in terms of the
familiar $h_0$. Were such a star also biaxial, rotations about $z = Z$ would be
axisymmetric, and there would be no CW emission.

Different conventions exist for defining an overall initial phase. For the
biaxial case, Eq.~\eqref{eq:cw-amp-biax-PhiC} (Table~\ref{tab:cw-amp}) relates
$\phiEuler$~\cite[Eqs.~(A9), (A12), and~(A16)]{PitkEtAl2015-FRFPrDlhSrGrvWSpNS}
to the $\Phi_0$ of~\cite{JaraEtAl1998-DAnGrvSgSpNSSDtc}. Note that
$\phiEuler + \pi/2$ and $\Phi_0$ differ by a factor of $\pi$, consistent with
the overall negative sign of Eqs.~\eqref{eq:hoft-plus} and~\eqref{eq:hoft-cross}
compared to~\cite[Eqs.~(21) and~(22)]{JaraEtAl1998-DAnGrvSgSpNSSDtc}.  For the
triaxial aligned case, Eq.~\eqref{eq:cw-amp-trialign-PhiC} relates $\phiEuler$ to
the $\phi_0$ of \cite{Prix2007-SrCnGrvWMMltFs}.

Equations for the CW amplitudes are often conveniently written with fiducial
values for their respective factors. In this review I include a comprehensive
set of such expressions, with consistent fiducial values for each factor across
all expressions.\footnote{Due to the similarities between Eqs.~\eqref{eq:h0}
  and~\eqref{eq:hp}, and the consistent choice of fiducial values (e.g.\
  $\frot = 150$~Hz, $f = 2\frot = 300$~Hz), the fiducial equations for
  $h_0, \epsilon$ will usually be similar to those for $h_p, \epsilon_p$;
  nevertheless I include both sets of expressions for clarity.} (A Python
package~\cite{cweqgen} also exists for creating and manipulating these type of
equations.)  Due to typical rounding of fiducial values to a few significant
figures, these expressions are less accurate than their exact equivalent
equations.  Fiducial equations in terms of $D, \Izz, f, \frot$ are given for
$h_0, h_p$:
\begin{align}
\begin{split}
\label{eq:GW_amplitudes_Triaxial_aligned_v_h0_D_f_Izz_epsilon}
h_0 &= 10^{-25}\left(\frac{1\text{ kpc}}{D}\right)\left(\frac{\epsilon }{1.05{\times}10^{-6}}\right)\\
   &\times \left(\frac{\Izz}{10^{38}\text{ kg m${}^2$}}\right)\left(\frac{f}{300\text{ Hz}}\right)^2\,,
\end{split}
 \\
\begin{split}
\label{eq:GW_amplitudes_Biaxial_v_hp_D_frot_Izz_epsilon_p}
h_p &= 10^{-25}\left(\frac{1\text{ kpc}}{D}\right)\left(\frac{\epsilon_p}{1.05{\times}10^{-6}}\right)\\
   &\times \left(\frac{\Izz}{10^{38}\text{ kg m${}^2$}}\right)\left(\frac{\frot}{150\text{ Hz}}\right)^2\,;
\end{split}
 %semicolon
\end{align}
for $\epsilon, \epsilon_p$:
\begin{align}
\begin{split}
\label{eq:GW_amplitudes_Triaxial_aligned_v_elleq_D_f_h_0_Izz}
\epsilon &= 1.05{\times}10^{-6}\left(\frac{300\text{ Hz}}{f}\right)^2\left(\frac{10^{38}\text{ kg m${}^2$}}{\Izz}\right)\\
   &\times \left(\frac{D}{1\text{ kpc}}\right)\left(\frac{h_0}{10^{-25}}\right)\,,
\end{split}
 \\
\begin{split}
\label{eq:GW_amplitudes_Biaxial_v_ellpol_D_frot_h_p_Izz}
\epsilon_p &= 1.05{\times}10^{-6}\left(\frac{150\text{ Hz}}{\frot}\right)^2\\
   &\times \left(\frac{10^{38}\text{ kg m${}^2$}}{\Izz}\right)\left(\frac{D}{1\text{ kpc}}\right)\left(\frac{h_p}{10^{-25}}\right)\,;
\end{split}
 %semicolon
\end{align}
and for the gravitational wave luminosity:
\begin{align}
\begin{split}
\label{eq:Upper_limits_Triaxial_aligned_v_GW_luminosity_triax_f_Izz_epsilon}
\Lgw &= 1.37{\times}10^{30}\left(\frac{\epsilon }{1.05{\times}10^{-6}}\right)^2\\
   &\times \left(\frac{\Izz}{10^{38}\text{ kg m${}^2$}}\right)^2\left(\frac{f}{300\text{ Hz}}\right)^6 \text{W}\,,
\end{split}
 \\
\begin{split}
\label{eq:Upper_limits_Biaxial_v_GW_luminosity_biax_frot_Izz_epsilon_p_Theta}
\Lgw &= 1.37{\times}10^{30}\left(\frac{\Theta }{1}\right)\left(\frac{\epsilon_p}{1.05{\times}10^{-6}}\right)^2\\
   &\times \left(\frac{\Izz}{10^{38}\text{ kg m${}^2$}}\right)^2\left(\frac{\frot}{150\text{ Hz}}\right)^6 \text{W}\,.
\end{split}
\end{align}

Upper limits on the gravitational wave amplitude may be derived by assuming that
gravitational radiation is driven by the rotation of the
star~\cite{CondRans2016-EssRdAst}. As angular momentum is carried away by
gravitational waves, the rotational kinetic energy of the star decreases, and
hence its rotation frequency decreases over time. It follows that the
gravitational wave luminosity can be no greater than the time derivative of the
star's rotational kinetic energy, and this therefore limits the gravitational
wave amplitude.

The rotational kinetic energy of a triaxial star is
\begin{multline}
  \label{eq:Erot}
  \Erot = 2 \pi^2 \Big[ \big( \Ixx \sin^2\psiEuler + \Iyy \cos^2\psiEuler \big) \sin^2\theta \\
  + \Izz \cos^2\theta \Big] \frot^2 \,.
\end{multline}
In the limit of small $\epsilon, \epsilon_p$, however, we may approximate
$\Ixx \approx \Iyy \approx \Izz$, and hence
$\Erot \approx 2 \pi^2 \Izz \frot^2$. The upper limit is then implied by
\begin{align}
  \label{eq:spindown-ul}
  \Lgw &\leq -\dErot = -4 \pi^2 \Izz \frot \fdrot \,,
\end{align}
where $\fdrot = d\frot/dt$ is the (first) \emph{spindown} or time derivative of
$\frot$. Equation~\eqref{eq:spindown-ul} defines the \emph{spindown upper limit}
on gravitational wave amplitude. Substituting
Eqs.~\eqref{eq:Upper_limits_Triaxial_aligned_v_GW_luminosity_triax_f_Izz_epsilon}
and~\eqref{eq:Upper_limits_Biaxial_v_GW_luminosity_biax_frot_Izz_epsilon_p_Theta}
into Eq.~\eqref{eq:spindown-ul} gives fiducial equations in terms of $\fdrot$
and $\fd = 2 \fdrot$ for $h_0, h_p$:
\begin{align}
\begin{split}
\label{eq:Upper_limits_Triaxial_aligned_v_h0_D_f_dot_f_Izz}
h_0 &\leq 10^{-25}\left(\frac{1\text{ kpc}}{D}\right)\left(\frac{300\text{ Hz}}{f}\right)^{\frac{1}{2}}\\
   &\times \left(\frac{\dot{f}}{-4.61{\times}10^{-12}\text{ Hz s${}^{-1}$}}\right)^{\frac{1}{2}}\left(\frac{\Izz}{10^{38}\text{ kg m${}^2$}}\right)^{\frac{1}{2}}\,,
\end{split}
 \\
\begin{split}
\label{eq:Upper_limits_Biaxial_v_hp_D_fdrot_frot_Izz_Theta}
h_p &\leq 10^{-25}\left(\frac{1\text{ kpc}}{D}\right)\left(\frac{150\text{ Hz}}{\frot}\right)^{\frac{1}{2}}\left(\frac{1}{\Theta }\right)^{\frac{1}{2}}\\
   &\times \left(\frac{\fdrot}{-2.31{\times}10^{-12}\text{ Hz s${}^{-1}$}}\right)^{\frac{1}{2}}\left(\frac{\Izz}{10^{38}\text{ kg m${}^2$}}\right)^{\frac{1}{2}}\,;
\end{split}
 %semicolon
\end{align}
and for $\epsilon, \epsilon_p$:
\begin{align}
\begin{split}
\label{eq:Upper_limits_Triaxial_aligned_v_elleq_f_dot_f_Izz}
\epsilon &\leq 1.05{\times}10^{-6}\left(\frac{300\text{ Hz}}{f}\right)^{\frac{5}{2}}\left(\frac{10^{38}\text{ kg m${}^2$}}{\Izz}\right)^{\frac{1}{2}}\\
   &\times \left(\frac{\dot{f}}{-4.61{\times}10^{-12}\text{ Hz s${}^{-1}$}}\right)^{\frac{1}{2}}\,,
\end{split}
 \\
\begin{split}
\label{eq:Upper_limits_Biaxial_v_ellpol_fdrot_frot_Izz_Theta}
\epsilon_p &\leq 1.05{\times}10^{-6}\left(\frac{150\text{ Hz}}{\frot}\right)^{\frac{5}{2}}\left(\frac{10^{38}\text{ kg m${}^2$}}{\Izz}\right)^{\frac{1}{2}}\\
   &\times \left(\frac{1}{\Theta }\right)^{\frac{1}{2}}\left(\frac{\fdrot}{-2.31{\times}10^{-12}\text{ Hz s${}^{-1}$}}\right)^{\frac{1}{2}}\,.
\end{split}
\end{align}
Given an observed value $h^{\mathrm{observed}}$ of either $h_0$ or $h_p$, the
maximum fraction of energy lost in gravitational waves is
\begin{equation}
  \frac{ \Lgw }{ \dErot } = \left( \frac{ h^{\mathrm{observed}} }{ h^{\mathrm{spindown}} } \right)^2 \,,
\end{equation}
where $h^{\mathrm{spindown}}$ is the right-hand side of either
Eqs.~\eqref{eq:Upper_limits_Triaxial_aligned_v_h0_D_f_dot_f_Izz}
or~\eqref{eq:Upper_limits_Biaxial_v_hp_D_fdrot_frot_Izz_Theta} respectively.

The fiducial equations above may be applied when $\fdrot$ is known, e.g.\
through electromagnetic observations of a pulsar. When $\fdrot$ is unknown, it
may be inferred from the following relations~\cite{CondRans2016-EssRdAst,
  WettEtAl2008-SrGrvWvCssLI}:
\begin{align}
  \label{eq:tau-n-frot}
  \tau &= \frac{1}{n - 1} \left( \frac{\frot}{-\fdrot} \right) \,, & n &= \frac{ \frot \fddrot }{ \fdrot^2 } \,, \\
  \label{eq:tau-n-f}
       &= \frac{1}{n - 1} \left( \frac{f}{-\fd} \right) \,, & n &= \frac{ f \fdd }{ \fd^2 } \,.
\end{align}
The \emph{characteristic age} $\tau$ is usually approximated by the astronomical
age of the star in question, e.g.\ a neutron star born in a supernova. The
\emph{braking index} $n$ appears when Eq.~\eqref{eq:spindown-ul} is rearranged
to solve for $\fdrot \propto \frot^{n}$. From theory, we expect $n = 5$ when
energy is lost purely through for gravitational wave emission from a mass
quadrupole [Eq.~\eqref{eq:cw-amp-general-Lgw}] and $n = 3$ when energy is lost
solely through electromagnetic radiation. (In practice, values of $n$ measured
for known pulsars vary widely, as discussed in~\cite{HobbEtAl2004-LngTmObs374Pl,
  HobbEtAl2010-AnlTmIrr366Pl, PonsEtAl2012-PlTIrrImMgFEvl,
  ZhanXie2012-WDBrkInPlSRnM100Mll}.)

Substituting Eq.~\eqref{eq:tau-n-frot},~\eqref{eq:tau-n-f} into
Eqs.~\eqref{eq:Upper_limits_Triaxial_aligned_v_h0_D_f_dot_f_Izz}%
--\eqref{eq:Upper_limits_Biaxial_v_ellpol_fdrot_frot_Izz_Theta} gives fiducial
equations in terms of $\tau$ and $n$ for $h_0, h_p$:
\begin{align}
\begin{split}
\label{eq:Upper_limits_Triaxial_aligned_v_h0_D_n_1_Izz_tau}
h_0 &\leq 10^{-25}\left(\frac{1\text{ kpc}}{D}\right)\left(\frac{516\text{ kyr}}{\tau }\right)^{\frac{1}{2}}\\
   &\times \left(\frac{4}{\text{n}-1}\right)^{\frac{1}{2}}\left(\frac{\Izz}{10^{38}\text{ kg m${}^2$}}\right)^{\frac{1}{2}}\,,
\end{split}
 \\
\begin{split}
\label{eq:Upper_limits_Biaxial_v_hp_D_n_1_Izz_Theta_tau}
h_p &\leq 10^{-25}\left(\frac{1\text{ kpc}}{D}\right)\left(\frac{516\text{ kyr}}{\tau }\right)^{\frac{1}{2}}\\
   &\times \left(\frac{4}{\text{n}-1}\right)^{\frac{1}{2}}\left(\frac{1}{\Theta }\right)^{\frac{1}{2}}\left(\frac{\Izz}{10^{38}\text{ kg m${}^2$}}\right)^{\frac{1}{2}}\,;
\end{split}
 %semicolon
\end{align}
and for $\epsilon, \epsilon_p$:
\begin{align}
\begin{split}
\label{eq:Upper_limits_Triaxial_aligned_v_elleq_f_n_1_Izz_tau}
\epsilon &\leq 1.05{\times}10^{-6}\left(\frac{300\text{ Hz}}{f}\right)^2\left(\frac{516\text{ kyr}}{\tau }\right)^{\frac{1}{2}}\\
   &\times \left(\frac{10^{38}\text{ kg m${}^2$}}{\Izz}\right)^{\frac{1}{2}}\left(\frac{4}{\text{n}-1}\right)^{\frac{1}{2}}\,,
\end{split}
 \\
\begin{split}
\label{eq:Upper_limits_Biaxial_v_ellpol_n_1_frot_Izz_Theta_tau}
\epsilon_p &\leq 1.05{\times}10^{-6}\left(\frac{150\text{ Hz}}{\frot}\right)^2\left(\frac{516\text{ kyr}}{\tau }\right)^{\frac{1}{2}}\\
   &\times \left(\frac{10^{38}\text{ kg m${}^2$}}{\Izz}\right)^{\frac{1}{2}}\left(\frac{4}{\text{n}-1}\right)^{\frac{1}{2}}\left(\frac{1}{\Theta }\right)^{\frac{1}{2}}\,.
\end{split}
\end{align}

\subsubsection{Free precession}\label{sec:free-precession}

Free precession of a biaxial neutron star occurs when the star's moment of
inertia changes with time, which (assuming conservation of angular momentum)
causes its angular velocity to also change with time.  With reference to
Fig.~\ref{fig:cw-signal-model}, while the angular momentum vector remains
aligned with the $Z$ axis, the angular velocity vector does not, and moreover
exhibits a superimposed rotation about the star's axis of symmetry (the $z$
axis). The general effect of this rotation on the CW
waveform~\cite{ZimmSzed1979-GrvWRtPrRBSMAppPl, JoneAnde2001-FrPrcNtStMObs,
  JoneAnde2002-GrvWvFrPrcNtS, JaraEtAl1998-DAnGrvSgSpNSSDtc,
  VanD2005-GrvWSpNnxFPrNtS} is the addition of the (generally small) precession
frequency to the CW harmonic frequencies $\frot$ and $f$. The precessional
rotation does not modify the mass quadrupole, and hence does not appear in the
harmonic amplitudes (Table~\ref{tab:cw-amp}).

\subsubsection{$r$-mode oscillations}\label{sec:r-mode-oscillations}

Neutron stars may exhibit various perturbations from their equilibrium
shape. The energy dissipated as the perturbation decays is described by the
quasi-normal modes of the star~\cite{KokkSchm1999-QsNMdsStBlHl}. $r$-mode oscillations are
one of a family of quasi-normal modes, which exist only when the star is
rotating, as it is the Coriolis force acting to restore the star to its
equilibrium shape.

Long-lived CW emission may by driven by $r$-modes~\cite{Ande1998-NClUnsMdRtRltS,
  LindEtAl1998-GrvRdInsHYnNtS} through the Chandrasekhar-Friedman-Schutz
instability~\cite{Chan1970-SltTPrThGrvRd, FrieSchu1978-ScInsRtNwtSt}, as follows
(cf.~\cite{JonesCFS}). An oscillation may propagate either in the same direction
as the rotation of the star, or in the opposite direction. An oscillation
propagating counter to the rotation of the star, but with a slower absolute
angular frequency, will still appear to be co-rotating to a inertial
observer. Because the mode rotates counter to the star, it contributes
\emph{negatively} to the star's angular momentum in the co-rotating frame. On
the other hand, because the mode co-rotates with the star to a inertial
observer, it contributes \emph{positively} to the star's angular momentum in the
inertial frame, and will hence radiate gravitational wave with \emph{positive}
angular momentum. The \emph{negative} angular momentum of a mode which loses
\emph{positive} angular momentum in gravitational waves therefore become
\emph{more} negative over time, resulting in a positive feedback loop which
promotes the growth of the mode over time. Eventually, in real stars, the mode
is expected to saturate due to viscous or other effects, which have been
extensively studied in~\cite{OwenEtAl1998-GrvWvHYRpRttNtS,
  LindEtAl2000-EfNtrCrRmIns, BildUsho2000-VscBndDmRMNtS,
  RezzEtAl2001-PrpRMRtMNSIKnmSEMEvEq, RezzEtAl2001-PrprMRtMgNSIIEvlrMSMgF-II,
  WuEtAl2001-RNtSCTrbStrSpCMl, BondEtAl2009-SpNwNSNnDvlRIns,
  AndeEtAl2010-RmLTmClrSprQrS, BondWass2013-NnDvlRInsMxRtRNS,
  AlfoSchw2014-GrvWEmsSpYnPl, AlfoSchw2015-GrvWEmOscMlPl,
  FrieEtAl2016-DffRtUnNnlRm}.

Some important differences distinguish CW emission from $r$-modes (bottom panel,
Table~\ref{tab:cw-amp}) and from a rigidly rotating star. CW emission from
$r$-modes is predominately through a time-varying current quadrupole, as opposed
to a mass quadrupole; this leads to a different basis for the wave frame
$(\tilde{x}, \tilde{y}, \tilde{z})$ (Fig.~\ref{fig:cw-signal-model}) than for a
rigidly rotating star. Data analysis algorithms designed for the model of
Fig.~\ref{fig:cw-signal-model} are still applicable, however, with a simple
reinterpretation of the polarisation angle
$\psi$~\cite[Eq.~(18)]{Owen2010-HAdBrdGrvSrRm}. In addition, the gravitational
wave luminosity [Eq.~\eqref{eq:cw-amp-rmode-Lgw}] of a current quadrupole scales
more steeply with frequency, leading to a characteristic braking index of
$n = 7$ instead of $5$. Finally, the gravitational wave frequency is
approximately $\frmode \approx 4 \frot / 3$, with corrections depending on the
nuclear equation of state of the neutron
star~\cite{IdriEtAl2015-RMFrqSRtRltNSRlEqS, CariEtAl2019-HSrGrvWvRmKPls}.

Fiducial equations relevant to $r$-mode emission are given for: the CW
amplitude\footnote{The CW amplitude for $r$-mode emission is generally labelled
  $h_0$ in the literature; in this review I use $\hrmode$ for clarity.}
\begin{align}
\begin{split}
\label{eq:GW_amplitudes_R_mode_v_hrmode_D_M_R_f_alpha_tilde_J_alpha}
h_{\alpha } &= 10^{-25}\left(\frac{1\text{ kpc}}{D}\right)\left(\frac{\alpha }{3.44{\times}10^{-4}}\right)\\
   &\times \left(\frac{\tilde{J}}{1.64{\times}10^{-2}}\right)\left(\frac{M}{1.4 M_{\odot}}\right)\left(\frac{f_{\alpha }}{200\text{ Hz}}\right)^3\\
   &\times \left(\frac{R}{11.7\text{ km}}\right)^3\,;
\end{split}
 \\ %semicolon
  \intertext{a dimensionless amplitude $\alpha$ which sets the magnitude of the $r$-mode velocity perturbation field,}
\begin{split}
\label{eq:GW_amplitudes_R_mode_v_alpharmode_D_M_R_f_alpha_h_alpha_tilde_J}
\alpha &= 3.44{\times}10^{-4}\left(\frac{200\text{ Hz}}{f_{\alpha }}\right)^3\left(\frac{11.7\text{ km}}{R}\right)^3\\
   &\times \left(\frac{1.64{\times}10^{-2}}{\tilde{J}}\right)\left(\frac{1.4 M_{\odot}}{M}\right)\\
   &\times \left(\frac{D}{1\text{ kpc}}\right)\left(\frac{h_{\alpha }}{10^{-25}}\right)\,;
\end{split}
 \\ %semicolon
  \intertext{and the gravitational wave luminosity}
\begin{split}
\label{eq:Upper_limits_R_mode_v_GW_luminosity_rmode_M_R_f_alpha_tilde_J_alpha}
\Lgw &= 6.07{\times}10^{29}\left(\frac{\alpha }{3.44{\times}10^{-4}}\right)^2\left(\frac{\tilde{J}}{1.64{\times}10^{-2}}\right)^2\\
   &\times \left(\frac{M}{1.4 M_{\odot}}\right)^2\left(\frac{R}{11.7\text{ km}}\right)^6\left(\frac{f_{\alpha }}{200\text{ Hz}}\right)^8 \text{W}\,.
\end{split}
\end{align}
In these equations, $M$ and $R$ are the mass and radius of the neutron star, and
$\tilde{J}$ is determined by the neutron star equation of
state~\cite{Owen2010-HAdBrdGrvSrRm}.  Further expressions give
$\hrmode, \alpha$ in terms of the spindown $\fdrmode = 4 \fdrot / 3$:
\begin{align}
\begin{split}
\label{eq:Upper_limits_R_mode_v_hrmode_D_f_alpha_Izz_dot_f_alpha}
h_{\alpha } &\leq 10^{-25}\left(\frac{1\text{ kpc}}{D}\right)\left(\frac{200\text{ Hz}}{f_{\alpha }}\right)^{\frac{1}{2}}\\
   &\times \left(\frac{\dot{f}_{\alpha }}{-1.37{\times}10^{-12}\text{ Hz s${}^{-1}$}}\right)^{\frac{1}{2}}\left(\frac{\Izz}{10^{38}\text{ kg m${}^2$}}\right)^{\frac{1}{2}}\,,
\end{split}
 \\
\begin{split}
\label{eq:Upper_limits_R_mode_v_alpharmode_M_R_f_alpha_Izz_dot_f_alpha_tilde_J}
\alpha &\leq 3.44{\times}10^{-4}\left(\frac{200\text{ Hz}}{f_{\alpha }}\right)^{\frac{7}{2}}\left(\frac{11.7\text{ km}}{R}\right)^3\\
   &\times \left(\frac{1.64{\times}10^{-2}}{\tilde{J}}\right)\left(\frac{1.4 M_{\odot}}{M}\right)\\
   &\times \left(\frac{\dot{f}_{\alpha }}{-1.37{\times}10^{-12}\text{ Hz s${}^{-1}$}}\right)^{\frac{1}{2}}\left(\frac{\Izz}{10^{38}\text{ kg m${}^2$}}\right)^{\frac{1}{2}}\,;
\end{split}
 %semicolon
\end{align}
and in terms of braking index $n = 7$ and characteristic age $\tau = (\frmode / -\fdrmode) / (n - 1)$:
\begin{align}
\begin{split}
\label{eq:Upper_limits_R_mode_v_hrmode_D_n_1_Izz_tau}
h_{\alpha } &\leq 10^{-25}\left(\frac{1\text{ kpc}}{D}\right)\left(\frac{773\text{ kyr}}{\tau }\right)^{\frac{1}{2}}\\
   &\times \left(\frac{6}{\text{n}-1}\right)^{\frac{1}{2}}\left(\frac{\Izz}{10^{38}\text{ kg m${}^2$}}\right)^{\frac{1}{2}}\,,
\end{split}
 \\
\begin{split}
\label{eq:Upper_limits_R_mode_v_alpharmode_M_n_1_R_f_alpha_Izz_tilde_J_tau}
\alpha &\leq 3.44{\times}10^{-4}\left(\frac{200\text{ Hz}}{f_{\alpha }}\right)^3\left(\frac{11.7\text{ km}}{R}\right)^3\\
   &\times \left(\frac{1.64{\times}10^{-2}}{\tilde{J}}\right)\left(\frac{1.4 M_{\odot}}{M}\right)\left(\frac{773\text{ kyr}}{\tau }\right)^{\frac{1}{2}}\\
   &\times \left(\frac{6}{\text{n}-1}\right)^{\frac{1}{2}}\left(\frac{\Izz}{10^{38}\text{ kg m${}^2$}}\right)^{\frac{1}{2}}\,.
\end{split}
\end{align}

\subsubsection{Accretion from a binary companion}\label{sec:accr-from-binary}

If a neutron star is in orbit about a companion star, it may accrete matter from
its companion at certain times during the lifetime of the binary system. This
behaviour is observed e.g.\ in low-mass X-ray binaries. The accretion of matter
transfers angular momentum to the neutron star, causing it to spin
faster~\cite{Bhatvand1991-FrmEvBnMllRPl}. If, however, the accreted matter
builds a non\-/axisymmetric deformation -- a ``mountain'' -- on the neutron
star, it will radiate rotational energy as gravitational waves and spin down
again~\cite{UshoEtAl2000-DfrAcNtSCGrvWEm, PaynMela2006-FrSGrvRGHydOsMgCMANS,
  VigeMela2008-ThrStMgnCMnAcNS, VigeMela2009-IEDtcGrvRdMgCMAcNS,
  VigeMela2009-RsRlxMgnCnMAcNS, WettEtAl2010-SnMgnCnMnAccNS,
  PriyEtAl2011-QdrMMgnCnMnAcNSEES, HaskEtAl2015-DtcGrvWMnNSAdDtE}.  If these two
processes are in balance with each other, the neutron star will maintain a
near-constant spin frequency. This \emph{torque balance} equilibrium has been
proposed to explain the observed narrow distribution of the rotation periods of millisecond pulsars~\cite{Bild1998-GrvRdtRtAcNtS, AndeEtAl2005-MdSEqlNSLXBnWGrvRd,
  HaskPatr2011-SEqlWGrvWEmCXJ1SJ1, PatrEtAl2012-GrvWvMxSFrqNtS,
  HaskPatr2017-GrvWvSpPSJ10, GittAnde2019-PplSyAcNSEmGrvW,
  Bhat2020-PrmEllNtStPJ1, Chen2020-CnsEllMlPObSpR}.

CW emission from a mountain on an accreting neutron star is generally assumed to
follow the triaxial aligned model (Sec.~\ref{sec:rigidly-rotat-stars}). By
equating the rotational kinetic energy gained through accretion to the energy
lost through CW emission, the torque balance equilibrium implies an upper limit
on $h_0$, analogous to the spindown upper limit.
Following~\cite{ZhanEtAl2021-SrCntGrvWvScXLOD}, fiducial equations for the
torque balance upper limit on $h_0, \epsilon$ are
\begin{align}
\begin{split}
\label{eq:Upper_limits_Accreting_binary_system_v_h0_f_M_R_F_X_X_r_m}
h_0 &\leq 10^{-25}\left(\frac{300\text{ Hz}}{f}\right)^{\frac{1}{2}}\left(\frac{1.4 M_{\odot}}{M}\right)^{\frac{1}{4}}\left(\frac{r_m}{11.7\text{ km}}\right)^{\frac{1}{4}}\\
   &\times \left(\frac{R}{11.7\text{ km}}\right)^{\frac{1}{2}}\left(\frac{F_X\text{/X}}{1.3{\times}10^{-6}\text{ erg cm${}^{-2}$ s${}^{-1}$}}\right)^{\frac{1}{2}}\,,
\end{split}
 \\
\begin{split}
\label{eq:Upper_limits_Accreting_binary_system_v_elleq_D_f_M_R_F_X_X_Izz_r_m}
\epsilon &\leq 1.05{\times}10^{-6}\left(\frac{300\text{ Hz}}{f}\right)^{\frac{5}{2}}\left(\frac{10^{38}\text{ kg m${}^2$}}{\Izz}\right)\\
   &\times \left(\frac{1.4 M_{\odot}}{M}\right)^{\frac{1}{4}}\left(\frac{r_m}{11.7\text{ km}}\right)^{\frac{1}{4}}\left(\frac{R}{11.7\text{ km}}\right)^{\frac{1}{2}}\\
   &\times \left(\frac{F_X\text{/X}}{1.3{\times}10^{-6}\text{ erg cm${}^{-2}$ s${}^{-1}$}}\right)^{\frac{1}{2}}\left(\frac{D}{1\text{ kpc}}\right)\,.
\end{split}
\end{align}
Here $r_m$ is the radius at which the spin-up torque due to accretion is
applied to the neutron star, $X$ is the fraction of the maximum accretion
luminosity which is radiated away as X-rays, and $F_X$ is the observed X-ray
flux.

\subsection{Phase parameters}\label{sec:cw-signal-phase}

The phase of a CW signal is the function $\phi\rot(t)$ that appears in
Eqs.~\eqref{eq:hoft-plus} and~\eqref{eq:hoft-cross}.  Given the common
assumption of a triaxial aligned star (Sec.~\ref{sec:cw-signal-amplitude}),
where CW emission is at the $f$ harmonic only, the phase is most commonly
written as $\phi(t) = 2 \phi\rot(t)$. The instantaneous frequency (of the $f$
harmonic) of the CW signal $f(t)$ is
\begin{equation}
  \label{eq:foft}
  f(t) = \frac{1}{2\pi} \frac{d \phi(t)}{dt} \,.
\end{equation}
Determining the CW signal phase involves considering time standards at three
locations: at the neutron star (Sec.~\ref{sec:time-at-neutron}, at the Solar
System barycentre (SSB; Sec.~\ref{sec:time-at-solar}), and at the detector
(Sec.~\ref{sec:time-at-detector}.

\subsubsection{Time at the neutron star}\label{sec:time-at-neutron}

At the neutron star, where time is measured by $t\NS$, the CW signal frequency
$f(t\NS)$ is generally modelled as a Taylor series, truncated to terms of order
$t\NS^{s\umax}$:
\begin{equation}
  \label{eq:frequency-at-NS}
  \begin{split}
    f(t\NS) &= \sum_{s=0}^{s\umax} f\ndot{s} \frac{ t\NS^{s} }{ s! } \, \\
    &= f + \fd t\NS + \frac{1}{2} \fdd t\NS^2 + \dots
  \end{split}
\end{equation}
Note that the \emph{instantaneous} frequency $f(t\NS)$ is distinct from the
frequency \emph{parameter} $f \equiv f(t\NS = 0)$. It is common to write
$\fd \equiv f\ndot{1}$, $\fdd \equiv f\ndot{2}$, etc.

The $f\ndot{s}$ coefficients represent the intrinsic frequency evolution of the
star, e.g.\ as it spins down due to energy lost through CW emission. The CW
phase at the neutron star is then given by
\begin{equation}
  \label{eq:phase-at-NS}
  \begin{split}
    \phi(t\NS) &= 2\pi \int dt\NS \, f(t\NS) \\
    &= 2\pi \sum_{s=0}^{s\umax} f\ndot{s} \frac{ t\NS^{s+1} }{ (s+1)! } \,.
  \end{split}
\end{equation}
Note that Eq.~\eqref{eq:phase-at-NS} does not include the initial phase
$\phi_0$, which (counter-intuitively) is considered an \emph{amplitude}
parameter; see Sec.~\ref{sec:cw-signal-amplitude}, and also
Sec.~\ref{sec:chall-cont-wave}, Eq.~\eqref{eq:eq:hoft-four-Ampi}.

\subsubsection{Time at the Solar System barycentre}\label{sec:time-at-solar}

The reference frame associated with the SSB takes the standard celestial sphere
reference frame but centres it at the SSB (Fig.~\ref{fig:cw-signal-model}). The
unit vector $\uvec{n}$ points from the SSB to the neutron star, or to the binary
barycentre (BB) if a binary companion is present, and defines the sky position
of the source. It may be written in terms of the source's right ascension
$\alpha$ and declination $\delta$:
\begin{equation}
  \label{eq:n}
  \uvec{n} = \big( \, \cos\alpha \cos\delta,\, \sin\alpha \cos\delta,\, \sin\delta \, \big) \,.
\end{equation}
At the SSB, where time is measured by $t\SSB$, we must account for the motion of
the neutron star relative to the SSB; given a gravitational wavefront emitted by
the neutron star at time $t\NS$, at which time $t\SSB$ does the same wavefront
arrive at the SSB? The relationship between the two timescales may be written as
\begin{multline}
  \label{eq:tNS-to-tSSB}
  t\SSB = \Delay{R}(t\NS) + \Delay{\dist}(t\NS) + \Delay{E}(t\NS) + \Delay{S}(t\NS) \,.
\end{multline}

The term $\Delay{R}(t\NS)$, also known as the R{\o}mer delay, accounts for the
changing distance the gravitational wave must travel due to the orbit of the
neutron star in a binary system around the BB. The effect of the R{\o}mer delay
is to Doppler-shift the CW signal frequency as the neutron star moves first
towards, and then away from the detector. For circular
orbits~\cite{LIGO2007-SrcPrGrvWUIsSScXRSLSR, LeacPrix2015-DrSCnGrvWBSPrmMOSXSn},
\begin{equation}
  \label{eq:Delay-R-BB}
  \Delay{R}(t\NS) = a_p \sin \frac{ 2\pi ( t\NS - \tasc ) }{ P } \,.
\end{equation}
Here $a_p = (a / c) \sin i$ is the semi-major axis of the orbit (in units of
$1/c$) projected onto the line of sight by $i$, the inclination angle of the
orbit; $P$ is the orbital period; and $\tasc$ the time at which the neutron star
passes through the ascending node\footnote{The ascending node is the point in
  the neutron star's orbit where it passes through the plane of the sky (which
  intersects the BB, perpendicular to $\uvec{n}$), in the direction away from
  Earth; see Fig.~\ref{fig:cw-signal-model}.} of the orbit. For eccentric
orbits, see the expressions given in~\cite{LeacPrix2015-DrSCnGrvWBSPrmMOSXSn}.

The term $\Delay{\dist}(t\NS)$ accounts for the distance from the SSB to the
neutron star (or BB if a binary companion if present), as well as any relative
motion of the neutron star (or BB) relative to the SSB. For a simple linear
motion, where the neutron star (or BB) is initially at a distance $\dist$ from
the SSB and moves with velocity $\vec{v}\NS$, we
have~\cite{JaraEtAl1998-DAnGrvSgSpNSSDtc}
\begin{equation}
  \label{eq:Delay-D-BB}
  \Delay{\dist}(t\NS) = \gamma\NS t\NS + \frac{1}{c} \big| \dist \uvec{n} + \gamma\NS \vec{v}\NS t\NS \big| \,,
\end{equation}
where $\gamma\NS = (1 - |\vec{v}\NS|/c)^{-1/2}$. If $|\vec{v}\NS|$ is small
compared to $c$, $\gamma\NS \approx 1$ and
\begin{equation}
  \label{eq:Delay-D-small-vNS}
  \Delay{\dist}(t\NS) \approx t\NS + \frac{ \dist }{ c } \,,
\end{equation}
i.e. the CW arrives at the the SSB a time $\dist/c$ after it was emitted. Most
CW searches assume that $|\vec{v}\NS|$ is small enough so that this
approximation holds.  The effect of radial motion
($\vec{v}\NS \parallel \uvec{n}$) is additional Doppler motion of the signal;
this effect may be accounted for by redefining the frequency parameters
[Eq.~\eqref{eq:frequency-at-NS}] as being observed at the SSB, instead of being
intrinsic to the star~\cite{JaraEtAl1998-DAnGrvSgSpNSSDtc}. As a consequence,
for sufficiently large radial motion, the intrinsic spindown of the neutron star
may be observed at the SSB as a \emph{spinup}, i.e. $\fd > 0$. To account for
this possibility, many CW searches cover a small positive range of frequency
derivatives~\cite[e.g.][]{LIGOEtAl2021-AlSCntGrvWIsNtSEOLD,
  LIGOEtAl2021-SrCntGrvWYSpRETObRALV, LIGOEtAl2022-ASCntGrvWIsNSUAdLAdVOD}.  The
robustness of CW searches to proper motion ($\vec{v}\NS \perp \uvec{n}$) is
studied in~\cite{Cova2021-EfPMNtSCntGrvSr}.

The terms $\Delay{E}(t\NS)$ and $\Delay{S}(t\NS)$ denote the Einstein and
Shapiro delays, respectively, associated with the binary
system~\cite{TaylWeis1989-FrExpTRltGrUBnPP191316}. The Einstein delay collects
the effects of gravitational red-shift and time dilation due to motions of the
binary stars; the Shapiro delay accounts for the delayed propagation of the
gravitational wave through the curved space-time of the binary system. These
effects are absent for \emph{isolated} (i.e.\ single) neutron stars; for binary
systems with circular orbits,
$\Delay{E}(t\NS) = 0$~\cite{TaylWeis1989-FrExpTRltGrUBnPP191316}, and
$\Delay{S}(t\NS)$ is small enough to be
negligible~\cite{LIGO2007-SrcPrGrvWUIsSScXRSLSR}.

\subsubsection{Time at the detector}\label{sec:time-at-detector}

Similar to Eq.~\eqref{eq:tNS-to-tSSB}, we can relate $t\SSB$ to the time
measured at the detector on Earth, $t\ifo$:
\begin{equation}
  \label{eq:tSSB-to-tifo}
  t\SSB = t\ifo + \Delay{R\solar}(t\ifo) + \Delay{E\solar}(t\ifo) - \Delay{S\solar}(t\ifo) \,.
\end{equation}
Here the Solar System R{\o}mer delay $\Delay{R\solar}(t\ifo)$ accounts for the
propagation time of the gravitational wave from the SSB to the detector:
\begin{equation}
  \label{eq:Delay-R-SSB}
  \Delay{R\solar}(t\ifo) = \frac{ \vec{r}(t\ifo) \cdot \uvec{n} }{c} \,,
\end{equation}
where $\vec{r}(t\ifo)$ is the vector from the SSB to the detector, accounting
for both the sidereal and orbital motion of the Earth
(Fig.~\ref{fig:cw-signal-model}). The Solar System Einstein and Shapiro delays
$\Delay{E\solar}(t\ifo)$ and $\Delay{S\solar}(t\ifo)$ quantify, respectively,
the effect of gravitational red-shift and time dilation due to motions of the
Earth and other Solar System bodies, and the delayed propagation of the
gravitational wave through the curved space-time of the
Sun~\cite{TaylWeis1989-FrExpTRltGrUBnPP191316, IrwiFuku1999-NmrTmEphEr,
  EdwaEtAl2006-TENPlTPcIITMPrcEst-II}.

\subsubsection{Determining the phase}\label{sec:determining-phase}

\begin{figure}[t]%
  \centering%
  \includegraphics[width=0.95\linewidth]{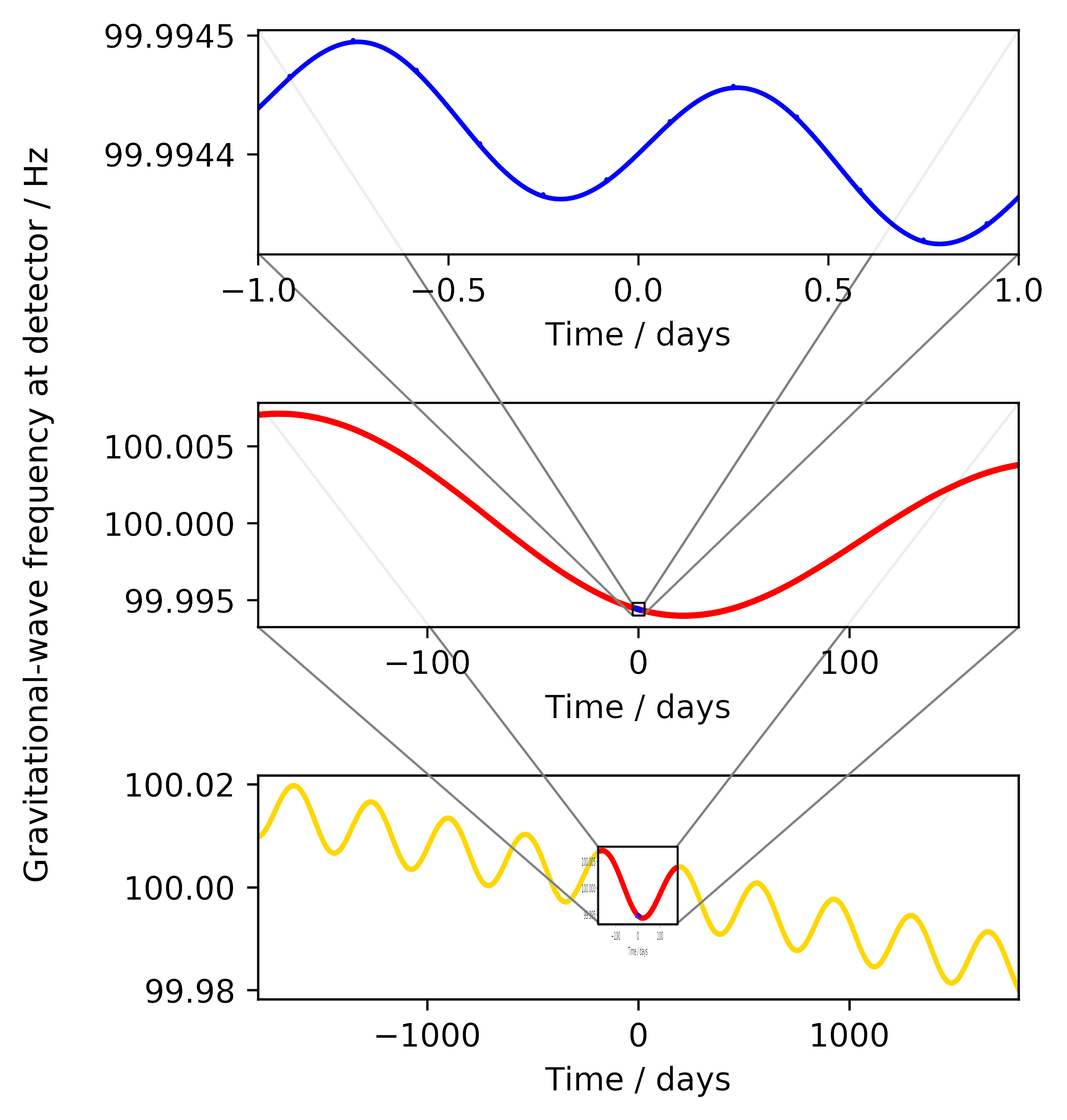}%
  \caption{\label{fig:cw-signal-frequency} Typical modulations of the CW signal
    frequency for an isolated neutron star as a function of time. Top: daily
    modulation (blue) due to the rotation of the Earth. Middle: yearly
    modulation (red) due to the orbit of the Earth. Bottom: long-term decrease
    (yellow) due to the spindown of the star.%
  }%
\end{figure}

Given gravitational wave data is timestamped by $t\ifo$, we must use the
equations in Secs.~\ref{sec:time-at-neutron}--~\ref{sec:time-at-detector} to
determine $t\NS$ as a function of $t\ifo$.  For negligible linear motion of the
neutron star ($|\vec{v}\NS| \ll c$), and circular binary orbits,
Eq.~\eqref{eq:tNS-to-tSSB} simplifies to
\begin{equation}
  \label{eq:tNS-to-tSSB-simp}
  t\SSB = t_0 + t\NS + \Delay{R}(t\NS) \,,
\end{equation}
where the constant $\dist/c$ has been absorbed into a reference time $t_0$.
Equating to Eq.~\eqref{eq:tSSB-to-tifo} gives
\begin{multline}
  \label{eq:tNS-to-tifo}
  t\NS + \Delay{R}(t\NS) = t\ifo - t_0 + \Delay{R\solar}(t\ifo) \\
  + \Delay{E\solar}(t\ifo) - \Delay{S\solar}(t\ifo) \,.
\end{multline}
For isolated stars, $\Delay{R}(t\NS) = 0$ and Eq.~\eqref{eq:tNS-to-tifo} gives
$t\NS$ directly in terms of functions of $t\ifo$; otherwise,
Eq.~\eqref{eq:tNS-to-tifo} must be numerically inverted to determine the
function $t\NS(t\ifo)$. Finally, the CW phase is
\begin{equation}
  \label{eq:phase}
  \phi(t\ifo) = 2\pi \sum_{s=0}^{s\umax} f\ndot{s} \frac{ t\NS(t\ifo)^{s+1} }{ (s+1)! } \,.
\end{equation}

Figure~\ref{fig:cw-signal-frequency} illustrates the typical timescales of
modulations for the CW phase of an isolated star. On timescales of a day, the
dominant modulation is from Doppler modulation due to the Earth's sidereal
rotation. Over the course of a year, the dominant modulation is Doppler
modulation due to the Earth's orbit. And over many years of observation, we
expect a steady spindown in frequency as the neutron star loses energy.

\subsubsection{Approximate phase}\label{sec:approximate-phase}

While the full phase expression of Eq.~\eqref{eq:phase} is required to
accurately track the CW phase over long observation times, it is often useful
(see e.g. Sec.~\ref{sec:param-space-breadth}) to consider a simpler, approximate
form of the phase~\cite{JaraEtAl1998-DAnGrvSgSpNSSDtc,
  WettPrix2013-FPrmMtASrGrvPl, LeacPrix2015-DrSCnGrvWBSPrmMOSXSn}.

We first discard the Einstein and Shapiro delay terms
$\Delay{E\solar}(t\ifo), \Delay{S\solar}(t\ifo)$, as they are always small
compared to the other terms in Eq.~\eqref{eq:tNS-to-tifo}. We then expand
Eq.~\eqref{eq:phase}, and discard any terms of order
$f\ndot{s} t\NS^{s-n+1} \Delay{R\cdots}^{n}$, where $n > 1$, and
$\Delay{R\cdots}$ is any of the R{\o}mer delay terms. This is because, over the
time-span of an observation $\Tspan$, $f\ndot{s}$ typically scales with
$\Tspan^{-s-1}$, $t\NS^{s-n+1}$ scales with $\Tspan^{s-n+1}$, but the
oscillatory R{\o}mer terms remain of order unity. Terms of order
$f\ndot{s} t\NS^{s-n+1} \Delay{R\cdots}^{n}$, therefore, scale as $\Tspan^{-n}$,
and are small enough to be neglected when $n > 1$. Finally, if $\Delay{R}(t\NS)$
is present, we assume that the orbital motion is slow compared to gravitational
wave transit time across the orbit, and we can therefore approximate
$\Delay{R}(t\NS) \approx \Delay{R}(t\ifo)$. Applying these approximations
yields:
\begin{align}
  \label{eq:phase-appx1}
  \begin{split}
    \phi(t\ifo) &\approx 2\pi \sum_{s=0}^{s\umax} f\ndot{s} \bigg\{ \frac{ (t\ifo - t_0)^{s+1} }{ (s+1)! } \\
    &\quad + \frac{ (t\ifo - t_0)^{s} }{ s! } \Big[ \Delay{R\solar}(t\ifo) - \Delay{R}(t\ifo) \Big] \bigg\}
  \end{split} \\
  \label{eq:phase-appx2}
  \begin{split}
    &= 2\pi f(t\ifo - t_0) \Big[ \Delay{R\solar}(t\ifo) - \Delay{R}(t\ifo) \Big] \\
    &\quad + 2\pi \bigg[ f t\ifo + \frac{1}{2} \fd t\ifo^2 + \frac{1}{6} \fdd t\ifo^3 + \dots \bigg] \,.
  \end{split}
\end{align}
In Eq.~\eqref{eq:phase-appx2}, the instantaneous frequency $f(t\ifo - t_0)$ is
usually replaced by a constant $f\umax$, the maximum frequency of the signal
over the observation, thereby giving the maximum modulation from the R{\o}mer
terms.

\section{Challenges of continuous wave searches}\label{sec:chall-cont-wave}

The fundamental challenge of CW searches lies in extracting a very weak signal
from comparatively noisy data. While gravitational wave signals from the mergers
of binary black holes and neutron star are strong enough to, on occasion, be
discernible to the naked eye~\cite[e.g.][]{LIGOVirg2017-BsPhyBnBHMrGW1,
  LIGOVirg2017-GWObsGrvWBNtSIn}, CW signals are comparatively much weaker. We
must therefore apply data analysis techniques to the data. All such techniques
rely on the idea of \emph{matched filtering}: we formulate a model for the CW
signal (Sec.~\ref{sec:cw-signal-model}), apply that model to the data, and
compute a \emph{detection statistic} which tells us which of two hypotheses are
favoured: the \emph{signal hypothesis}, that the data contains a CW signal
matching our model; or the \emph{noise hypothesis}, that it does not.

A first challenge is breaking the following circular dependency: how we can
detect an unknown CW signal, when we must first know its model parameters, in
order to apply the model to the data, in order to detect the signal in the first
place? It is here that the distinction between \emph{amplitude} and
\emph{phase} parameters made in Sec.~\ref{sec:cw-signal-model} becomes
important. To see why, we first express the detector response functions
$\Fplus(t), \Fcross(t)$ using two new functions\footnote{Note that, in
  contract to the definitions given in~\cite{JaraEtAl1998-DAnGrvSgSpNSSDtc}, I
  have absorbed the factor of $\sin\zeta$ into the definitions $a(t), b(t)$.},
$a(t)$ and $b(t$)~\cite{JaraEtAl1998-DAnGrvSgSpNSSDtc}:
\begin{equation}
  \begin{split}
    \label{eq:aoft-boft}
    \Fplus(t)  &= a(t) \cos2\psi + b(t) \sin2\psi \,, \\
    \Fcross(t) &= b(t) \cos2\psi - a(t) \sin2\psi \,.
  \end{split}
\end{equation}
By combining Eqs.~\eqref{eq:hoft},~\eqref{eq:hoft-plus},~\eqref{eq:hoft-cross}
and~\eqref{eq:aoft-boft}, we see that the component $h^{2m}(t)$ of $h(t)$
associated with each harmonic ($\frot$, and $f$) may be written as the linear product of four
amplitudes $\Amp^{2m}_{i}$ and four basis functions
$h^{2m}_{i}(t)$~\cite{JaraEtAl1998-DAnGrvSgSpNSSDtc}:
\begin{equation}
  \label{eq:hoft-four-Ampi-hi}
  h(t) = \sum_{m=1}^{2} h^{2m}(t) = \sum_{m=1}^{2} \sum_{i=1}^{4} \Amp^{2m}_{i} h^{2m}_{i}(t) \,.
\end{equation}
The basis functions are:
\begin{equation}
  \label{eq:hoft-four-hi}
  \begin{aligned}
    h^{2m}_{1}(t) &= a(t) \cos m\phi\rot(t) \,, &
    h^{2m}_{2}(t) &= b(t) \cos m\phi\rot(t) \,, \\
    h^{2m}_{3}(t) &= a(t) \sin m\phi\rot(t) \,, &
    h^{2m}_{4}(t) &= b(t) \sin m\phi\rot(t) \,.
  \end{aligned}
\end{equation}
The amplitudes are:
\begin{equation}
  \label{eq:eq:hoft-four-Ampi}
  \begin{split}
    \Amp^{2m}_{1} &=   \Aplus^{2m} \cos\PhiC_{2m} \cos2\psi - \Across^{2m} \sin\PhiC_{2m} \sin2\psi \,, \\
    \Amp^{2m}_{2} &=   \Aplus^{2m} \cos\PhiC_{2m} \sin2\psi + \Across^{2m} \sin\PhiC_{2m} \cos2\psi \,, \\
    \Amp^{2m}_{3} &= - \Aplus^{2m} \sin\PhiC_{2m} \cos2\psi - \Across^{2m} \cos\PhiC_{2m} \sin2\psi \,, \\
    \Amp^{2m}_{4} &= - \Aplus^{2m} \sin\PhiC_{2m} \sin2\psi + \Across^{2m} \cos\PhiC_{2m} \cos2\psi \,,
  \end{split}
\end{equation}
where
\begin{equation}
  \label{eq:Aplus-Across}
  \begin{aligned}
    \Aplus^{22} &= -C_{22} ( 1 + \cos^2\iota ) \,, & \Aplus^{21} &= -\frac{1}{2} C_{21} \sin\iota \cos\iota \,, \\
    \Across^{22} &= -2 C_{22} \cos\iota \,, & \Across^{21} &= -\frac{1}{2} C_{21} \sin\iota \,.
  \end{aligned}
\end{equation}
The $\Amp^{2m}_{i}$ are functions of the four model parameters introduced in
Sec.~\ref{sec:cw-signal-amplitude}: the CW amplitude $C_{2m}$ and phase
$\PhiC_{2m}$, inclination angle $\iota$, and polarisation angle $\psi$.  Because
$h(t)$ is \emph{linear} in the $\Amp^{2m}_{i}$, we conjecture that we can
\emph{analytically} determine best-fit values for the amplitudes, without having
to know them \emph{a priori}.

When only one harmonic is present, the four amplitudes may indeed be found
analytically, by choosing the values which maximise the \emph{likelihood
  function}.  This function gives the probability of the observed data, given a
choice of CW signal model; it is generally computing by subtracting the CW model
waveform from the data, and then computing the probability that what remains is
pure noise following an assumed distribution (e.g. a Gaussian).  The matched
filter computed using the maximum likelihood estimators of the four amplitudes
is known as the \Fstat~\cite{JaraEtAl1998-DAnGrvSgSpNSSDtc}~(see also
Sec.~\ref{sec:fstat}). In terms of detection power -- the ability of a statistic
to correctly pick the signal and/or noise hypotheses where appropriate -- the
maximum likelihood approach of the \Fstat\ compares favourably to the
theoretically more powerful approach of Bayesian
marginalisation~\cite{Sear2008-MntByTcGrvWBDAn,PrixKris2009-TrSCnGrvWBVMxmSt}.
When both harmonics are present, we may apply the \Fstat\ to each harmonic
individually~\cite{JaraEtAl1998-DAnGrvSgSpNSSDtc}, although this ignores
degeneracies between the $\Amp^{2m}_{i}$, e.g.\ the common dependence on $\iota$
and $\psi$~\cite{BejgKrol2014-SrcGrvWvKPlTSFr, Jone2015-PrCRCntGrvWSrStSNS,
  PitkEtAl2015-FRFPrDlhSrGrvWSpNS}.

While we are fortunate in that we can find best-fit values for the
\emph{amplitude} parameters (via the $\Amp^{2m}_{i}$) without prior knowledge,
we are out of luck when it comes to the \emph{phase}
parameters~(Sec.~\ref{sec:cw-signal-phase}): the neutron star's intrinsic
frequency $f$ and spindowns $\fd, \fdd$, etc.; its position on the sky
$\alpha, \delta$; and, if relevant, its velocity $\vec{v}\NS$ and binary
(circular) orbital parameters $a_p, P, \tasc$. Given the nonlinear dependence of
$h(t)$ on these parameters, there is little prospect of analytically finding
their best-fit values. We must therefore resort to numerical methods, as
follows.

A CW \emph{search} is, in essence, a process of numerical maximisation over the
phase parameters. We pick values for the phase parameters, apply the CW signal
model given by those parameters to the data, and compute a detection statistic
e.g.\ the \Fstat. We continue to pick sets of phase parameter values until, by
trial and error, we find a combination where the detection statistic strongly
favours the signal hypothesis. While this process sounds straightforward enough,
in reality we must now confront several further challenges:
\begin{enumerate}

\item What values for the phase parameters should we pick? In other words, what
  is the \emph{parameter space} -- the subset of the space spanned by the phase parameters -- from which we should sample
  vectors of phase parameter values? What motivates the choice of a particular
  parameter space?  For example, should we search for CW signals at low $\frot$,
  where most of the known pulsar reside~\cite{Kram2005-Pls}; or at high $\frot$,
  where the gravitational wave amplitude $\propto \frot^2$ is largest?

\item How many phase parameter vectors should we sample? We assume that, given
  the parameter space we have selected, at least one vector within that space
  will yield a model that fits any signal present in the data very well --
  otherwise the model is at fault -- but we do not know the best-fit vector
  \emph{a priori}. Moreover, the probably of us picking the \emph{exact}
  best-fit vector is vanishingly small. But can we assume that we will
  eventually, given enough trials, pick a vector which is ``close enough'' to
  the best-fit vector, such that the detection statistic will still favour the
  signal hypothesis? If so, how do we quantify what we mean by ``close enough''?
  And how many vectors do we need to sample in order to guarantee that we'll
  eventually pick one ``close enough'' to the best-fit vector?

\item Suppose we know how many phase parameter vectors to sample from the
  parameter space. Is it realistic to compute a detection statistic for every
  vector? Each computation must take a finite amount of time. Given that CW
  signals are weak, we expect that we will need to analyse as much available
  data as possible in order to accumulate signal power versus noise. Can we
  compute all the required detection statistics in a finite amount of time,
  using currently-available computer technology? If not, does there exist a
  \emph{sub-optimal} detection statistic which is either computationally cheaper
  to compute, or required the computation of few values (i.e.\ allows us to
  relax what we require as ``close enough'' to the best-fit vector)? Will this
  sub-optimal statistic still allow us to make a detection?

\item Suppose that we can compute (possibly sub-optimal) detection statistics
  for all the phase parameter vector we are required to sample (in order to
  guarantee that at least one is ``close enough'' to the best-fit vector).
  Since we have computed multiple detection statistics, we are no longer
  considering a binary choice of \emph{one} signal hypothesis versus one noise
  hypothesis; instead we must weigh \emph{multiple} signal hypothesis, one for
  each detection statistic.  How do we determine whether any of the signal
  hypotheses are favoured strongly enough to claim a detection? We might choose
  e.g.\ to set a \emph{threshold} value on the detection statistics, and only
  consider detection statistics above the threshold as signifying a
  detection. We must then contend with the law of large numbers, however: as the
  number of detection statistics increases, eventually one is guaranteed to pass
  the threshold, even if there is no signal present in the data. How can we make
  decisions using the detection statistics, e.g. by setting a threshold, while
  avoiding (with high confidence) the possibilities of falsely claiming a
  detection, or of falsely rejecting one?

\end{enumerate}

\section{Continuous wave search performance measures}\label{sec:cont-wave-perf-meas}

One of the challenges of CW searches, as outlined in
Sec.~\ref{sec:chall-cont-wave}, is deciding on the parameter space of signals to
cover, and the detection statistic to compute. As we will see in
Sec.~\ref{sec:cont-wave-search}, these aspects of CW search design must
typically be balanced against each other. Due to limited computational
resources, we must typically either choose a sensitive (but computationally
expensive) detection statistic over a limited (but computationally cheaper)
parameter space; or a wide (but expensive) parameter space and a sub-optimal
(but cheaper) detection statistic.

In Sec.~\ref{sec:cont-wave-search}, I examine the performance of CW searches
through the lens of this trade-off. I quantify searches according to the
following metrics: sensitivity \emph{depth} (Sec.~\ref{sec:sensitivity-depth}),
and parameter-space \emph{breadth} (Sec.~\ref{sec:param-space-breadth}).

\subsection{Sensitivity depth}\label{sec:sensitivity-depth}

The CW search sensitivity depth~\cite{BehnEtAl2015-PstMtUSCnGrvSGlC,
  DreiEtAl2018-FAcSnsEsCntSr} is a signal-to-noise ratio.  As a measure of
signal strength, an \emph{upper limit at confidence $C$} is computed on the
gravitational wave amplitude of the signal, typically $h_0$. The upper limit
$h_0^C$ is computed using a variety of methods~\cite{LIGO2008-AlSrPrdGrvWvLSD,
  Wett2012-EsSnWdpSrGrvP, DreiEtAl2018-FAcSnsEsCntSr}, which generally follow
this procedure:
\begin{enumerate}

\item\label{item:ul-criteria} Decide on a criterion for detecting CW signals
  using a given search algorithm. Common practise is to take the maximum
  detection statistic $\DSmax$ found by a search as indicating the most
  promising detection candidate. Then, apply the algorithm to many synthesised
  data sets containing only noise, take the $\DSmax$ found from each data set,
  and set a threshold $\DSthr$ such that only a fraction $A$ of the $\DSmax$
  satisfy $\DSmax > \DSthr$. Here $A$ is the \emph{false alarm probability}: the
  probability that we would falsely claim a detection by finding a
  $\DSmax > \DSthr$, despite the data containing no signal. It is typically
  small; conventionally CW searches establish a $A = 1\%$ false alarm
  threshold. (Note that this threshold is established using the maximum of many
  detection statistics computed by the search; the probability of a
  \emph{single} detection statistic exceeding the threshold is therefore
  $\ll A$.)

\item\label{item:ul-sample} Given a fixed $h_0$, draw random sets of values for
  the CW signal model parameters. Amplitude parameters (other than $h_0$) are
  drawn from their natural priors; phase parameters are drawn from the search
  parameter space.

\item\label{item:ul-synth} For each set of model parameter values, synthesise a
  CW signal time series $h(t)$ in software using the given parameter values. Add
  to each $h(t)$ a realistic representation of the detector noise, either by
  using real detector data, or by synthesising stationary Gaussian noise with
  power spectral density $\Sh$ (which may be assumed constant over the narrow
  bandwidth of CW signals).

\item\label{item:ul-search} Analyse each noisy $h(t)$ time series using the CW
  search algorithm in question. For each noisy $h(t)$, decide whether the
  synthesised CW signal would have been detected using the criteria established
  previously, i.e.\ is the $\DSmax$ found from the search of each noisy $h(t)$
  greater than $\DSthr$?  The fraction of CW signals considered detected is the
  \emph{detection confidence} $C$. (The quantity $1 - C$ is the \emph{false
    dismissal probability}: the probability that we would dismiss claiming a
  detection because $\DSmax < \DSthr$, despite the data containing a signal.)
  The $h_0$ fixed in step~\ref{item:ul-sample} is then interpreted as an upper
  limit $h_0^C$ with $C$ confidence; if the data contains a signal with
  amplitude $h_0$, we would have confidence $C$ of detecting it.

\item\label{item:repeat} Adjust $h_0$ and repeat
  steps~\ref{item:ul-sample}--~\ref{item:ul-search} above until $C$ converges to
  the desired confidence; typical choices are 90\% or 95\%. Given that CW
  signals have yet to be detected, published CW searches typically quote upper
  limits $h_0^C$ as a function of search frequency $f$ as their primary
  scientific result.

\end{enumerate}

The sensitivity depth takes the ratio of an upper limit, given by $h_0^C$, to an
estimate of the noise in the detector, given by $S_h$. It is defined
as:\footnote{The depth is conventionally defined as $\sqrt{\Sh}/h_0$ and
  therefore has the same units as $\sqrt{\Sh}$, typically Hz${}^{-1/2}$. In this
  review I explicitly normalise $\Sh$ by units of Hz${}^{-1}$, so that $\depth$ is a
  dimensionless quantity.}
\begin{equation}
  \label{eq:depth}
  \depth = \frac{ \sqrt{ \Sh \cdot \mathrm{Hz} } }{ h_0^C } \,.
\end{equation}
Note that $\Sh$ is defined as the \emph{single}-sided power spectrum, over the
same frequency band as the CW signals used to determine $h_0^C$, and taking the
harmonic mean of the data over time (and over multiple detectors). Averaging
over time gives a representative value of the detector power spectrum, whose
value over short time periods may vary over the course of an
observation. Non-Gaussian ``glitches'' in the noise of short ($\lesssim 1$~s)
duration generally do not impact long-duration continuous wave searches, unless
they are sufficiently short and loud (resembling a Dirac delta function) that
their Fourier transform contaminates a wide frequecy
band~\cite{ZweiRile2020-InfSlHUsOCntSr,StelEtAl2022-IdnRNnNTrGrvSr}.

Since a smaller $h_0^C$ implies a more sensitive search, $\depth$ increases with
search sensitivity.  By factoring out the performance of the detector, i.e.\ its
noise power spectrum, $\depth$ quantifies the contribution of the CW search
algorithm to the overall search sensitivity.  Typically, $\depth$ is found to be
approximately constant for a given CW search, provided we exclude frequency
bands where $\Sh$ is degraded by excessive detector noise, e.g.\ instrumental
line artefacts.

Sensitivity depth serves as a useful measure for qualitatively comparing the
sensitivities achieved by different CW searches and algorithms. This comes with
the following caveat, however: a strictly ``apples-to-apples'' comparison is
difficult to achieve in practise, for the following reasons:
\begin{itemize}

\item When computing $h_0^C$, the criteria for considering a CW signal detected
  (step~\ref{item:ul-criteria} of the upper limit procedure) often varies
  between searches. For example, a CW signal may be considered detected only
  after more sensitive follow-up studies of candidates found by the initial
  search. This makes it challenging to model the statistical properties of the
  detection criteria (as discussed in~\cite{DreiEtAl2018-FAcSnsEsCntSr}) and
  thereby determine its false alarm probability $A$. It is generally assumed
  that $A$ is both small and weakly dependent on the choice of detection
  criteria, e.g.\ $\DSthr$ scales weakly with $A$ and the number of computed
  detection statistics~\cite{Wett2012-EsSnWdpSrGrvP}. The overall $A$ of a CW
  search is rarely quantified explicitly, however, which makes a strictly
  equitable comparison of search sensitivities at equal false alarm rates
  difficult to achieve.

\item The population of signals sampled from in determining $h_0^C$
  (step~\ref{item:ul-sample} of the upper limit procedure) also varies. For
  example, searches using the PowerFlux algorithm
  (Sec.~\ref{sec:powerflux-algorithms}) traditionally fix the inclination angle
  $\iota$, in addition to $h_0$, and report two values of $h_0^C$ for circular
  (most sensitive/best case) and linear (least sensitive/worst case)
  polarisations. Analytic scaling to convert upper limits of this type to
  so-called \emph{population-averaged} upper limits (where $\cos\iota$ is
  sampled uniformly from $[-1, 1]$) were proposed
  in~\cite{Wett2012-EsSnWdpSrGrvP}, while~\cite{LIGOVirg2017-AlSrPrdGrvWvOLD}
  suggests that such scaling must depend on the data being analysed.

\item There are different approaches to determining the confidence $C$
  (steps~\ref{item:ul-synth}--~\ref{item:ul-search} of the upper limit
  procedure). Searches using the PowerFlux algorithm determine an upper limit on
  $h_0$ on individual signals, rather than over a population of signals, using
  Feldman-Cousins confidence intervals; the largest (worst case) upper limit
  found for an individual signal is then selected to represent the population as
  a whole (see discussion in~\cite{Wett2012-EsSnWdpSrGrvP}). Even when the
  population-averaged procedure is followed, various approaches are used to find
  $h_0^C$ at the desired $C$, usually motivated to reduce computational cost:
  e.g. linear or spline interpolation of $h_0^C$ as a function of C. (A Bayesian
  approach, which fits a sigmoid curve to a Boolean array of detections as a
  function of $h_0$, is proposed in~\cite{Whel2015-ByEstPrmEf}.)

\item The CW data analysis community has historically not converged to a
  consistent choice of confidence $C$ at which to set upper limits
  (step~\ref{item:repeat} of the upper limit procedure), with both 90\% and 95\%
  being common choices. (Though perhaps a consensus is now emerging; of the
  \NOHTHREEPAPERS\ papers reviewed in Sec.~\ref{sec:cont-wave-search} which
  analysed data from the LIGO-Virgo 3rd observing run, only one
  paper~\cite{LIGOEtAl2021-CnsLODGrvEmDRGlPPJ0} set upper limits at 90\%
  confidence, with the remainder choosing 95\%.)

\item Finally, choosing an appropriate value for $\Sh$ is not entirely
  straightforward~\cite{DreiEtAl2018-FAcSnsEsCntSr}. Unless we have performed
  the analysis ourselves, we are unlikely to have access to the original data
  set used in the analysis, and it may be impractical to reconstruct the
  original data set without access to the software and configuration details
  used in its preparation. We must therefore rely on generic sensitivity
  curves~\cite{S5sens, S6VSR23sens, VSR4sens, O1sens, O2sens, O3sens, O3Vsens}
  which give representative $\Sh$ for each data set. In addition, computation of
  $\Sh$ over an appropriate frequency band can be sensitive to choices of
  windowing and/or averaging over frequency.

\end{itemize}
For these reasons, comparisons of the $\depth$ achieved by different CW searches
or algorithms should not be taken too seriously beyond the first one or two
significant figures.

\subsection{Parameter-space breadth}\label{sec:param-space-breadth}

To complement $\depth$ as a measure of CW search sensitivity, in this review I
introduce the breadth $\MV$ as a measure of CW parameter-space coverage:
\begin{equation}
  \label{eq:breadth}
  \MV = \int_{\paramsp} d\vec{p} \, \sqrt{ g(\vec{p}) } \,.
\end{equation}
where $\paramsp$ represents that CW search parameter space, from which parameter
vectors $\vec{p} \in \paramsp$ are drawn.  The function $g(\vec{p})$ is the
determinant of the \emph{parameter-space metric}
$g_{ij}(\vec{p})$~\cite{BalaEtAl1996-GrvWClsBDStMCEsPr,
  Owen1996-STmGrvWInsBnCTmS}. The metric $g_{ij}(\vec{p})$ provides a distance
measure on the parameter space, as follows: suppose a CW signal is present in
the data with best-fit parameters $\vec{p}$, and we attempt to match that signal
to a model waveform with parameters $\vec{p} + \Delta\vec{p}$. The
\emph{mismatch} $\mu = g_{ij}(\vec{p}) \Delta p_{i} \Delta p_{j}$
measures how much signal power we expect to lose; if $\Delta\vec{p} = \vec{0}$,
we have perfectly matched the signal and expect the detection statistic to be at
its optimal maximum; for $\Delta\vec{p} \ne \vec{0}$ we expect the detection
statistic to be reduced by a factor $\approx 1 - \mu$ for small
$\mu$~\cite{Prix2007-SrCnGrvWMMltFs, WettPrix2013-FPrmMtASrGrvPl}.

The metric gives a quantitative measure of how ``close''
(cf. Sec.~\ref{sec:chall-cont-wave}) signals are to each other in parameter
space, in the context of recovering signal power. Its determinant quantifies the
``density'' of parameter space: denser regions being where signals are
``closer'' to each other than in sparser regions. The breadth $\MV$ is therefore
a measure of the number of CW model waveforms (or \emph{templates}) needed to
properly search a given parameter space. That said, I do not attempt to rescale
$\MV$ to give actual template counts for a given CW search, as I consider that
to be an implementation detail of the algorithm in question. Ideally, the
algorithm would use the minimum number of templates needed to cover the
parameter space, while guaranteeing that the mismatch remains under a
pre-established maximum~\cite{Prix2007-TmpSrGrvWEfLCFPrS,
  Wett2014-LTmPlcChASrGrvP}.

It has been shown~\cite{Prix2007-SrCnGrvWMMltFs} that, for observing time-spans
longer than a day, the CW parameter-space metric is largely independent of the
amplitude parameters of the CW signal. We therefore require only the CW phase to
compute the metric; indeed, it is sufficient to use the simplified, approximate
phase given by Eq.~\eqref{eq:phase-appx2}
(Sec.~\ref{sec:approximate-phase}). Using the \emph{phase metric}
approximation~\cite{BradEtAl1998-SrcPrdSrLI, Prix2007-SrCnGrvWMMltFs}, the
metric is then computed from
\begin{multline}
  \label{eq:phase-metric}
  g_{ij}(\vec{p}) = \frac{1}{\Tspan} \int_{\Tspan} dt \,
  \frac{\partial \phi(t)}{\partial p_{i}}
  \frac{\partial \phi(t)}{\partial p_{j}} \bigg|_{\vec{p}}
  \\ - \frac{1}{\Tspan^2}
  \bigg[ \int_{\Tspan} dt \,
  \frac{\partial \phi(t)}{\partial p_{i}} \bigg|_{\vec{p}} \bigg]
  \bigg[ \int_{\Tspan} dt \,
  \frac{\partial \phi(t)}{\partial p_{j}} \bigg|_{\vec{p}} \bigg] \,,
\end{multline}
where we integrate over the observation spanned by $\Tspan$.

In computing the parameter-space breadth, I will assume the parameter-space
metric of a \emph{fully-coherent} CW signal, i.e. one where $\phi(t\ifo)$ is
matched to the data across the entire observation.  Not all detection statistics
obey this criteria, however; indeed a common trade-off made by sub-optimal
detection statistics is to relax this restriction.  The purpose of $\MV$,
however, is to quantify the size of the parameter space of the CW signal model,
whereas the size of the CW search (in terms of the number of templates it
searches) is considered an implementation detail of the search.  As discussed in
Sec.~\ref{sec:chall-cont-wave}, a CW data analyst must often decide whether to
prioritise sensitivity or parameter-space coverage in their CW search design; a
particular choice of trade-off is then reflected in the $\depth$ and $\MV$
achieved for that search.  An optimal choice could be to choose a detection
statistic which balances sensitivity (thereby increasing depth) with
computational efficiency, allowing a wider parameter space to be searched
(thereby increasing breadth).

In the following sections I give formulas for $\MV$ over the frequency and
spindown (Sec.~\ref{sec:freq-spind-param}), sky
(Sec.~\ref{sec:sky-parameter-space}), and binary orbital parameter spaces
(Sec.~\ref{sec:binary-orbit-param}). I use Eq.~\eqref{eq:phase-appx2} as a
suitable approximation to the metric, with $f(t\ifo - t_0)$ set to $f$, so that
$\MV$ is correct when integrated over frequency. This means, however, that the
breadth of the sky and binary orbital parameter spaces scale with $f$, and
therefore we must defer integration over $f$ until the end. I therefore define
\emph{raw breadths} $\rawMV$, which have \emph{not} been integrated over
$f$, in the following sections. These factors will be assembled to find the
overall breadth $\MV$ in Sec.~\ref{sec:assembl-over-breadth}.

\subsubsection{Frequency and spindown parameter space}\label{sec:freq-spind-param}

The phase metric over the frequency and spindown parameters
is~\cite{Whit2006-ObsCnGrvWEmSpCS, WettEtAl2008-SrGrvWvCssLI}
\begin{equation}
  \label{eq:phase-metric-fndot}
  g_{f\ndot{r}, f\ndot{s}} =
  \frac{ 4 \pi^2 (r + 1) (s + 1) }{ (r + 2)! (s + 2)! (r + s + 3)!} \Tspan^{r + s + 2} \,,
\end{equation}
where $f\ndot{r}$, $f\ndot{s}$ are the $r$th, $s$th spindown parameters
respectively.  Note that $g_{f\ndot{r}, f\ndot{s}}$ is independent of the
$f\ndot{s}$. The square root of the determinants of this metric, up to the 2nd
spindown, are
\begin{align}
  \label{eq:rawMVfreq}
  \rawMVfreq = \sqrt{ g(f) } &= \frac{\pi T}{\sqrt{3}} \,, \\
  \label{eq:sqrtdet-g-fd}
  \sqrt{ g(f, \fd) } &= \frac{\pi^2 T^3}{6 \sqrt{15}} \,, \\
  \label{eq:sqrtdet-g-fdd}
  \sqrt{ g(f, \fd, \fdd) } &= \frac{\pi^3 T^6}{360 \sqrt{105}} \,,
\end{align}
Note that Eq.~\eqref{eq:phase-metric-fndot} assumes the observation runs over
$t\ifo \in [0, \Tspan]$, while
Eqs.~\eqref{eq:rawMVfreq}--~\eqref{eq:sqrtdet-g-fdd} are independent of this
choice.  Equation~\eqref{eq:rawMVfreq} immediately gives the raw frequency
breadth $\rawMVfreq$.  In order to separate out the contributions from each
spindown parameter, I define their (raw) breadths as ratios to the breadth of
the preceding spindown, as follows:
\begin{align}
  \label{eq:rawMVfd}
  \rawMVfd &= \frac{1}{\rawMVfreq} \int d\fd \, \sqrt{ g(f, \fd) }
          = \frac{\pi T^2}{6 \sqrt{5}} \MVrange{\fd} \,, \\
  \label{eq:rawMVfdd}
  \rawMVfdd &= \frac{1}{\rawMVfd} \int d\fd \int d\fdd \, \sqrt{ g(f, \fd, \fdd) }
           = \frac{\pi  T^3}{60 \sqrt{7}} \MVrange{\fdd} \,.
\end{align}
Here I adopt the notation of~\cite{LeacPrix2015-DrSCnGrvWBSPrmMOSXSn}: for a
parameter $p$ with parameter space $[p_0, p_1] \cup [p_2, p_3] \cup \dots$,
define
\begin{equation}
  \label{eq:MVrange}
  \MVrange{p^{q}} \equiv p_1^{q} - p_0^{q} + p_3^{q} - p_2^{q} + \dots \,.
\end{equation}
The breadths in $f\ndot{s}$ [Eqs.~\eqref{eq:rawMVfreq}, \eqref{eq:rawMVfd},
\eqref{eq:rawMVfdd}] scale as $\Tspan^{s+1}$, consistent with the typical
spacings $\Delta f\ndot{s} \propto \Tspan^{-s-1}$ used to construct a
rectangular search grid in these parameters.

\subsubsection{Sky parameter space}\label{sec:sky-parameter-space}

The metric over the sky arises from the Solar System R{\o}mer delay
$\Delay{R\solar}(t\ifo)$ [Eq.~\eqref{eq:Delay-R-SSB}]. I approximate the
detector position vector $\vec{r}(t\ifo)$ with a Ptolomaic-like orbit
(cf.~\cite{Whit2006-ObsCnGrvWEmSpCS}), where the Earth's orbit is circular and
co-planar with its equator. I also ignore the relative phase differences between
the Earth's sidereal and orbital motions, which should be immaterial over
observation times much greater than a day. With these assumptions the sky
component of the CW phase simplifies to
\begin{multline}
  \label{eq:phase-sky}
  \phi\usky(t\ifo) = 2 \pi  f \cos\delta \Big[ \tauS \cos (\alpha - \OmegaS t\ifo) \\
  + \tauO \cos (\alpha - \OmegaO t\ifo) \Big] \,,
\end{multline}
where $\tauS \approx 2.13{\times}10^{-2}$~lt-s is the radius of the Earth,
$\tauO \approx 4.99{\times}10^{2}$~lt-s is the orbital radius of the Earth
around the Sun, and $\OmegaS \approx 7.27{\times}10^{-5}$~Hz,
$\OmegaO \approx 1.99{\times}10^{-7}$~Hz are the respective sidereal and orbital
angular frequencies. Combining Eqs.~\eqref{eq:breadth}, \eqref{eq:phase-metric},
and \eqref{eq:phase-sky}, and noting that $\tauS \ll \tauO$ and
$\OmegaO \ll \OmegaS$, we arrive at the following expression for
\begin{multline}
  \label{eq:rawMVsky}
  \rawMVsky = \frac{8}{3} \pi ^3 f^2 \tauO^2 \bigg\{
  1
  - \sinc^2 \OmegaO \Tspan
  \\
  - 2 \sinc^2 \frac{\OmegaO \Tspan}{2}
  + 2 \sinc(\OmegaO \Tspan) \sinc^2 \frac{\OmegaO \Tspan}{2}
  \\
  + \frac{4 \tauS}{\tauO} \sinc \frac{\OmegaS \Tspan}{2}
  + \order \bigg[ \bigg(\frac{\tauS}{\tauO}\bigg)^2 \bigg]
  \bigg\}^{1/2} \,,
\end{multline}
where $\sinc(x) = \sin(x)/x$. Note that $\rawMVsky$ converges to
$8 \pi ^3 f^2 \tauO^2 / 3$ for $\OmegaO \Tspan \gg 1$, i.e.\ once the
observation time spans many years. For CW searches which cover patches of the
sky, I rescale $\rawMVsky$ by the fraction of the sky covered, thereby assuming
that the template density is isotropic over the sky for long observations.

\subsubsection{Binary orbital parameter space}\label{sec:binary-orbit-param}

The phase metric over the binary orbital parameters is derived
in~\cite{Mess2011-SmcSStKCntWSrBSy, LeacPrix2015-DrSCnGrvWBSPrmMOSXSn}. In
calculating the breadth, I use the ``long-segment'' approximation to the metric,
where $\Tspan \gg P$, and thus assume that the CW signal is observed over many
orbital periods. The metric is most conveniently parameterised by $a_p$, the
angular frequency $\Omega_p = 2\pi/P$ of the orbit, and $\tasc$. It is given by
Eq.~(71) of~\cite{LeacPrix2015-DrSCnGrvWBSPrmMOSXSn}, and the square root of its
determinant is
\begin{equation}
  \label{eq:sqrtdet-g-ap-OmegaP-tasc}
  \sqrt{ g(a_p, \Omega_P, \tasc) } = \sqrt{\frac{2}{3}} \pi^3 f^3 a_p^2 \Omega_P T \,.
\end{equation}
When $\tasc$ is known, the (raw) breadth $\rawMVbin$ is
\begin{equation}
  \label{eq:rawMVbin-known-tasc}
  \begin{split}
    \rawMVbin &= \int da_p \int d\Omega_P \int d\tasc \, \sqrt{ g(a_p, \Omega_P, \tasc) } \\
    &= -\left(\frac{2}{3}\right)^{3/2} \pi^5 f^3 \MVrange{a_p^3} \MVrange{P^{-2}} \MVrange{\tasc}  T \,,
  \end{split}
\end{equation}
where the bounds of the integral over $\Omega_P$ are expressed in terms of $P$
for convenience.\footnote{Note that the sign of $\MVrange{P^{-2}}$,
  $\MVrange{P^{-1}}$ is negative, hence the overall minus sign in
  Eqs.~\eqref{eq:rawMVbin-known-tasc},~\eqref{eq:rawMVbin-unknown-tasc}
  respectively.} When $\tasc$ is unknown, we must search over its full range
$[0, P]$; the breadth is then
\begin{equation}
  \label{eq:rawMVbin-unknown-tasc}
  \begin{split}
    \rawMVbin &= \int da_p \int d\Omega_P \int_{0}^{2\pi/\Omega_P} \hspace{-2.4em} d\tasc \, \sqrt{ g(a_p, \Omega_P, \tasc) } \\
    &= -2 \left(\frac{2}{3}\right)^{3/2} \pi^5 f^3 \MVrange{a_p^3} \MVrange{P^{-1}} T \,.
  \end{split}
\end{equation}

The TwoSpect algorithm (Sec.~\ref{sec:twospect-algorithm}) conventionally
searches over a fixed range of frequency modulation depth
$\Delta f\uobs = f a_p \Omega_P$ instead of a fixed range of $a_p$.  The
appropriate formula for its breadth is found by substituting
$a_p = \Delta f\uobs / f \Omega_P$ in Eq.~\eqref{eq:sqrtdet-g-ap-OmegaP-tasc}
and integrating over $\Delta f\uobs$:
\begin{equation}
  \label{eq:rawMVbin-unknown-tasc-fixed-freq-mod-dpth}
  \begin{split}
    \rawMVbin &= \int d\Delta f\uobs \int d\Omega_P \int_{0}^{2\pi/\Omega_P} \hspace{-2.4em} d\tasc \, \frac{\partial a_p}{\partial \Delta f\uobs} \sqrt{ g(a_p, \Omega_P, \tasc) } \\
    &= \left(\frac{1}{6}\right)^{3/2} \pi^2 \MVrange{\Delta f\uobs^3} \MVrange{P^2} T \,.
  \end{split}
\end{equation}

\subsubsection{Assembling the overall breadth}\label{sec:assembl-over-breadth}

Given the raw breaths defined in
Sec.~\ref{sec:freq-spind-param}--~\ref{sec:binary-orbit-param}, the overall
breath is then found by multiplying these factors and integrating over $f$:
\begin{equation}
  \label{eq:MV-from-rawMV}
  \MV = \int df \, \prod \rawMVany \,,
\end{equation}
where the product is taken over those $\rawMVany$ relevant to a particular search.

It is informative to divide $\MV$ into factors which arise from each component
of the parameter space (frequency, spindown, sky, binary orbit), so that we can
see the relative contribution of each component to the overall breadth. I define
the factors of $\MV$ as follows. Let $k$ be the exponent of $f$ in each raw
breadth $\rawMVany$ relevant for a search -- except for $\rawMVfreq$, where we set
$k = 1$ to account for integration over $f$. Let $\kappa$ be the sum over all
$k$. When integrated over $f$, $\MV$ will contain a factor
$K = \MVrange{f^{\kappa}} / \kappa$. We now define
\begin{align}
  \label{eq:MVany-from-rawMVany}
  \MVfreq &= \rawMVfreq K^{1 / \kappa} \,, &
  \MVany &= \frac{ \rawMVany }{ f^{k} } K^{k / \kappa} \,;
\end{align}
each $\MVany$ weights $\rawMVany$ by its contribution to the integration over
$f$. When multiplied together, these factors will give $\MV$ consistent with
Eq.~\eqref{eq:MV-from-rawMV}.

As an example, consider a CW search over the frequency, sky, and binary orbital
parameters; we therefore require $\rawMVfreq$~[Eq.~\eqref{eq:rawMVfreq}],
$\rawMVsky$~[Eq.~\eqref{eq:rawMVsky}], and $\rawMVbin$~[either
Eq.~\eqref{eq:rawMVbin-known-tasc} or~\eqref{eq:rawMVbin-unknown-tasc}]. The
exponents of $f$ that appears in $\rawMVsky$ and $\rawMVbin$ are $k = 2$,
$k = 3$ respectively, and for $\rawMVfreq$ we set $k = 1$; hence $\kappa =
6$. The component breadths are therefore
\begin{equation}
  \label{eq:MV-example-1}
  \begin{aligned}
    \MVfreq = \rawMVfreq \left( \MVrange{f^{6}} / 6 \right)^{1 / 6} \,, \\
    \MVsky = \frac{ \rawMVsky }{ f^{2} } \left( \MVrange{f^{6}} / 6 \right)^{2 / 6} \,, \\
    \MVbin = \frac{ \rawMVbin }{ f^{3} } \left( \MVrange{f^{6}} / 6 \right)^{3 / 6} \,,
  \end{aligned}
\end{equation}
and $\MV$ is given equivalently by
\begin{equation}
  \label{eq:MV-example-2}
  \MV = \int df \, \rawMVfreq \rawMVsky \rawMVbin = \MVfreq \MVsky \MVbin \,.
\end{equation}

\subsubsection{Hidden Markov models}\label{sec:hidden-markov-models}

The parameter-space breadth [Eq.~\eqref{eq:breadth}] is intended to be agnostic
to the implementation details of a CW search, to enable a fair comparison of CW
searches which cover the same parameter space using different algorithms. One
particular CW search algorithm, however, requires special consideration.

In place of the Taylor series representation of the CW signal frequency $f(t)$
[Eq.~\eqref{eq:frequency-at-NS}], Hidden Markov models (HMMs) represent the
frequency $f(t)$ as an unobserved (hidden) state variable over a time-frequency
representation of $h(t)$, typically discretised into $N$ frequency bins and $M$
time steps. The Viterbi algorithm is then used to determine the most likely
sequence of hidden states -- i.e.\ the CW frequency as a function of time --
based on a set of observables -- a detection statistic computed for each
frequency bin and time step.

A fuller description of the Viterbi algorithm is deferred until
Sec.~\ref{sec:viterbi-algorithms}.  For now, we note that the Viterbi algorithm
is efficient at considering a \emph{very} large number of possible CW frequency
paths: $N \delta n^{M}$, where $\delta n$ is the number of possible paths the CW
frequency may take between successive time steps. Given that $M$ is typically of
order 10--$10^4$, by this measure the parameter space covered by a CW search
using the Viterbi algorithm may be \emph{thousands} of orders of magnitude
greater than a comparable search which models the CW frequency as a Taylor
series. On the other hand, many of these paths represent small deviations -- by
a frequency bin here or there -- from a CW frequency path that otherwise follows
a Taylor series, and one might argue whether it is fair to count these
deviations as completely different paths.

As a compromise, I account for the increased parameter space coverage of HMM
searches using the Viterbi algorithm as follows. Suppose e.g.\ that the CW
frequency is allowed to jump by $\pm 1$ bins per time step; i.e.
$\delta n = 3$. Given a frequency bin $n$ at time step $m$, at the previous time
step $m - 1$ the CW frequency may pass through 3 frequency bins
$n - 1, n, n + 1$; at the next previous time step $m - 2$ the CW frequency may
pass through 5 frequency bins $n - 2, n - 1, n, n + 1, n + 2$; and so on. In
general, the parameter-space volume encompassing the possible CW frequency paths
increases by $\delta n - 1$ between time steps. Calculating this volume over all
$M$ time steps and normalising by $M$ defines the HMM breadth factor
\begin{equation}
  \label{eq:MVHMM}
  \begin{split}
    \MVHMM &= \frac{1}{M} \sum_{m=0}^{M-1} 1 + m (\delta n - 1) \\
    &= \frac{1}{2} (M \delta n - M - \delta n + 3) \,.
  \end{split}
\end{equation}
This definition is consistent with the spirit of the parameter-space breadth, in
that it seeks to quantify the volume of the parameter space, while the placement
of templates within that space is considered an implementation choice. Note that
$\MVHMM = 1$ when either $M = 1$ or $\delta n = 1$, i.e. when the CW frequency
follows a single path consistent with the Taylor series model.

\section{A brief history of gravitational wave data}\label{sec:brief-hist-grav}

\begin{table}[t]%
  \centering%
  \caption{\label{tab:observing_runs} Data collection runs of the LIGO and Virgo
    gravitational wave detectors, 2002--2020. Columns are: detector generation,
    run label, start date, end date, time-span, and number of CW searches
    reviewed in Sec.~\ref{sec:cont-wave-search} which used data from the
    run. (Some CW searches use data from more than one run, which is accounted
    for in column~6.) %
  }%
\begin{tabular*}{\linewidth}{@{}l @{\extracolsep{\fill}} l @{\extracolsep{\fill}} r @{\extracolsep{\fill}} r @{\extracolsep{\fill}} r @{\extracolsep{\fill}} r@{}}
\hline
\hline
Gen. & Obs. & Start & End & $\Tspan$ & CW \\
 &  &  &  & days & \# \\
\hline
\hline
0.5G & S1 & 23 Aug 2002 & 9 Sep 2002 & 17 & 2 \\
0.5G & S2 & 14 Feb 2003 & 14 Apr 2003 & 59 & 4 \\
0.5G & S3 & 31 Oct 2003 & 9 Jan 2004 & 70 & 1 \\
0.5G & S4 & 22 Feb 2005 & 23 Mar 2005 & 29 & 5 \\
1G & S5 & 4 Nov 2005 & 1 Oct 2007 & 696 & 14 \\
1G & VSR1 & 18 May 2007 & 1 Oct 2007 & 136 & 1 \\
1.5G & VSR2 & 7 Jul 2009 & 8 Jan 2010 & 185 & 9 \\
1.5G & S6 & 7 Jul 2009 & 20 Oct 2010 & 470 & 22 \\
1.5G & VSR3 & 11 Aug 2010 & 19 Oct 2010 & 69 & 2 \\
1.5G & VSR4 & 3 Jun 2011 & 5 Sep 2011 & 94 & 6 \\
2G & O1 & 12 Sep 2015 & 19 Jan 2016 & 129 & 57 \\
2G & O2 & 30 Nov 2016 & 25 Aug 2017 & 268 & 79 \\
2G & O3 & 1 Apr 2019 & 27 Mar 2020 & 361 & 116 \\
 & -- a & 1 Apr 2019 & 1 Oct 2019 & 183 & \\
 & -- b & 1 Nov 2019 & 27 Mar 2020 & 147 & \\
\hline
\hline
\end{tabular*}
\end{table}

This section summarises the development of interferometric gravitational wave
detectors, and the data collected by them, over the last two decades.

Kilometre-scale gravitational wave observatories have been in operation for
nearly 20 years, and have completed 13 data collection runs to date
(Table~\ref{tab:observing_runs}). The first few runs of the Initial
LIGO~\cite{AbraEtAl1992-LILsIntGrvOb} detectors -- the 0.5 generation (0.5G) in
Table~\ref{tab:observing_runs} -- were generally short (a few months or less)
with their primary aim being to fully commission the instruments and gain
experience in analysing their data. This effort culminated in the 1st generation
(1G) of the Initial LIGO and Virgo~\cite{AccaEtAl2012-VrLsIntDtGrvW} detectors
at their inaugural design sensitivities. A period of further sensitivity
improvements followed -- the 1.5 generation (1.5G) Enhanced
LIGO~\cite{LIGO2009-LILsIntGrvOb} and Virgo+~\cite{AccaEtAl2011-SttVrPrj}
detectors -- before an extended shutdown period for significant upgrades, in
order to achieve sensitivities capable of detecting binary black hole/neutron
star mergers.

The commencement of the 1st observing run (O1) of Advanced
LIGO~\cite{LIGO2015-AdvLIG} -- and the first detection, the binary black hole
merger GW~150914~\cite{LIGOVirg2016-ObsGrvWvBnBHMr} two days into the run --
began the current era of 2nd-generation (2G) detectors and the beginning of
gravitational wave astronomy. Advanced Virgo~\cite{Virg2015-AdVScnIntGrWDt}
joined the end of the 2nd observing run (O2), in time to detect the first binary
neutron star merger GW~170817~\cite{LIGOVirg2017-GWObsGrvWBNtSIn}.  LIGO and
Virgo commenced joint observations with the 3rd observing run (O3).

KAGRA~\cite{KAGR2021-OvrKDtDsCnsHs}, the first kilometre-scale interferometer to
use cryogenic cooling to reduce detector noise, first collected data in 2020,
and has recently joined the 4th observing run (O4), currently underway,
alongside LIGO and Virgo. A third instrument of the LIGO Observatory is under
construction in India~\cite{Indi2011-LIPrIntGrvOb}.

Several sub-kilometre-scale detectors have also existed at various times; given
their limited sensitivity to gravitational wave signals, their primary focus has
been technology development. Of these detectors,
GEO-600~\cite{LIGO2010-GEO600Stt, DoolEtAl2016-G600GEUpPrSccChl} has been in
operation the longest at $\approx 16$~years, and has opportunistically collected
$\approx 10$~years of data\footnote{See
  \url{https://gwosc.org/timeline/show/history/G1_SCI/770000000/501462418/}} in
case of a spectacular gravitational wave event.

\section{Continuous wave searches, \MINYEAR--\MAXYEAR}\label{sec:cont-wave-search}

\begin{figure*}[!t]%
  \centering%
  \includegraphics[width=0.95\linewidth]{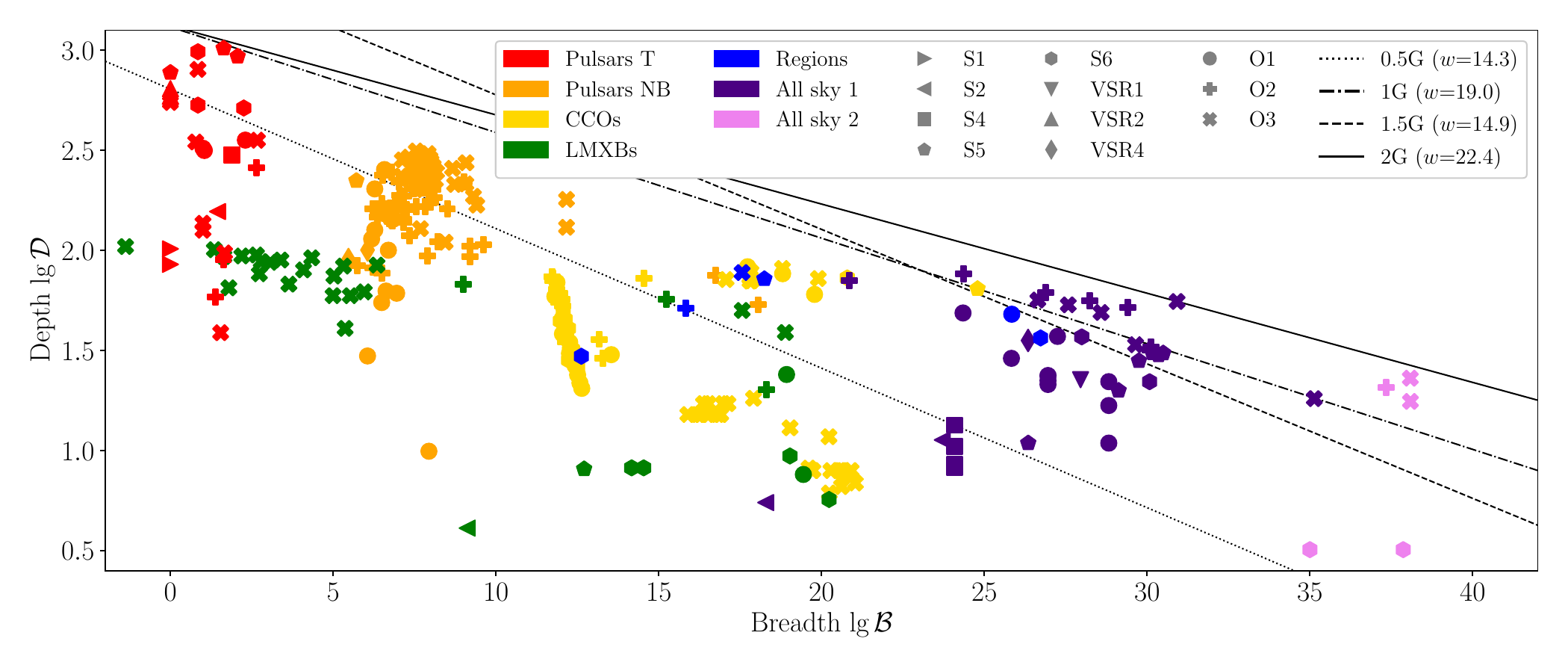}%
  \caption{\label{fig:analysis_overview} Sensitivity depth $\lg\depth$ versus
    parameter-space breadth $\lg\MV$ for \NSEARCHES\ CW searches. Colours denote
    classification into seven search categories (see
    Sec.~\ref{sec:cont-wave-search}): targeted (``Pulsars T'') and narrow-band
    (``Pulsars NB'') searches for known pulsars; directed searches for central
    compact objects (CCOs) and low-mass X-ray binaries (LMXBs); searches
    directed at interesting regions of the sky (``Regions''); and all sky
    searches for isolated neutron stars (``All sky 1'') and neutron stars in
    binary systems (``All sky 2'').  Markers denote the data used by each search
    (Table~\ref{tab:observing_runs}). The lines represent the weight $w$ used to
    calculate the weighted volume $\WVol$ for each detector generation; see
    Sec.~\ref{sec:cont-wave-search} for details. %
  }%
\end{figure*}

In this section, I review searches for CW signals in the LIGO and Virgo data. I
select searches for review that satisfy the following criteria:
\begin{enumerate}

\item Per the subject of this review, I review only searches for CWs from
  rapidly-spinning neutron stars. This excludes several recent
  searches~\cite[e.g.][]{LIGOEtAl2022-AlSrGrvWEmsSBCASpBHLOD,
    LIGOEtAl2022-CnsDPhDMtUDLVrTObsR} for beyond Standard Model particles such
  as ultra-light bosons and dark photons. These phenomena are expected to
  produce signals that follow the same CW signal morphology as for
  rapidly-spinning neutron stars, and are therefore amenable to the same CW
  search techniques.

\item I select only searches which generally assume the CW signal model
  described in Sec.~\ref{sec:cw-signal-model}. I have included searches which
  use an HMM to tracking a wandering CW signal frequency, but excluded searches
  which assume an unmodeled, stochastic
  signal~\cite[e.g.][]{LIGOEtAl2021-SAnsGrvBcUDAdLAVFTObR,
    LIGOEtAl2022-AAllDrSPrGrvWALAVFTObR}.

\item Finally, I select only searches which set upper limits on $h_0$ or an
  equivalent amplitude, so that the sensitivity depth $\depth$ may be computed.

\end{enumerate}
These criteria select \NSEARCHES\ searches from \NPAPERS\ published articles.
They include searches performed by the joint Continuous Wave Working Group of
the LIGO Scientific Collaboration, Virgo Collaboration, and KAGRA Collaboration,
as well as by other CW research groups.

The intention of this review is to show the development of the field of CW data
analysis over time, and illustrate how the challenges outlined in
Sec.~\ref{sec:chall-cont-wave} have been addressed. It does not attempt an
``apples-to-apples'' comparison between CW search designs or algorithms,
i.e. comparing one choice of search design or algorithm against another, while
keeping all other choices the same. As the field has developed, CW analysts will
have faced constraints on resources -- e.g. time, people power, computing power
-- and made practical choices of search design and algorithm within those
constraints. Controlling for those choices equitably is impractical. As
discussed in Sec.~\ref{sec:sensitivity-depth}, the various differences in
determine $h_0^C$ upper limits does not easily allow a precise comparison of
sensitivities. Comparing searches simply by breadth assumes that there is a
uniform probability of CW detection per unit parameter space, and therefore
broader searches should rank higher. In fact, there are good reasons to target
more limited parameter spaces based on promising astrophysically-motivated
priors for CW detection.

Figure~\ref{fig:analysis_overview} plots, for the selected CW searches, their
sensitivity depths $\depth$ against their parameter-space breadths $\MV$. (The
raw data for this plot is provided in Table~\ref{tab:analysis_data}.)  The
searches are divided in seven categories, commonly used in the literature, based
on their astronomical targets:
\begin{enumerate}

\item Targeted searches for known pulsars (``Pulsars T'' in
  Fig.~\ref{fig:analysis_overview}). These searches\footnote{In this review I
    count a survey of known pulsars from a particular paper and analysis
    pipeline as one ``search'', and define its breadth as equal to the number of
    upper limits it outputs. (I do not count upper limits which assume
    restricted priors on $\iota$ and/or $\psi$ based on electromagnetic
    observations.) For example, the most recent known pulsar
    survey~\cite{LIGOEtAl2022-SrGrvWKPTHrmSTLIObR} used three analysis
    pipelines, which are counted as separate searches. The Bayesian analysis
    pipeline (Sec.~\ref{sec:bayesian-inference}) produced 470 upper limits from
    236 pulsars at 2 harmonics (with 2 upper limits excluded), and hence has a
    breadth of 470.}  assume that the CW signal is phase-locked to the
  electromagnetic emission from the pulsar, and so that $\phi\rot(t)$ is given
  by the pulsar's electromagnetic ephemeris. The phase parameter space is
  therefore a single point. Canonical targets are the Crab and Vela pulsars, and
  PSR J0537$-$6910~\cite{LIGOEtAl2020-GrvCnsEqElMlP,
    LIGOEtAl2021-DBSpLCnsGrvWEnYPPJ0, LIGOEtAl2021-CnsLODGrvEmDRGlPPJ0,
    LIGOEtAl2022-SrGrvWKPTHrmSTLIObR}.

\item Narrow-band searches for known pulsars (``Pulsars NB'' in
  Fig.~\ref{fig:analysis_overview}). These searches also target known pulsars,
  but relax the assumption that the electromagnetic and CW phases are
  phase-locked, and allow $\frot$, $\fdrot$, etc.\ to deviate from their
  electromagnetically\-/measured values by a small
  fraction~\cite{LIGOEtAl2022-NrSCnLngTrGrvWKPLTOR}. Narrow-band searches cover
  small parameter spaces in the frequency and spindown parameters.

\item Directed searches for central compact objects (CCOs), suspected to be
  young neutron stars born in core-collapse supernovae. These searches typically
  require only a single point in the sky parameter space: some CCOs are observed
  as bright, well-localised X-ray emission from the centre of the remnant, and
  in any case the size of the remnant can usually be considered small compared
  to the sky parameter resolution. CCOs are not, however, observed as pulsars,
  and their rotational frequency evolution is therefore unknown. The searches
  must therefore cover broad ranges of frequency and (1st, sometimes 2nd)
  spindowns. The most promising CCOs for CW detection -- based on their likely
  ages (young) and distances (close), and the likelihood that the CCO is a
  neutron star -- are in the supernova remnants Cassiopeia A (Cas A) and
  Vela~Jr.~\cite{LIGOEtAl2021-SrCntGrvWYSpRETObRALV,
    LIGOEtAl2022-SEOLDCntGrvWCsVJSpRm}

\item Directed searches for low-mass X-ray binaries (LMXBs). Here, the accretion
  of matter onto the neutron star from a binary companion may build up an
  observable non\-/axisymmetry (Sec.~\ref{sec:accr-from-binary}). Based on the
  torque balance upper limit
  [Eq.~\eqref{eq:Upper_limits_Accreting_binary_system_v_h0_f_M_R_F_X_X_r_m}],
  which scales with the observed X-ray flux, Scorpius X-1 (Sco X-1) is the most
  promising LMXB for CW emission, and has been the primary target of searches to
  date~\cite{LIGOEtAl2022-SrGrvWScrXHdMrMOLD,
    LIGOEtAl2022-MdCrsSGrvWLwXBScXLOD}. Its sky position is well defined, but
  its spin frequency is unknown~\cite{GalaEtAl2022-DSrXPlsScXCXSCnGrvWS}, and
  searches must cover (roughly in order of increasing range) $P$, $\tasc$,
  $a_p$, and $f$.  Other LMXBs, where the spin frequency is known from X-ray
  outbursts, require smaller parameter
  spaces~\cite{LIGOEtAl2022-SCnGrvW20AcMllXPlOLD}. An additional challenge is
  that, due to the time-varying accretion torque, $f$ is likely to wander
  stochastically over long timescales~\cite{MukhEtAl2018-AccSpnENSSXImCnGrWS}.

\item Searches directed at interesting regions of the sky (``Regions'' in
  Fig.~\ref{fig:analysis_overview}). These searches focus on multiple
  astronomical objects concentrated in a particular region of space, generally
  chosen for its prospects of containing young neutron
  stars~\cite{ChenLori2014-GlcCnPlPpl}. These include the Galactic
  centre~\cite{LIGOEtAl2022-SrCntGrvWEmMWCOLID}, globular clusters, and
  star-forming regions. Searches must cover a wide range of frequencies and
  spindowns, and possibly multiple sky positions.

\item All sky searches for isolated neutron stars (``All sky 1'' in
  Fig.~\ref{fig:analysis_overview}). These searches target unknown, isolated
  neutron stars in the Galaxy. There are expected to be $10^{8}$--$10^{9}$ such
  stars~\cite{SartEtAl2010-GlcNtStISVlDstDH}, of which only $\approx 10^{3}$ are
  observed as pulsars~\cite{MancEtAl2005-AstTlsNtFcPCt}. It is hoped that a
  sub-population of these neutron stars will be strong gravitational wave
  emitters, known as \emph{gravitars}~\cite{Palo2005-SmlPpIsNSEvTEGrvW,
    KnisAlle2008-BlnArStCnGrvWS, WadeEtAl2012-CntGrvWIsGlNSAdDE,
    CiesEtAl2021-DtcCnGrvWINSMWPpSyA}. Searches cover a 4-dimensional parameter
  space: sky, frequency, and spindowns~\cite{LIGOEtAl2021-AlSCntGrvWIsNtSEOLD,
    LIGOEtAl2022-ASCntGrvWIsNSUAdLAdVOD}.

\item All-sky searches for neutron stars in binary systems (``All sky 2'' in
  Fig.~\ref{fig:analysis_overview}). These searches target unknown Galactic
  neutron stars in binary systems, and cover a 6-dimensional parameter space of
  the sky, frequency, and binary orbital
  parameters~\cite{LIGOVirg2021-ASEOLDCntGrvSgUnNtSBS,
    CovaEtAl2022-CnsRMnMllNtSBSy}. It is assumed, based on known pulsar
  observations, that the orbits will be close to circular, and so we can avoid
  the additional computational expense of searching over the higher-dimensional
  eccentric orbital parameters.

\end{enumerate}
Figure~\ref{fig:analysis_overview} illustrates that CW searches generally
trade-off between high sensitivity, or broad parameter-space coverage. The seven
search categories group searches along a spectrum of trade-offs between depth
and breadth: targeted searches achieve the greatest sensitivity depth over a
limited parameter space, which all-sky searches for binary systems cover the
widest parameter spaces at limited sensitivity depths, and generally the other
categories fall somewhere in between. (An exception are the LMXB searches
performed in~\cite{LIGOEtAl2022-SCnGrvW20AcMllXPlOLD} where, due to the known
spin frequencies of the targets, the parameters spaces are relatively small and
more closely resemble the narrow-band pulsar searches.)  Noting the data used by
each search (Table~\ref{tab:observing_runs}), we see that search depth has
typically increased with time; given that $\depth$ factors out improvements in
detector sensitivity (i.e.\ $\Sh$), this represents an improvement in CW search
techniques themselves. To date, the deepest CW search is a targeted pulsar
search of S5 data~\cite{PitkEtAl2015-FRFPrDlhSrGrvWSpNS}, while the broadest
search is an all-sky search for binary neutron stars in O3
data~\cite{CovaEtAl2022-CnsRMnMllNtSBSy}.

The parameter spaces of the (targeted, narrow-band) known pulsar searches have
remained similar from run to run (Fig.~\ref{fig:analysis_overview}), as they
scale with the number of known pulsars within the sensitive band of the
detectors $\frot \gtrsim 50$~Hz; this population of relatively fast-spinning
pulsars has grown modestly over time. Similarly, searches for the LMXBs Sco X-1
and XTE J1751$-$305 (green markers with $\lg\MV \gtrsim 9$) have covered
generally the same parameter spaces, as e.g.\ for Sco X-1 the uncertainties in
its orbital parameters from electromagnetic observations have remained broadly
similar over time~\cite{GallEtAl2014-PrEphGrvSrIScX,
  PremEtAl2016-PrEphGrvWvSrIICX2-II, WangEtAl2018-PrEphGrvSrIIIRSPrScoX1-III,
  KillEtAl2023-PrEphGrvSIVCrREpSX1-IV}.  In contrast, parameter-space coverage
of all-sky isolated searches initially saw dramatic increases, from
$\lg\MV \approx 18$ in the S2 run to $\approx 30$ by the 1G detector era
(Tables~\ref{tab:observing_runs},~\ref{tab:analysis_data}), where searches in
subsequent runs have clustered. The CCO, regional, and all-sky binary searches
have seen more mixed evolution in parameter-space coverage, due to different
targets and search design choices; e.g.\ CCOs have typically been targeted using
either fully-coherent searches over a limited data sets, or searches of all the
data from a run using computationally cheaper, less sensitive algorithms.

\begin{figure*}[!t]%
  \centering%
  \includegraphics[width=0.95\linewidth]{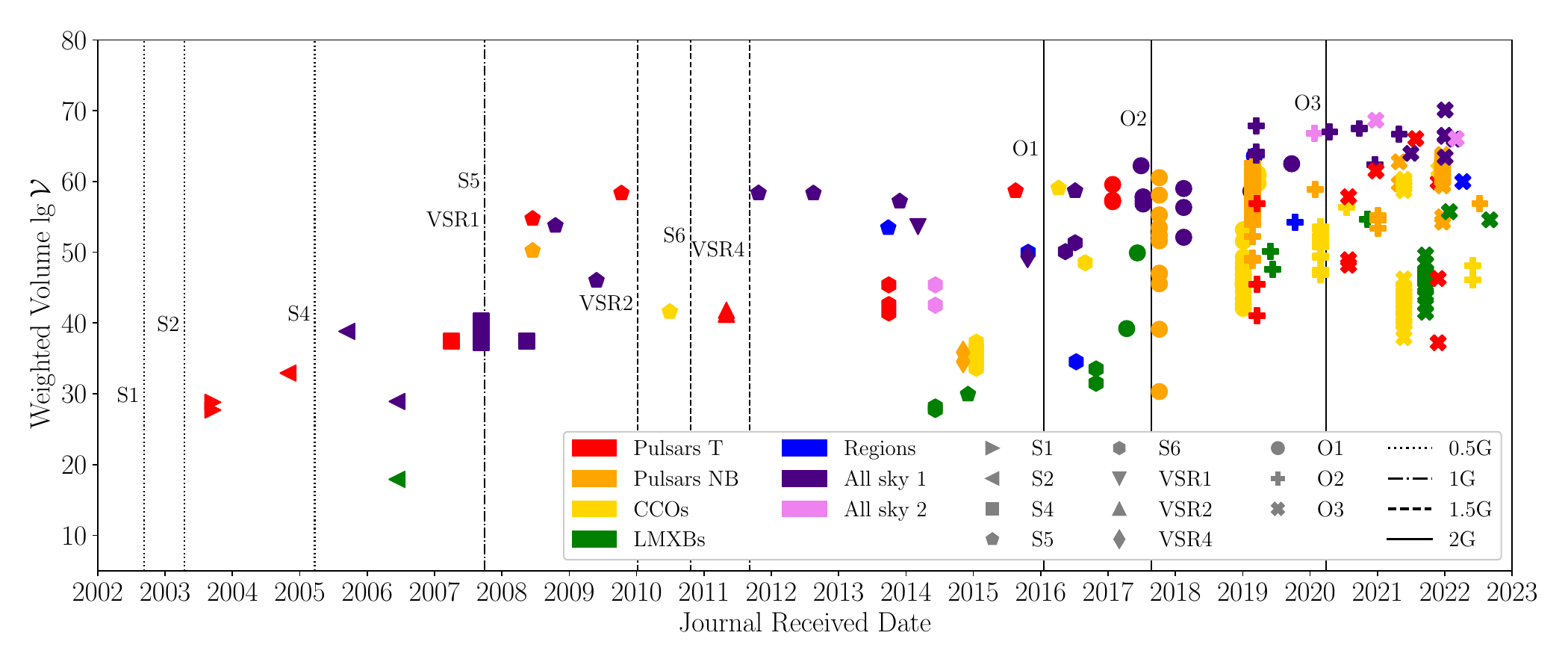}%
  \caption{\label{fig:analysis_timeline} Weighted sensitivity depth-parameter
    space volume $\WVol$ of \NSEARCHES\ CW searches versus time (given by the
    journal submission date of the article presenting each search). Colours and
    markers have the same meanings as in
    Fig.~\ref{fig:analysis_overview}. Vertical lines (with labels on the left)
    indicate the end date of each data collection run listed in
    Table~\ref{tab:observing_runs}. %
  }%
\end{figure*}

Given the inverse correlation between depth and breadth, we might suppose that
their product -- a sensitivity depth-parameter space \emph{volume} -- might
serve as a useful figure of merit for CW searches. Given the very different
scales of $\depth$ and $\MV$, however, it seems naive to simply multiply
them. Instead, I define a weighted volume
\begin{equation}
  \WVol = \depth^{w} \MV \,.
\end{equation}
The weight $w$ is chosen separately for each detector generation (as defined in
Table~\ref{tab:observing_runs}) as follows. For a given detector generation,
select the two searches with: the maximum $\depth$, and the maximum $\MV$. Then,
define $w$ as
\begin{equation}
  w = -\frac{
    \log\MV^{\text{max $\depth$}} - \log\MV^{\text{max $\MV$}}
  }{
    \log\depth^{\text{max $\depth$}} - \log\depth^{\text{max $\MV$}}
  } \,,
\end{equation}
where $\cdot^{\text{max $\depth$}}$ and $\cdot^{\text{max $\MV$}}$ denote a
quantity from the search with the maximum $\depth$ and $\MV$
respectively. Geometrically, a line drawn on Fig.~\ref{fig:analysis_overview}
with slope $-1/w$ and appropriate intercept would pass through both the maximum
$\depth$ and maximum $\MV$ searches selected to calculate $w$.

For each detector generation, lines are drawn in
Fig.~\ref{fig:analysis_overview} with slopes $-1/w$. The intercepts of each line
are chosen such that the line passes through the search with the maximum $\WVol$
for that detector generation; note that this search is not necessarily either of
the searches with the maximum $\depth$ and $\MV$ used in compute $w$. The
weights found for each detector generation are given in the legend of
Fig.~\ref{fig:analysis_overview}. We see that $w$ increases with each
generation, reflecting the general trend that CW searches have increased more in
depth than in breadth over time.

We should be cautious about using $\WVol$ as a single ranking measure to decide
which CW search is ``best''. In addition to the many difficulties to fairly
comparing searches (as discussed above in this section, and in
Sec.~\ref{sec:sensitivity-depth}), we note the following properties of
$w$. First, $w$ is defined separately for each detector generation, and hence
follows changes in CW search design and algorithm over time, rather than giving
a time-independent ``best'' ranking. Second, $w$ is defined by two extreme
searches within a detector generation cohort -- those with the maximum $\depth$
and maximum $\MV$. The $\WVol$ computed for each search, therefore, depends on
which other searches are present in the cohort; addition of new searches with
greater $\depth$ or $\MV$ would change $w$ and therefore the $\WVol$ of all
searches. Consistent with the spirit of this review, we interpret $\WVol$ as
representing the general trend in CW search performance over time and within
detector generation cohorts.

Figure~\ref{fig:analysis_timeline} plots the weighted volume $\WVol$ of each CW
search against time. We can see a marked increase in volume from 2003 to 2010,
corresponding to the initial 0.5--1.5G detectors, followed by a plateau from
2010 to 2016 as the detectors upgraded to the 2G generation. Since 2016 $\WVol$
has continued to increase, albeit more moderately. Together with
Fig.~\ref{fig:analysis_overview}, this suggest that CW searches were initially
driven by increasing parameter-space coverage in the 1G detector
era, but have increasingly been driven by improving sensitivity depth (albeit at
a slower rate) in the 2G detector era.

We also see an increased concentration in number of searches performed in the 2G
detector era. This reflects increased activity within the Continuous Wave
Working Group and from other CW research groups, as well as a growing diversity
of CW search algorithms and astronomical targets. It also suggests a shorter
cadence from the acquisition of data to the submission of searches for
publication, driven by deepening experience of the field with all aspects of
search design and execution. Higher quality data with greater incentives to
perform analyses, more frequent observing runs, and shorter proprietary periods
for LIGO and Virgo data sets may also be relevant factors.

In addition to an increasing density of searches, we also see a broader range in
the $\WVol$ of searches performed in the 2G detector era. In the early years of
the field, many CW searches were the first of their kind, and therefore faced
little competition. As the number of searches increases, there is likely a
desire for each search to differentiate itself from other searches in terms of
design and algorithm choices. There is a tradition of multiple CW pipelines
covering similar parameter spaces~\cite[e.g.][]{LIGO2008-AlSrPrdGrvWvLSD}, in
order to guard against software bugs in one pipeline preventing a
detection. This essential redundancy must nonetheless be balanced against
putting too many resources into analysing identical parameter spaces, and
practical concerns such as being able to publish searches which are sufficiently
novel.

The broadened range of $\WVol$ in the 2G detector era suggests that, in seeking
to differentiate searches from each other, CW analysts are not solely driven by
maximising the obvious metrics of sensitivity depth and parameter-space breadth,
but also by other considerations. These include focusing on parameter spaces
which are interesting for astrophysical reasons, if not for their size. For
example, both the Crab pulsar~\cite{RajbEtAl2021-FrSrcGrvWvRMCPl} and PSR
J0537$-$6910~\cite{FesiPapa2020-FrSrRmGrvWPJ05,
  LIGOEtAl2021-CnsLODGrvEmDRGlPPJ0} has received special interest in recent
years as potential sources of $r$-mode emission; in particular, observations of
PSR J0537$-$6910 suggest that, in between frequent glitches, its spindown rate
is consistent with a braking index of $n = 7$ which would be consistent with
$r$-modes~\cite{AndeEtAl2018-EngSEvPJ0RGrvWCCnT,
  HoEtAl2020-RtBGltNTmGlPJ05}. Another factor is the expansion of the CW signal
space beyond the model of Sec.~\ref{sec:cw-signal-model}, e.g. by using HMMs
(Sec.~\ref{sec:viterbi-algorithms}) which permit the CW signal frequency to
randomly wander. Such considerations are not easily captured by the simple
metrics used in this section.

\section{Continuous wave search algorithms}\label{sec:cont-wave-alg}

In this section, I review the algorithms and pipelines used in the CW searches
examined in Sec.~\ref{sec:cont-wave-search}. Following the spirit of the
previous section, I aim to qualitatively compare the different algorithms to
illustrate general principles and trends. I refer the reader to the cited
references for technical descriptions of each algorithm.

Given a model for the CW signal waveform (Sec.~\ref{sec:cw-signal-model}), the
optimal algorithm is to simply match filter the whole data set against the model
over the parameter space of interest. While this approach is possible when all
phase parameters of the signal are well known, such as for targeted and
narrow-band searches for known pulsars, it is otherwise not achievable over any
wide parameter space. Let us demonstrate this concretely by attempting to
perform an all-sky search of 1~year of data, covering typical ranges of
$f \in [50, 1000] \text{ Hz}$ and $\fd \in [-10^{-8}, 0] \text{ Hz
  s}^{-1}$. Applying for formulae for parameter-space breadth given in
Sec.~\ref{sec:param-space-breadth}, we calculate $\MVsky = 9.9{\times}10^{12}$,
$\MVfreq = 4.0{\times}10^{10}$, $\MVfd = 2.3{\times}10^{06}$, and
$\MV = 9.2{\times}10^{29}$. Let us assume that $\MV$ gives the rough order of
magnitude for the number of matched filters we must apply to the data to cover
this parameter space. On contemporary computer hardware, it takes
$\approx 8.1{\times}10^{-4}$~s to apply one matched filter to a year's worth of
data~\cite{Prix2017-ChTMmrFsImpLA}. The computational cost of our search is
therefore $\approx 2.3{\times}10^{19}$~yr. We could then complete the search
with $\sim 5$~\emph{billion} computers running for $\sim 5$~\emph{billion} years
-- just in time before the Sun becomes a red giant and engulfs the Earth\dots

Given this obvious impracticality, CW search algorithms designed to cover wide
parameter spaces all follow a \emph{hierarchical}
structure~\cite{BradEtAl1998-SrcPrdSrLI, SchuPapa1999-EnAlHrrASrLngGSG600,
  BradCrei2000-SrcPrSrLIIHrrSr-II}. To start, the whole data set (with time-span
$\Tspan$) are partitioned in time into $N$ \emph{segments}, each of which span a
\emph{coherence time} $\Tcoh \ll \Tspan$. In the first, \emph{coherent stage} of
the hierarchical pipeline, matched filters covering the phase parameter space of
interest are applied to each segment individually. The number of matched filters
required generally scales with $\Tcoh^{\tilde{\delta}}$, with
$\tilde{\delta} \gtrsim 6$~\cite{PrixShal2012-SCntGrvWOpStMFCmC}. (Consider the
powers of $\Tspan$ in the breadth formulae given in
Sec.~\ref{sec:param-space-breadth}.) This steep scaling is too computationally
expensive for a fully-coherent search, but we can choose $\Tcoh$ to be short
enough to make the computational cost manageable.

In the second, \emph{semi-coherent stage} of the hierarchical
pipeline,\footnote{This stage is sometimes referred to as the ``incoherent''
  stage; the term ``semi-coherent'' is also often used to describe a
  hierarchical pipeline as a whole. In this review I use ``semi-coherent'' to
  refer specifically to the algorithm used in the second stage, and
  ``hierarchical'' to refer to the pipeline as a
  whole. See~\cite{CutlEtAl2005-ImStcSrGrvPl} for a study of hierarchical
  pipelines with more than two stages.} an algorithm is selected to apply to the
matched filter results from the $N$ segments. Over the same phase parameter
space as the coherent stage, this algorithm adds together $N$ matched filter
results from the $N$ segments in a manner consistent (to a degree specified by
the algorithm) with the CW signal waveform over the whole data set. In general,
semi-coherent algorithms do not require the \emph{phase} of the CW signal
$\phi(t\NS)$ [Eq.~\eqref{eq:phase-at-NS}] to be consistent over all segments;
instead, they require only that the \emph{frequency} of the CW signal $f(t\NS)$
[Eq.~\eqref{eq:frequency-at-NS}] be consistent. Put another way, the phase is
allowed to jump by some random offset from segment to segment, while the
frequency remains consistent across segments:
\begin{equation}
  \label{eq:semicoh-consistency}
  \begin{aligned}
    \phi(t)_{N+1} &= \phi(t)_{N} + \phi_{\mathrm{random}} \,, \\
    2\pi f(t)_{N+1} &= \frac{d\phi(t)_{N}}{dt} + \frac{d\phi_{\mathrm{random}}}{dt} \\
    &= \frac{d\phi(t)_{N}}{dt} + 0 = 2\pi f(t)_{N} \,,
  \end{aligned}
\end{equation}
where $\cdot_{N+1}$ and $\cdot_{N}$ label the $(N{+}1)$th and $N$th segments
respectively.

Relaxing phase consistency between segments reduces sensitivity, and makes the
detection statistic computed by the semi-coherent algorithm more susceptible to
instrumental line artefacts at near-constant
frequencies~\cite{CovaEtAl2018-IdnMtNSArDSPrGrvWFTORAL}. The significant benefit
of this approach, however, is that the computational cost of the semi-coherent
stage no longer scales as
$\Tspan^{\tilde{\delta}} = N^{\tilde{\delta}} \Tcoh^{\tilde{\delta}}$, but as
$N^{\hat{\eta}} \Tcoh^{\hat{\delta}}$ with
$\hat{\eta} \ll \hat{\delta} \lesssim
\tilde{\delta}$~\cite{PrixShal2012-SCntGrvWOpStMFCmC}. This saves a factor
$\sim N^{\tilde{\delta} - \hat{\eta}} \Tcoh^{\tilde{\delta} - \hat{\delta}}$ in
computational cost compared to a fully-coherent analysis, which is generally
substantial enough to make hierarchical searches of year-long data sets
tractable.  Indeed, given the constraint of a fixed computing
budget~\cite{PrixShal2012-SCntGrvWOpStMFCmC}, a hierarchical search is often
\emph{more sensitive} than a fully-coherent analysis, simply because it is
computationally cheap enough to be able to analyse all available data; a
fully-coherent search, on the other hand, would be forced to analyse only a
subset of the data, due to its steep computational cost scaling with $\Tspan$,
and would thereby degrade in sensitivity.

In the remainder of this section I review the algorithms typically employed at
the coherent (Sec.~\ref{sec:coher-match-filt}) and semi-coherent
(Sec.~\ref{sec:semic-algor}) stages of a hierarchical pipeline. Other algorithms
and applications are briefly mentioned in Sec.~\ref{sec:other-algorithms}.

\subsection{Coherent matched filtering algorithms}\label{sec:coher-match-filt}

\begin{figure}[!t]%
  \centering%
  \includegraphics[width=0.95\linewidth]{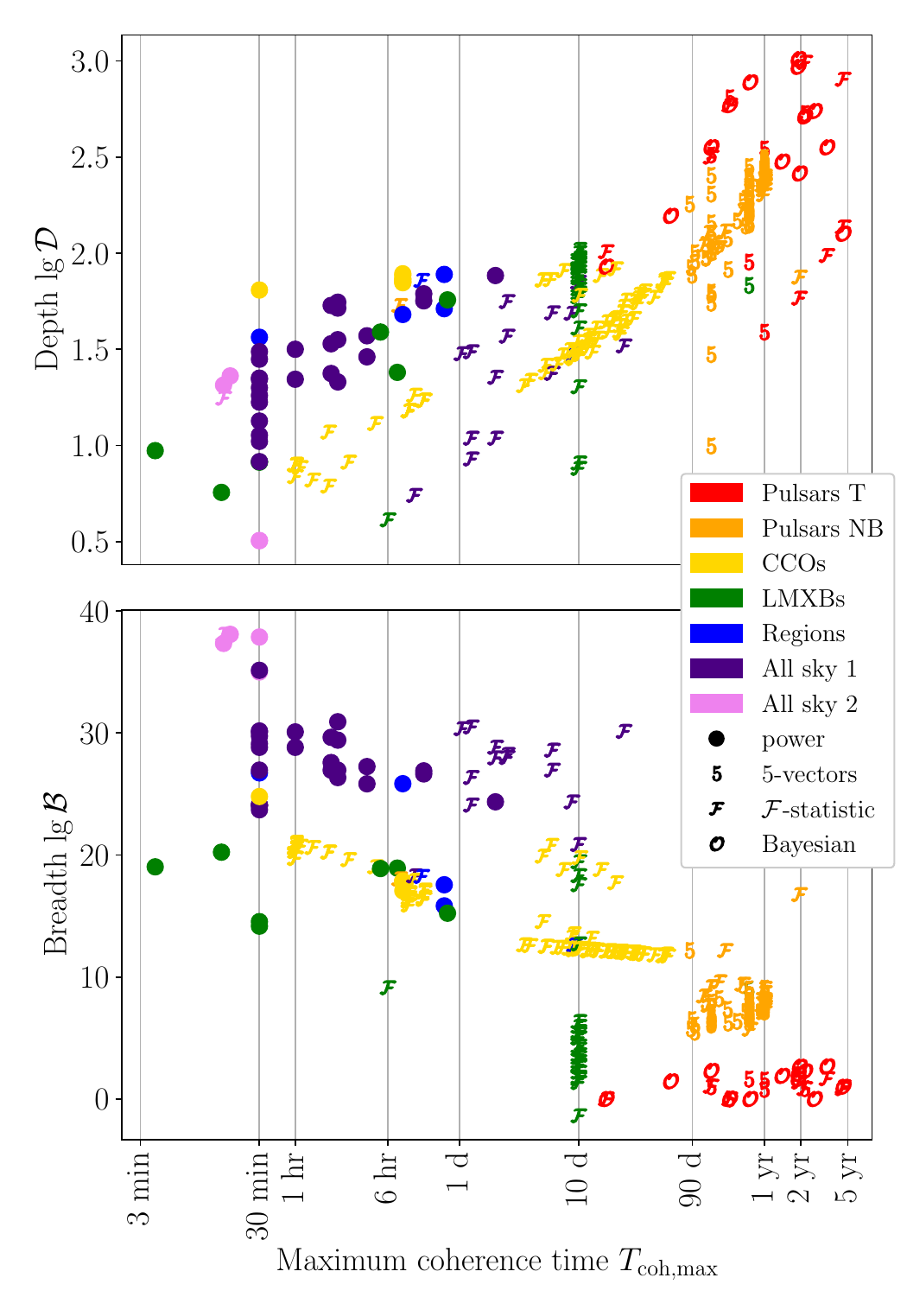}%
  \caption{\label{fig:analysis_coherence_times} Sensitivity depth $\depth$ (top
    panel) and parameter-space breadth $\MV$ (bottom panel) of \NSEARCHES\ CW
    searches, plotted against the maximum coherence time $\Tcohmax$ used in each
    search. Colours have the same meanings as in
    Fig.~\ref{fig:analysis_overview}; markers denote the coherent algorithm
    employed by each search (described in Sec.~\ref{sec:coher-match-filt}). %
  }%
\end{figure}

Figure~\ref{fig:analysis_coherence_times} plots $\depth$ and $\MV$ for each
search against the coherence time $\Tcoh$ used in the first coherence
stage. (Where a search uses more than one coherence time, the maximum
$\Tcohmax = \max \Tcoh$ is shown.) We immediately see that search sensitivity
increases with longer coherence times, but parameter-space coverage decreases
due to the steep increase in computational cost with $\Tcoh$. The four coherent
algorithms labelled in the figure are described in the remainder of this
section.

\subsubsection{Frequency-domain power}\label{sec:freq-doma-power}

Over a ``short enough'' time $\Tcoh$, the CW signal frequency is approximately
constant. In this case, the simplest coherent algorithm is to compute the
discrete Fourier transform of a data segment, and then compute the power
(i.e. the sum of the squares of the real and imaginary Fourier components) of
the bin where the CW signal is expected to be. The condition for $\Tcoh$ to be
``short enough'' so that the CW signal power is concentrated in one bin is, more
precisely,\footnote{Equation~\eqref{eq:Tsft-max} is derived
  in~\cite{LeacPrix2015-DrSCnGrvWBSPrmMOSXSn} for any R{\o}mer delay, and
  implemented as given in the function \texttt{XLALFstatMaximumSFTLength()} of
  the LALSuite~\cite{lalsuite} software package.}
\begin{equation}
  \label{eq:Tsft-max}
  \Tcoh^2 \le \frac{ 6 \sqrt{ 5 \mu_{\Tcoh} } }{ \pi \max \{ f \} \max \{ \tauS, a_p \} \max \{ \OmegaS^2, \Omega_P^2 \}  }
\end{equation}
where: $\mu_{\Tcoh}$ sets the fraction of signal power we are prepared to lose
for a particular choice of $\Tcoh$; $\max \{ f \}$ is the maximum frequency over
the parameter space; $\max \{ \tauS, a_p \}$ is the maximum of the Earth's
radius (Sec.~\ref{sec:sky-parameter-space}) and the searched binary orbit
projected semi-major axis (if any); and $\max \{ \OmegaS^2, \Omega_P^2 \}$
maximises the square of either the Earth's sidereal angular frequency or (if
any) the searched angular frequency of a binary orbit.

Gravitational wave data, divided into segments of time-span $\Tcoh$ as above and
Fourier transformed, serves as a common input data product, not only for
semi-coherent algorithms that sum power in each segment, but also for algorithms
such as the \Fstat\ (Sec.~\ref{sec:fstat}). Common file formats are the Short
Fourier Transform (SFT;~\cite{AlleEtAl2022-SFDtFrVrs23Spc}) format, and the
Short FFT DataBase (SFDB;~\cite{AstoEtAl2005-SFDtbPMHrrSrPrSr}).

\subsubsection{5-vectors algorithm}\label{sec:5-vectors-algorithm}

The 5-vectors method~\cite{AstoEtAl2010-MDtKSrCnGrvWSNnsD,
  AstoEtAl2012-ChSCntGrvWSEx5MNDt} is based on the following property of the CW
signal; when the phase parameters of the signal are fully specified (i.e.\ we
know precisely the function $\hplus(t), \hcross(t)$ in Eq.~\eqref{eq:hoft}), one
is left with only the modulation from the detector responses
$\Fplus(t), \Fcross(t)$. These functions are periodic, with angular frequencies
of $\OmegaS$ and $2\OmegaS$. Their effect on a CW signal with known angular
frequency $2\pi f(t)$ is to generate four side-bands, at $2\pi f(t) \pm \OmegaS$,
$2\pi f(t) \pm 2 \OmegaS$, resulting in five harmonics in total. The 5-vectors
method sums these 5 harmonics, with appropriate weights derived from
$\Fplus(t), \Fcross(t)$.

To date, the 5-vectors method has been mostly used as a fully-coherent algorithm
for targeted and narrow-band searches for known pulsars. The phase demodulation
of the signal using the known $\hplus(t), \hcross(t)$ is accomplished
efficiently using the Band-Sampled Data
(BSD;~\cite{PiccEtAl2019-NDAnFrSCntGrvWSg}) framework, which provides
band-limited data, heterodyned at regular intervals (typically 10~Hz), and
down-sampled to speed up further computations. The method was extended
in~\cite{SingEtAl2019-RsmAlDCnGrvSgNSBS} to perform directed searches for LMXBs.

\subsubsection{\Fstat}\label{sec:fstat}

The \Fstat\ is the log-likelihood function of the observed data given a CW phase
modulation, maximised over the four amplitudes of
Eq.~\eqref{eq:hoft-four-Ampi-hi}; see Sec.~\ref{sec:chall-cont-wave}.  It is
extensively studied in a series of papers~\cite{JaraEtAl1998-DAnGrvSgSpNSSDtc,
  JaraKrol1999-DAnGrvSSpNSIIAEstPr-II, JaraKrol2000-DAGrvSSNSIIIDtStCmpRq-III,
  AstoEtAl2002-DAnlGrvSgSpNtSIVAS-IV, AstoEtAl2010-DAnGrvSgSpNSVNrrAS-V}, the
first of which is colloquially referred to by CW analysts as ``JKS''. It has
been employed in a wide variety of CW searches: for known pulsars (as a
full-coherent search), directional targets such as CCOs and LMXBs (both
fully-coherently and as the first stage of a hierarchical pipeline), and all-sky
searches (as part of a hierarchical pipeline).

The \Fstat\ has the following statistical properties. In the absence of a
signal, and assuming Gaussian noise, the value of $\twoF$ follows a central
chi-squared distribution with four degrees of freedom. When a signal is present,
$\twoF$ follows a non-central chi-squared distribution with four degrees of
freedom and non-centrality parameter $\rho^2$. Here, $\rho^2$ is the optimal
signal-to-noise ratio when signal and template are perfectly
matched~\cite{JaraEtAl1998-DAnGrvSgSpNSSDtc}. Due to these useful properties,
explicit values of the \Fstat\ quoted in the literature are usually values of
$\twoF$ and \emph{not} values of $\oneF$.

Several implementations of software to compute the \Fstat\ have been
developed. A time-domain implementation~\cite{PoghEtAl2015-ArImpPrlSfSPGrWS}
uses the ``resampling'' technique first proposed
in~\cite{JaraEtAl1998-DAnGrvSgSpNSSDtc}. This technique uses the Fast Fourier
Transform to simultaneously compute $N$ values of $\twoF$ at regularly-spaced
frequencies. The computational cost of resampling scales as $\order(\log N)$,
whereas computing each value of $\twoF$ individually would scale as $\order(N)$.
The LALSuite~\cite{lalsuite} software package contains an independent
implementation of the ``resampling''
technique~\cite{PateEtAl2010-ImpBrRsCnWSGrvWD, Prix2017-RsmFFImpFs,
  Prix2017-ChTMmrFsImpLA, DunnEtAl2022-GPrUImpFsCnGrvWS} which takes
frequency-domain SFT data files as input.  LALSuite also contains another
frequency-domain technique, known as ``demodulation'', which efficiently
computes individual $\twoF$ values~\cite{WillSchu2000-EfMFlAlDCntGrvWS,
  Prix2018-FsImpCmp, Prix2017-ChTMmrFsImpLA}.

For neutron star sources in binary orbits, we must demodulate the signal phase
according to $\Delay{R}(t\NS)$ (Sec.~\ref{sec:time-at-solar}). An alternative
technique~\cite{SuvoEtAl2017-HMMTrCntGrvWBNSWnSIIBOPTr-II} is to compute the
\Fstat\ omitting $\Delay{R}(t\NS)$ from the phase; the signal is then split into
multiple side-bands according to the Jacobi-Anger
expansion~\cite[cf.][]{SammEtAl2014-ImpFrqSdSMGrWLMXBn}. A subset of the
side-bands may then be added together to recover the \Fstat.

\subsubsection{Bayesian inference}\label{sec:bayesian-inference}

Bayes' theorem gives the probability of a model, given some observed data -- the
posterior probability -- from the following ingredients: a prior probability for
the model, the likelihood of the data given the model, and the evidence (or
marginal likelihood). Provided that the prior probability distributions are
consistent with the population of signals one expects to detect, Bayesian
inference provides the most powerful detection
statistic~\cite{Sear2008-MntByTcGrvWBDAn}. The \Fstat, for example, does not
satisfy this criteria; the maximisation over the $\Amp^i$ amplitudes implicitly
assumes unphysical prior distributions for the underlying model parameters $h_0$
and $\cos\iota$~\cite{PrixKris2009-TrSCnGrvWBVMxmSt}.

A Bayesian inference pipeline is used to perform surveys of the known
pulsars. The relative flexibility of Bayesian inference, compared to ad-hoc
constructed statistics, permits the use of the complicated pulsar timing
model~\cite{HobbEtAl2006-TENwPlsPcIOvr, EdwaEtAl2006-TENPlTPcIITMPrcEst-II},
including eccentric binary orbits and irregular timing
noise~\cite{PitkWoan2004-SrcGrvWCPlPrTmN, PitkWoan2007-BnSyDTNSrcGrvWKPl}. The
pipeline first heterodynes the data at the prescribed CW phase inferred from the
pulsar ephemeris, then computed posterior probabilities on the four physical
amplitude parameters $h_0, \cos\iota, \psi, \phi_0$.

The first implementation of the pipeline~\cite{DupuWoan2005-ByEstPlPrGrvWD}
computed the posterior probability using a Markov Chain Monte Carlo (MCMC). A
more efficient implementation~\cite{PitkEtAl2017-NSmCTrSrCntGrvWP} uses nested
sampling~\cite{Skil2006-NsSmpGnByCmp}. It is able to search over small ranges of
the phase parameters, quantify detection significance, and perform model
comparison. It can also search at both harmonics of the CW
signal~\cite{PitkEtAl2015-FRFPrDlhSrGrvWSpNS}, and model gravitational wave
polarisations predicted by theories of gravity other than general
relativity~\cite{IsiEtAl2017-PrDyGPlrCnGrvW}. This implementation is part of
LALSuite~\cite{lalsuite}. A third generation of the
pipeline~\cite{Pitk2022-CPPcInCntGrvSgP} is written in Python, and can use a
variety of Bayesian inference solvers via~\cite{AshtEtAl2019-BUsrByInLGrvAs}.

\subsection{Semi-coherent algorithms}\label{sec:semic-algor}

\begin{figure*}[!t]%
  \centering%
  \includegraphics[width=0.95\linewidth]{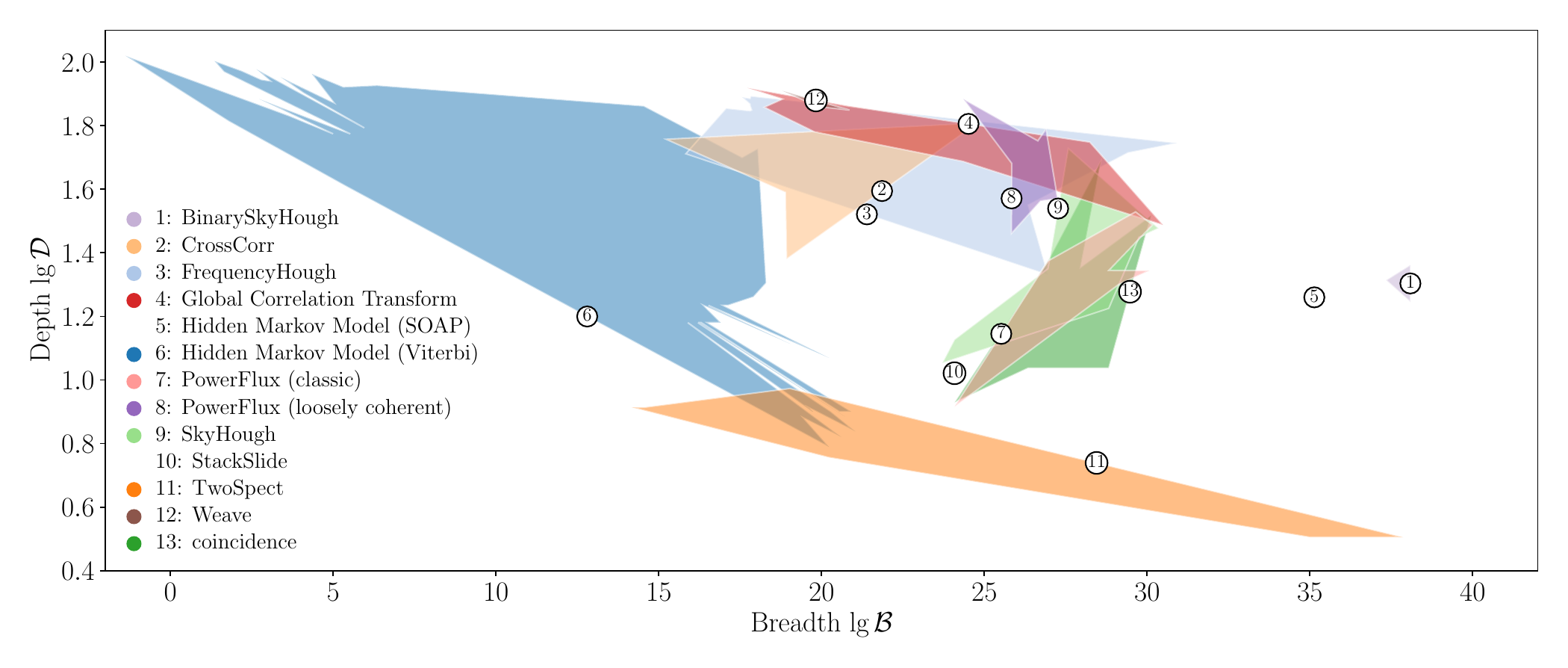}%
  \caption{\label{fig:analysis_inc_depth_breadth_cov} Sensitivity depth
    $\lg\depth$ versus parameter-space breadth $\lg\MV$ for \NSEARCHES\ CW
    searches. Plotted are polygons, each of whose vertices indicates a CW search
    using a given semi-coherent algorithm. Names of all semi-coherent algorithms
    are listed in the legend. Each polygon is given a numeric label, placed on
    an arbitrary edge, corresponding to the legend. Where only one CW search
    uses a particular semi-coherent algorithm, no polygon is plotted and instead
    the numeric label marks the search $(\lg\MV, \lg\depth)$.  Colours are
    arbitrary.%
  }%
\end{figure*}

\begin{figure}[!t]%
  \centering%
  \includegraphics[width=0.95\linewidth]{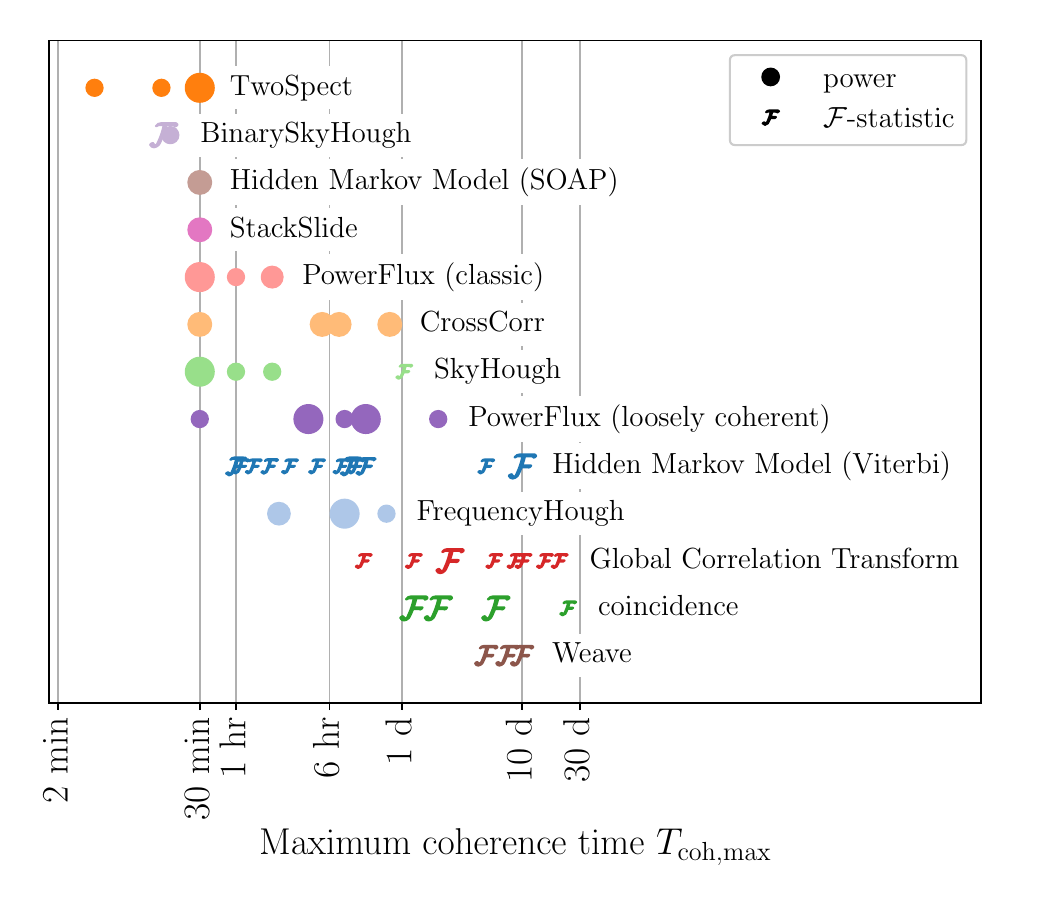}%
  \caption{\label{fig:analysis_incoherent_times} Maximum coherence times
    $\Tcohmax$ chosen by the semi-coherent algorithms used in \NSEARCHES\ CW
    searches. Colours are arbitrary but are the same as in
    Fig.~\ref{fig:analysis_inc_depth_breadth_cov}; markers denote the coherent
    algorithm used by the hierarchical pipelines; marker size indicates the
    relative popularity of a particular choice of $\Tcohmax$ for a given
    algorithm. %
  }%
\end{figure}

Figure~\ref{fig:analysis_inc_depth_breadth_cov} plots $\depth$ versus $\MV$, as
in Fig.~\ref{fig:analysis_overview}; here, however, we highlight the 13
semi-coherent algorithms used by the searches, by plotting polygons whose
vertices are the searches. The polygons illustrate the area in $\MV$--$\depth$
space each algorithm is demonstrably capable of operating over; this gives some
idea of the typical configurations, and the flexibility thereof, for each
algorithm. For example, the Viterbi HMM algorithm
(Sec.~\ref{sec:viterbi-algorithms}) is capable of a wide variety of
configurations due to its computational efficiency; while the TwoSpect algorithm
(Sec.~\ref{sec:twospect-algorithm}), which is specialised for neutron stars in
binary systems, can be configured for both all-sky surveys (higher $\MV$) and
directed targets (lower $\MV$). Many algorithms have a traditional focus on a
particular parameter space. The following algorithms have typically been used
for all-sky searches for isolated neutron stars: coincidence
(Sec.~\ref{sec:coinc-based-algor}), SkyHough and FrequencyHough
(Sec.~\ref{sec:hough-transf-algor}), StackSlide, Global Correlation Transform
(Sec.~\ref{sec:stacksl-algor}), and PowerFlux (and its loosely coherent
extension; Sec.~\ref{sec:powerflux-algorithms}). CrossCorr
(Sec.~\ref{sec:cross-corr-algor}) focuses on LMXBs, Sco X-1 in particular, while
BinarySkyHough (Sec.~\ref{sec:hough-transf-algor}) is designed for all-sky
searches for neutron star in binaries. Relatively newer algorithms, such as
Weave (Sec.~\ref{sec:stacksl-algor}) and the SOAP HMM algorithm
(Sec.~\ref{sec:viterbi-algorithms}) have only been used in a few searches to
date.

Figure~\ref{fig:analysis_incoherent_times} plots the maximum coherence time
$\Tcohmax$ typically chosen for each semi-coherent algorithm. (A reminder that
$\Tcohmax$ is the maximum $\Tcoh$ used by a given search, in cases where more
than one coherence length is employed.) Seven of the 13 semi-coherent algorithms
use power (Sec.~\ref{sec:freq-doma-power}) as the first-stage coherent
algorithm; four of the 13 use the \Fstat\ (Sec.~\ref{sec:fstat}); and two of the
13 may use either. For semi-coherent algorithms using power, $\Tcoh = 30$~min is
a popular choice of coherence time as it satisfies Eq.~\eqref{eq:Tsft-max} up to
$f \lesssim 2$~kHz,\footnote{This is the upper frequency limit on CW searches
  imposed by the 4~kHz sampling rate of the calibrated $h(t)$ data prepared by
  the LIGO Scientific Collaboration and Virgo Collaboration.} and for that
reason is the standard choice of time-span for SFTs.  Longer coherence times are
possible when summing power at lower frequencies; on the other hand, searches
over binary orbit parameters may require shorter SFTs if $\Omega_P < \OmegaS$
[Eq.~\eqref{eq:Tsft-max}]. Semi-coherent algorithms which use the \Fstat\ are
generally not restricted by Eq.~\eqref{eq:Tsft-max}, except that computational
cost scales steeply with $\Tcoh$.

The remainder of this section outlines the semi-coherent algorithms referenced
in Figs.~\ref{fig:analysis_inc_depth_breadth_cov}
and~\ref{fig:analysis_incoherent_times}. In essence, each algorithm ultimately
performs the same task: to correctly demodulate the CW signal frequency, while
allowing some flexibility in the CW signal phase per
Eq.~\eqref{eq:semicoh-consistency}. That said, the algorithms vary considerably
in their conceptual basis and design choices.

\subsubsection{Coincidence-based algorithms}\label{sec:coinc-based-algor}

The simplest method of stitching together the $N$ segments is to not do so;
coincidence-based algorithms do not attempt to add together results across
segments from the coherent stage. Instead, each segment is treated as an
independent search, and coincidences are sought between significant candidates
from each segment~\cite{PoghEtAl2015-ArImpPrlSfSPGrWS}. Candidates may be
considered coincident e.g.\ by binning their parameters over a grid, thereby
collecting candidates with similar parameters~\cite{LIGO2009-EnsSrPrGrvWvLSD}.

The coincidence method is potentially less sensitive than other semi-coherent
methods. For example, a weaker signal may become detectable by building up
signal power over segments, whereas it could be lost if it fails to pass a
per-segment threshold. This method does have the advantage, however, of sparing
the computational expense the second semi-coherent stage. Moreover, it can be
applied to data in (relatively) real time -- as soon as $\Tcoh$ worth of data is
collected and prepared -- rather than having to wait until the end of the run
for all the data to be collected. The coincidence method has been used in
all-sky searches for isolated pulsars.

\subsubsection{Hough transform algorithms}\label{sec:hough-transf-algor}

The Hough transform~\cite{Houg1959-McAnlBbChPct} is a pattern recognition
algorithm originally developed to find particle tracks in bubble chamber
photographs. Let us suppose we are analysing a two-dimensional image $(x, y)$ in
search of straight line patters, i.e. of the form $y = m x + c$. The Hough
transform is a one-to-many mapping from $(x, y)$ to the set $\{ (m, c) \}$ of
every line that passes through $(x, y)$. Now suppose there is a straight line
artefact $y = m_0 x + c$ in the image; as each point $(x, y)$ along the line is
mapped, their corresponding sets $\{ (m, c) \}$ will intersect at the true
parameters of the line $(m_0, c_0)$. Such intersections may be found by binning
mapped sets $\{ (m, c) \}$ over a grid in $(m, c)$, and seeing which bins
accumulate the most counts.

The Hough transform has generally been applied to all-sky CW searches. Because
it is a two-dimensional transform, one must select two dimensions from the (at
least) four dimensions of the all-sky parameter space to apply the transform
on. This had lead to two implementations:
SkyHough~\cite{KrisEtAl2004-HgTrSrCntGrvW, KrisLIGO2005-WPrmSrIsPlUHTrn}, which
applies the Hough transform to the sky parameter space at fixed frequency and
spindown; and FrequencyHough~\cite{AntoEtAl2008-DtcPrGrvWSrHTrFVFP,
  AstoEtAl2014-MASrCnGrvWSUFrqTr}, which applies the Hough transform to the
frequency--spindown parameter space at fixed sky position. Aside from this
choice, the two implementations operate in a similar manner, as follows.

A Hough transform pipeline starts with matched filter results, for each segment,
from the first coherent stage; these are usually power,
although~\cite{LIGOVirg2013-EnsAlSrPrGrvWLSD} used the \Fstat. They then require
each matched filter result pass a threshold and be a local maxima with respect
to neighbouring filters. This selects a collection of points in the domain of
the Hough transform, known as a \emph{peak map}. The Hough transform is then
applied to the peak map, and its results are binned into a grid called a
\emph{partial Hough map}; each grid cell is one if it contains a value of the
transform, or zero otherwise. Finally, the remaining dimensions of the search
parameter space are considered (frequency and spindown for SkyHough, sky
position for FrequencyHough): for each vector of parameters selected from these
dimensions, the partial Hough maps consistent with a CW signal with these
parameters are summed over segments to give a \emph{total Hough map}. The final
output of the SkyHough (FrequencyHough) pipelines are total Hough maps over the
sky (frequency--spindown) parameter spaces, for fixed values of frequency and
spindown (sky position). The detection statistic in each bin of the total Hough
map is the \emph{number count}; the number of segments (out of $N$) in which the
corresponding bin of the partial Hough maps registered a one.

Further developments of each implementation have included: weighted summing of
the partial Hough maps to account for the amplitude modulation of the CW
signal~\cite{PaloEtAl2005-AdpHTrnSrPrSr, KrisSint2007-HgSrImpSns};
BinarySkyHough, an extension of the SkyHough implementation to efficiently
search over binary orbital parameters~\cite{CovaSint2019-NMSCntGrvWUnNtSBSy};
extensions of the FrequencyHough implementation to perform directed searches
(i.e.\ at fixed sky position) efficiently using the BSD
framework~\cite{PiccEtAl2020-DrSCntGrvSGlCALSObR}, and to search over binary
orbital parameters~\cite{LeacEtAl2017-NDrSStDCnGrvWNSLHghBS}; and adaptions of
the Hough transform beyond the CW realm to search for long-duration but
transient gravitational wave signals~\cite{MillEtAl2018-MSLDGrvWTrINSUGnFrqTr,
  OlivEtAl2019-AdTrHMLngGrvWTr}.

\subsubsection{StackSlide algorithms}\label{sec:stacksl-algor}

The StackSlide algorithm~\cite{BradCrei2000-SrcPrSrLIIHrrSr-II,
  MendLand2005-StcHgSrSNStt} is based on the following mental picture. Imagine
taking a set of SFTs, computing their power spectra, and (taking each spectrum as
a vertical column) ``stacking'' them along a horizontal time axis; the result is
a time-frequency plane of power versus time. In this plane, a CW signal would
appear as a wavy horizontal line, due to the various modulations of its
frequency $f(t)$ [Eq.~\eqref{eq:frequency-at-NS}]. Now imagine, for a given
vector of sky and spindown parameters, ``sliding'' each spectra up and down so
as to demodulate the signal, i.e.\ so that the wavy horizontal line is now
perfectly straight. Finally, sum up the frequency bins of the spectra (in their
post-``sliding'' positions) over time; this accumulates CW signal power over
time for a range of frequencies. The summed StackSlide power is distributed
according to a chi-squared distribution with $2 N$ degrees of
freedom~\cite{MendLand2005-StcHgSrSNStt}. Only one StackSlide-on-power search
has been performed~\cite{LIGO2008-AlSrPrdGrvWvLSD}.

The StackSlide concept has been generalised to use the \Fstat\, instead of
power, as input~\cite{BradCrei2000-SrcPrSrLIIHrrSr-II}. Here, the concept of
``sliding'' the power spectra of each segment is made concrete by the
construction of \emph{coarse} and \emph{fine} grids, as follows. First, in the
coherent stage, for each segment, we compute the \Fstat\ over the coarse grid: a
grid of points in the search parameter space appropriate for an \Fstat\ search
over time-span $\Tcoh$, constructed e.g.\ by using the parameter-space metric
(Sec.~\ref{sec:param-space-breadth}) to decide on the spacings between grid
points. In the semi-coherent stage, we then construct the fine grid: here the
grid must be appropriate for an \emph{incoherent} sum of \Fstat\ values. This
requirement is easily satisfied using the parameter space metric; given the
\emph{coherent metrics} computed for the $N$ coarse grids in each segment, the
\emph{semi-coherent metric} is given by the average\footnote{By ``average'', we
  mean that the element $\hat{g}_{ij}$ of the semi-coherent metric is given by
  $\hat{g}_{ij} = \sum_{k=1}^{N} \tilde{g}^{N}_{ij}$, where $\tilde{g}^{N}_{ij}$
  are the equivalent elements of the $N$ coherent metrics. Note that coherent
  metrics \emph{must} be computed using consistent definitions of the phase
  parameters; for example, the metric elements for frequency $f$ must have been
  computed at the same reference time $t_0$.} of the $N$ coherent
metrics. Typically, the fine grid contains $\gamma$ more points per unit
parameter space volume than the average coarse grid; $\gamma$ is known as the
\emph{refinement factor}.

Next, each coarse grid is interpolated onto the fine grid: for each fine grid
point, and for each segment, we find and record the coarse grid point to which
the fine grid point is ``closest'' (in the sense of the coherent parameter-space
metric for that segment). This gives a one-to-$N$ mapping from each fine grid
point to its closest coarse grid points in each segment. Finally, for each fine
grid point, the $N$ values of $\twoF$ corresponding to its $N$ closest coarse
grid points are summed. The output detection statistic is distributed according
to a chi-squared distribution with $4 N$ degrees of
freedom~\cite{MendLand2005-StcHgSrSNStt}.

Implementation of the StackSlide algorithm with the \Fstat\ requires knowledge
of the coherent and semi-coherent parameter-space metrics. The coherent \Fstat\
metric, both its most general form and the phase metric approximation, are
detailed in~\cite{Prix2007-SrCnGrvWMMltFs}. A particularly desirable property is
for the metrics to be \emph{flat}, i.e.\ $g_{ij}(\vec{p}) = g_{ij}$ is constant
with respect to the phase parameters. This property facilitates generation of
\emph{template banks} -- i.e.\ grids -- which minimise the number of templates
and hence computational cost~\cite{Owen1996-STmGrvWInsBnCTmS,
  JaraKrol2005-GrvDAnFrSAppGsC, Prix2007-TmpSrGrvWEfLCFPrS,
  Wett2014-LTmPlcChASrGrvP}, although such template banks may not be optimal for
detection~\cite{Alle2021-OptTmpBn, AlleShoo2021-TmpBnBsZA_Ltt}.

An important limitation of the metric is that it is only a quadratic
approximation to the true loss of signal power, and therefore remains valid only
for relatively small mismatches
$\mu \lesssim 0.4$~\cite{Prix2007-SrCnGrvWMMltFs,
  WettPrix2013-FPrmMtASrGrvPl}. As discussed
in~\cite{Wett2016-EmExRVlPrmMASGrvP}, the semi-coherent mismatch \emph{rarely}
satisfies this requirement, due to computational constraints that limit the size
of the fine template bank. Nevertheless, empirical studies show that the loss of
signal power progresses slowly even at very large mismatches (beyond the domain
of validity of the metric), In practise, therefore, StackSlide searches can
still achieve good sensitivity; even if the fine grid must be constructed with a
large ($\gtrsim 1$) \emph{metric} mismatch $\mu$, the true loss of signal power
will still be reasonable. An alternative to the quadratic metric approximation
is proposed in~\cite{Alle2019-SphAnPrmMt}.

Building on studies of large-scale correlations in phase parameter
space~\cite{PrixItoh2005-GPrmCrChSCnGrvW, Plet2008-PrmCrOStCnGrvD} and of
simplified CW phase models~\cite{JaraKrol1999-DAnGrvSSpNSIIAEstPr-II}, the
Global Correlation Transform (GCT) was derived
in~\cite{PletAlle2009-ExLrCrrDCnGrvW, Plet2010-PrmMSmSrCnGrvW}. The GCT
introduces new phase parameters, where a Taylor expansion of the orbital motion
of the Earth is absorbed into new frequency and spindown coordinates. Because
such a Taylor expansion remains valid only for $\Tspan \ll 1$~year, the GCT
metric significantly underestimates the loss of signal power over realistic
observing times. Nonetheless, the GCT remains in use for all-sky CW searches.

The supersky metric~\cite{WettPrix2013-FPrmMtASrGrvPl, Wett2015-PrmMASmSrGrvPl}
expands upon some of the ideas of the GCT, while addressing its limitations. The
sky parameter space is projected from the 2-sphere to three dimensions, then
projected back onto a two-dimensional plane, which corresponds to either the
equatorial plane of the Earth's equator (for short $\Tspan$) or the ecliptic
plane of its orbit (for long $\Tspan$). The projection is accomplished by
absorbing linear and quadratic (with $\Tspan$) terms in a Taylor expansion of
the Earth's orbital motion into new frequency and spindown coordinates,
respectively, while retaining higher-order terms which become important at
longer $\Tspan$. Combined with an optimal lattice-based template
bank~\cite{Wett2014-LTmPlcChASrGrvP}, a StackSlide search pipeline based on the
supersky metric was implemented, known as
Weave~\cite{WettEtAl2018-ImpSmSCnGrvWUOCnTB}, and demonstrated improved
sensitivity compared to the GCT~\cite{WalsEtAl2019-OpCAnMASCnGrvWEns}.  Weave
requires sufficient computer memory to store the $\twoF$ values needed for the
one-to-many fine-to-coarse-grid mapping, which may limit its usage in some
circumstances~\cite{WalsEtAl2019-OpCAnMASCnGrvWEns}. To date Weave has been used
for narrow-band searches for known pulsars, and for directed searches for CCOs.

\subsubsection{PowerFlux algorithms}\label{sec:powerflux-algorithms}

PowerFlux~\cite{Derg2005-DscPwAlImp, LIGO2008-AlSrPrdGrvWvLSD} is an all-sky
search pipeline for isolated neutron stars. It may be seen as an extension of
the StackSlide-on-power semi-coherent paradigm. The principal difference is that
PowerFlux weights the power from each SFT by the detector response functions,
thereby emphasising times during the day when, for a given sky position, the
detector is most sensitive. PowerFlux also inversely weights the power from each
SFT by its noise, thereby de-weighting times where detector sensitivity is
degraded. The implementation of the method is highly
optimised~\cite{Derg2005-DscPwAlImp, Derg2011-DscPw2AlImp} and is often used to
perform ``quick-look'' searches of the early data from a run, taking advantage
of the typical step-up in detector sensitivity after upgrades and commissioning
between runs. PowerFlux has developed a procedure for computing $h_0^C$ upper
limits which are strictly conservative (i.e.\ worst case) even in the presence
of spectral artefacts~\cite{Derg2013-NUnStCmULIllBck}.

The loosely coherent extension to the classic PowerFlux
algorithm~\cite{Derg2010-BSrcNDmnSgLChAp, Derg2012-LsChrSrSWllSg} generalises
the concept of summation of SFT power over time. Instead, a two-dimension
summation over all pairs of SFTs at times $t_1, t_2$ is considered, where a
kernel function $K_{t_1, t_2}(\delta)$ with parameter $\delta$~\footnote{Note
  that this is \emph{not} the declination $\delta$ of the source's sky
  position.}  decides which pairs to add and with what weight. In this picture,
summation of SFT power is equivalent to a kernel which is one when $t_1 = t_2$
and zero otherwise. The kernel blurs the distinction in the hierarchical search
paradigm between the coherent and semi-coherent stages, and permits a smoother
transition between full phase coherence and phase-incoherent power summing over
a timescale determined by $\delta$. It bears similarities with the
cross-correlation method (Sec.~\ref{sec:cross-corr-algor}) as well as the
approaches of~\cite{Plet2011-SChWTcHrrDCnGrvW, Cutl2012-ImPhsFsGrvDAn}.  The
loosely coherent method was first developed to follow up candidates from an
initial PowerFlux search, as it allows longer effective coherence times $\Tcoh$
to be used, as seen in Fig.~\ref{fig:analysis_incoherent_times}. A recent, fast
implementation of the method called Falcon~\cite{DergPapa2019-SnImpSrPrGrvWUOLD}
enables loose coherence to also be used in the initial search.

\subsubsection{Cross-correlation algorithms}\label{sec:cross-corr-algor}

Cross-correlation is a well-established concept in signal processing; it
quantifies the similarity of two independent times series as a function of their
relative time offset. It was first applied to gravitational wave data as a
radiometer implemented in software~\cite{Ball2006-RdmStGrvWv} for finding unmodeled
stochastic gravitational waves. A modelled cross-correlation algorithm was then
developed to search for CW signals~\cite{DhurEtAl2008-CrsSrPrGrvWv}. Similar to
the kernel of loosely coherent PowerFlux, each pair of independent SFTs labelled
$I, J$ -- from either different times, or different detectors -- are
cross-correlated with a filter $Q_{IJ}$. The filter weights each pair of SFTs
according to how a given CW signal would appear in the two SFTs; it essentially
performs the role of demodulating the signal in order to maximise
signal-to-noise ratio. The filter can be tuned to select which SFTs to
cross-correlate; correlating only SFTs close to each other in time recovers a
power-like detection statistic, while correlating all SFTs recovers the
\Fstat. The cross-correlation method can therefore tune its degree of phase
coherence in a more flexible manner than a traditional two-stage hierarchical
search.

The first version of CrossCorr, an implementation of the cross-correlation
algorithm, initially targeted a CCO in the supernova remnant SN
1987A~\cite{ChunEtAl2011-DsCrsSCntGrRdNSSRS1}. A second version of CrossCorr has
been used to target the LMXB Sco X-1~\cite{WhelEtAl2015-MdCrsSrGrvWScX}. Recent
developments include the addition of ``resampling'' for efficient computations
over frequency~\cite{MeadEtAl2018-RsAcCrsSCnGrvWBS}, similar to the
\Fstat. Lattice template placement has been used to minimise computational cost,
in particular by a choice of coordinate transform in the $P$--$\tasc$ space
which reduces the template bank to a single point in
$P$~\cite{WagnEtAl2022-TmLtCrsSGrvWScX}. CrossCorr typically tunes its effective
coherence length $T\umax$ as a function of $f$, $a_p$, and $\tasc$.

\subsubsection{TwoSpect algorithm}\label{sec:twospect-algorithm}

TwoSpect~\cite{GoetRile2011-ASAlCntGrvWSpNtSBS, GoetRile2016-ChCmDBDASmcCnGrvWS}
is a specialised algorithm for CW signals from neutron stars with binary
companions. It starts, in a similar manner to the StackSlide-on-power method,
by stacking SFT power spectra and forming a time-frequency plane of SFT
frequency (in the vertical direction) versus time (in the horizontal
direction). A second series of Fourier transformations and power spectra are
then computed over the time (horizontal) plane, yielding a frequency-frequency
plane of SFT frequency (vertical) versus the 2nd Fourier transform frequency
(horizontal); see Fig.~1 of~\cite{GoetRile2011-ASAlCntGrvWSpNtSBS}. Due to the
various modulations of the CW signal from the orbits of the neutron star and
Earth, signal power will appear in the TwoSpect frequency-frequency plane at
regularly-spaced pixels, indicating the fundamental periods of the modulations
and their harmonics. A first stage of analysis incoherently sums power in pixels
and their harmonics, to identify promising candidates; a second stage then
construct templates which match the distinctive pattern of pixels expected for a
CW signal with given sky, frequency, and binary orbital parameters.

TwoSpect was used to perform the first all-sky search for neutron stars in
binary systems using S6 data (Table~\ref{tab:observing_runs}); to date this
remains the broadest CW search ever performed
(Fig.~\ref{fig:analysis_overview}). TwoSpect has also been used in a directed
search mode to target the LMXBs Sco X-1 and XTE
J1751$-$305~\cite{MeadEtAl2016-TnScXAdCntGrvSKBS}.

\subsubsection{Viterbi and SOAP algorithms}\label{sec:viterbi-algorithms}

An HMM (Sec.~\ref{sec:hidden-markov-models}) models the CW signal frequency as a
randomly-wandering path over a time-frequency plane, as opposed to a
pre-determined function. The Viterbi algorithm is used to recover the most
likely path of the signal frequency through the plane.

The effect of the Viterbi algorithm is often described as ``tracking'' the
signal (forward) in time. This is somewhat misleading, as in fact the Viterbi
algorithm works by looking \emph{backwards} in time. As implemented for CW
searches~\cite{SuvoEtAl2016-HMMTrCntGrvWNtSWnS,
  SuvoEtAl2017-HMMTrCntGrvWBNSWnSIIBOPTr-II, SunEtAl2018-HMMTrCntGrvWYSpRm,
  MelaEtAl2021-HMMTrCntGrvWBNSWnSIIIRtPT-III, BaylEtAl2019-GnrApVAlSCnGrvS}, the
Viterbi algorithm operates as follows. For each time step $m$ and frequency bin
$n$, it considers the frequency bins
$n - \delta n_{-}, n - \delta n_{-} + 1, \cdots, n + \delta n_{+}$ from the
previous time step $m - 1$. Here $\delta n_{-}, \delta n_{+} \ge 0$ are chosen
based on the expected properties of the CW frequency, e.g. whether it may
decrease ($\delta n_{-} > 0$) or increase ($\delta n_{+} > 0$) with time. Based
of the detection statistics computed at these
$\delta n = 1 + \delta n_{-} + \delta n_{+}$ frequency bins, the Viterbi
algorithm chooses the bin with the maximum detection statistic. This indicates
the most likely path from time step $m - 1$ to time step $m$ which ends at bin
$n$, out of the $\delta n$ possibilities.  The maximum detection statistic at
time step $m - 1$ is then added to the detection statistic computed at the
current time step $m$ and frequency bin $n$. In short, at every time step $m$,
the Viterbi algorithm find the $N$ most probable past path of the CW frequency
that intersects the $N$ frequency bins. At the final time step $M - 1$, the
Viterbi algorithm has found the $N$ most probable paths of the CW frequency
through the time-frequency plane that end at the $N$ frequency bins.

The Viterbi algorithm has many advantages, including low computational cost,
effective coverage of a vast space of possible signal frequency variations (see
Sec.~\ref{sec:hidden-markov-models}), and straightforward adaptability to a wide
variety of search targets. The first implementation of the algorithm for CW
searches~\cite{SuvoEtAl2016-HMMTrCntGrvWNtSWnS,
  SuvoEtAl2017-HMMTrCntGrvWBNSWnSIIBOPTr-II} targeted LMXBs such as Sco X-1
where, in contrast to templated CW algorithms, the Viterbi algorithm could
robustly handle the expected spin wandering of the signal frequency due to the
time-varying accretion torque. The same implementation has since been applied to
CCOs in young supernova remnants~\cite{SunEtAl2018-HMMTrCntGrvWYSpRm} and
long-duration transient gravitational waves from a neutron star born in a binary
neutron star merger~\cite{SunMela2019-ApHMMTrSLngTrGrvWRBNSMG}.

A variant of the algorithm, SOAP~\cite{BaylEtAl2019-GnrApVAlSCnGrvS}, augments
the basic Viterbi algorithm with a memory -- where the algorithm looks back
several time steps instead of just one -- to better tune the search toward
periodic-like signals. A post-processing step using convolutional neural
networks~\cite{BaylEtAl2020-RMcLrAlSCntGrvW} is added to improve robustness
towards spectral artefacts. SOAP is intended as a general-purpose,
``quick-look'' search method for CW signals, as well as a tool for identifying
instrumental line artefacts.

\subsection{Other algorithms and applications}\label{sec:other-algorithms}

CW searches over wide parameter spaces typically yield a large number of
candidate signals. Post-processing of these candidates may require: a robust
determination of their significance, but determining the statistical
distribution expected for the maximum detection
statistic~\cite{LIGO2010-FrSrGrvWYngKNtS, WettEtAl2021-DExpCnGrvW171172HLScObRD,
  TenoEtAl2022-EmEsDstLCnGrvS}; clustering of candidates with similar
parameters~\cite{SingEtAl2017-AdClsPrCnGrvWS, MoraEtAl2020-MLrClsCnGrvSCn,
  BehePapa2020-DLrClsCnGrvWCn, BehePapa2021-DLClCnGrvWCnIIIdnLC-II,
  TenoEtAl2021-TmfTDsCmCnGrvWS, PierEtAl2022-ISClWdSrCntGrvW,
  StelEtAl2022-DnsCnGrvWCnLSr}; vetoing of candidates due to instrumental
artefacts~\cite{Leac2015-MFOSpDsCntSGrvDt, ZhuEtAl2017-NVtCntGrvWvSr,
  IntiEtAl2020-DppBVDsFCnGrvCn} or use of statistics insensitive to such
artefacts~\cite{KeitPrix2015-LnStCnGrvWSCUDSns, KeitEtAl2014-SCnGrvWImRbVInsA,
  Keit2016-RSmcSCnGrvWNSMInHDLTr, AshtEtAl2018-SmcGlCntSrMt}; and the
performance of successive follow-up searches with increasing coherence times to
sieve out the most significant candidates~\cite{CutlEtAl2005-ImStcSrGrvPl,
  Shal2012-ChFCnGrvCnMRObsT, ShalPrix2013-FChFlCntGrvCn,
  AshtPrix2018-HrrMlMFCnGrvWCn, KeitEtAl2021-PyPPcCntGrvDAn,
  TenoEtAl2021-ApHrrMFALCnGrvCn}.

Recently there has been interest in addressing the challenges of CW searches
using deep learning techniques~\cite{DreiEtAl2019-DplCnGrvWv,
  MillEtAl2019-HEffMcLrDLTrGrvWNSRS, DreiPrix2020-DplCnGrvWMDtRlN,
  YamaTana2021-UExPMCnvNNtASCnGrvW}, and by engaging expertise from beyond
academia through competitive challenges~\cite{G2NetCWKaggleChallenge}. In
anticipation of a first CW detection, there is also growing interest in
quantifying what knowledge of neutron star physics we might be able to
learn~\cite{Jone2022-LrFrCnCntGrvWS, SienJone2022-GrvWSpNtSNtqSr,
  LuEtAl2023-InNtSPrpCnGrvW, SienEtAl2023-MsNtrDsPrGrvP}.

\section{Summary}\label{sec:summary}

Significant challenges stand in the way of making a first detection of CWs from
neutron stars: the very weak nature of the signal compared to contemporary
detector sensitivities, the vast breadth of the parameter space in which it may
exist, and severe computational prohibitions on using the optimal analysis
method. In this review we have seen that, in response to these challenges, CW
data analysts have applied a wide variety of algorithms, each with different
strengths and compromises, and performed a diverse number of searches of LIGO
and Virgo detector data encompassing broad swathes of parameter space.

The field of CW data analysis has developed considerably in the last twenty
years. Continued refinement of algorithms and search designs, combined with
ever-more sensitive detectors, may one day pay off in the initial thrill of a
first detection, followed by a unique and enduring perspective on the extreme
physics of neutron stars. Let us hope that Nature is so kind.

\paragraph{Further reading}

There are a growing number of informative review articles covering different
aspects of CW research. In addition to this review, the reader is encouraged to
consult: \cite{HaskellBejger2023, HaskSchw2020-IslNtrSt,
  HaskEtAl2015-GrvWvRpRttNtS, Lask2015-GrvWvNtrStRv} for reviews of CW sources
and astrophysics; \cite{TenoEtAl2021-SMtCnGrvSgUSAdvE, Prix2009-GrvWvSpnNtSt,
  JaraKrol2005-GrvDAnFrSAppGsC} for reviews with a focus on CW searches and
statistical techniques; and \cite{Rile2023-SrCntGrvRd, Picc2022-StPrsCnGrvWSr,
  SienBejg2019-CntGrvWNtSCrSPr, Rile2017-RcSrCntGrvWv,
  LeacEtAl2012-SrCntGrvWSgULVDt} for broad overviews of CW sources, algorithms,
and results.

\section*{Acknowledgements}

I thank Pep Covas, Paola Leaci, Andrew Miller, Ben Owen, Matt Pitkin, and Keith
Riles for helpful comments on the manuscript. The author is supported by the
Australian Research Council Centre of Excellence for Gravitational Wave
Discovery (OzGrav), project number CE170100004. This manuscript was prepared
using the following software: Adobe Convert PDF to Excel~\cite{AdobePDFToExcel},
Astropy~\cite{Astropy}, graphreader.com~\cite{graphreader},
Jupyter~\cite{Jupyter}, Mathematica~\cite{Mathematica}, and
Numpy~\cite{Numpy}. This manuscript has document number LIGO-P2300131-v4.

\section*{Corrigendum}

Equation~\ref{eq:depth} of version 2 of the arXiv
article\footnote{\url{https://arxiv.org/abs/2305.07106}} is in error. The power
spectral density $\Sh$ was incorrectly normalised by units of Hz, rather than
Hz${}^{-1}$. The equation is corrected in version 3.

Equation~\ref{eq:rawMVbin-unknown-tasc-fixed-freq-mod-dpth} of version 2 of the
arXiv article is in error. A factor of $\partial a_p / \partial \Delta f\uobs$
is required inside the integral to correctly transform the parameter space
metric coordinates from projected semi-major axis $a_p$ to frequency modulation
depth $\Delta f\uobs$. The equation is corrected in version 3. Corrected
versions of Fig.~\ref{fig:analysis_overview} and Tab.~\ref{tab:analysis_data}
may be obtained here\footnote{\url{https://cw-vista.streamlit.app/}}. I am
grateful to Evan Goetz for helping to discover this error.

\appendix

\onecolumn

\section{Table of continuous wave searches}

\begin{longtable}{@{\extracolsep{\fill}} l @{\extracolsep{\fill}} l @{\extracolsep{\fill}} l @{\extracolsep{\fill}} l @{\extracolsep{\fill}} r@{\extracolsep{0pt}}l @{\extracolsep{\fill}} r@{\extracolsep{0pt}}l @{\extracolsep{\fill}} r@{\extracolsep{0pt}}l @{\extracolsep{\fill}} r@{\extracolsep{0pt}}l @{\extracolsep{\fill}} r@{\extracolsep{0pt}}l @{\extracolsep{\fill}} r@{\extracolsep{0pt}}l @{\extracolsep{\fill}} r@{\extracolsep{0pt}}l @{\extracolsep{\fill}} r@{\extracolsep{0pt}}l @{\extracolsep{\fill}} r@{\extracolsep{0pt}}l @{\extracolsep{\fill}} r@{\extracolsep{0pt}}l @{\extracolsep{\fill}} l@{\extracolsep{\fill}}}
\caption{\label{tab:analysis_data}CW searches, 2003--2022, organised by target parameter space. Columns are: most recent observing run data used, astrophysical target, (coherent, incoherent) analysis algorithm used, parameter-space breadth, sensitivity depth, weighted search volume, and references. Notes to the table, labelled with superscript letters, are listed following the table (starting on page~\pageref{sec:analysis_data_notes}). A machine-readable version is provided in the supplementary material to this review.} \\
\hline
\hline
Obs. & Target${}^{\mathrm{a}}$ & \multicolumn{2}{c}{Algorithm${}^{\mathrm{b}}$}  & \multicolumn{14}{c}{Breadth}       & \multicolumn{4}{c}{Depth}  & \multicolumn{2}{c}{Vol.} & Ref. \\
 &  & \multicolumn{1}{c}{Coh.} & \multicolumn{1}{c}{Inc.} & \multicolumn{2}{c}{$\MVfreq$} & \multicolumn{2}{c}{$\MVfd$} & \multicolumn{2}{c}{$\MVfdd$} & \multicolumn{2}{c}{$\MVsky$} & \multicolumn{2}{c}{$\MVbin$} & \multicolumn{2}{c}{$\MVHMM$} & \multicolumn{2}{c}{$\MV$} & \multicolumn{2}{c}{$\depth$} & \multicolumn{2}{c}{$f_{\depth}$} & \multicolumn{2}{c}{$\WVol$} &  \\
 &  &  &  & \multicolumn{2}{c}{$\lg$} & \multicolumn{2}{c}{$\lg$} & \multicolumn{2}{c}{$\lg$} & \multicolumn{2}{c}{$\lg$} & \multicolumn{2}{c}{$\lg$} & \multicolumn{2}{c}{$\lg$} & \multicolumn{2}{c}{$\lg$} & \multicolumn{2}{c}{} & \multicolumn{2}{c}{Hz} & \multicolumn{2}{c}{$\lg$} &  \\
\hline
\hline
\endfirsthead
\caption*{Table~\ref{tab:analysis_data} continued.} \\
\hline
\hline
Obs. & Target${}^{\mathrm{a}}$ & \multicolumn{2}{c}{Algorithm${}^{\mathrm{b}}$}  & \multicolumn{14}{c}{Breadth}       & \multicolumn{4}{c}{Depth}  & \multicolumn{2}{c}{Vol.} & Ref. \\
 &  & \multicolumn{1}{c}{Coh.} & \multicolumn{1}{c}{Inc.} & \multicolumn{2}{c}{$\MVfreq$} & \multicolumn{2}{c}{$\MVfd$} & \multicolumn{2}{c}{$\MVfdd$} & \multicolumn{2}{c}{$\MVsky$} & \multicolumn{2}{c}{$\MVbin$} & \multicolumn{2}{c}{$\MVHMM$} & \multicolumn{2}{c}{$\MV$} & \multicolumn{2}{c}{$\depth$} & \multicolumn{2}{c}{$f_{\depth}$} & \multicolumn{2}{c}{$\WVol$} &  \\
 &  &  &  & \multicolumn{2}{c}{$\lg$} & \multicolumn{2}{c}{$\lg$} & \multicolumn{2}{c}{$\lg$} & \multicolumn{2}{c}{$\lg$} & \multicolumn{2}{c}{$\lg$} & \multicolumn{2}{c}{$\lg$} & \multicolumn{2}{c}{$\lg$} & \multicolumn{2}{c}{} & \multicolumn{2}{c}{Hz} & \multicolumn{2}{c}{$\lg$} &  \\
\hline
\hline
\endhead
\hline
\endfoot
\multicolumn{25}{c}{Targeted searches for known pulsars (``Pulsars T'')} \\
\hline
S1 & J1939 & Bayes & \multicolumn{1}{c}{--} & \multicolumn{2}{c}{--} & \multicolumn{2}{c}{--} & \multicolumn{2}{c}{--} & \multicolumn{2}{c}{--} & \multicolumn{2}{c}{--} & \multicolumn{2}{c}{--} & 0& & 85&.0${}^{\mathrm{c}}$ & \multicolumn{2}{c}{--} & 27&.7 & \citenum{LIGO2004-SULStPrGrvWPJ19UFSDG600LDt} \\
S1 & J1939 & $\mathcal{F}$-stat & \multicolumn{1}{c}{--} & \multicolumn{2}{c}{--} & \multicolumn{2}{c}{--} & \multicolumn{2}{c}{--} & \multicolumn{2}{c}{--} & \multicolumn{2}{c}{--} & \multicolumn{2}{c}{--} & 0& & 100&.0${}^{\mathrm{c}}$ & \multicolumn{2}{c}{--} & 28&.8 & \citenum{LIGO2004-SULStPrGrvWPJ19UFSDG600LDt} \\
S2 & 28 pulsars & Bayes & \multicolumn{1}{c}{--} & \multicolumn{2}{c}{--} & \multicolumn{2}{c}{--} & \multicolumn{2}{c}{--} & \multicolumn{2}{c}{--} & \multicolumn{2}{c}{--} & \multicolumn{2}{c}{--} & 1&.45 & 160&.0${}^{\mathrm{c}}$ & \multicolumn{2}{c}{--} & 32&.9 & \citenum{LIGOEtAl2005-LmGrvEmsSlPlULD} \\
O3 & 35 pulsars & 5-vec & \multicolumn{1}{c}{--} & \multicolumn{2}{c}{--} & \multicolumn{2}{c}{--} & \multicolumn{2}{c}{--} & \multicolumn{2}{c}{--} & \multicolumn{2}{c}{--} & \multicolumn{2}{c}{--} & 1&.54 & 39&.0${}^{\mathrm{d}}$ & 89&.6 & 37&.2 & \citenum{LIGOEtAl2022-SrGrvWKPTHrmSTLIObR} \\
S4 & 78 pulsars & Bayes & \multicolumn{1}{c}{--} & \multicolumn{2}{c}{--} & \multicolumn{2}{c}{--} & \multicolumn{2}{c}{--} & \multicolumn{2}{c}{--} & \multicolumn{2}{c}{--} & \multicolumn{2}{c}{--} & 1&.89 & 300&.0${}^{\mathrm{c}}$ & \multicolumn{2}{c}{--} & 37&.4 & \citenum{LIGOEtAl2007-UpLmGrvWEms78RPl} \\
O2 & 24 pulsars & $\mathcal{F}$-stat & \multicolumn{1}{c}{--} & \multicolumn{2}{c}{--} & \multicolumn{2}{c}{--} & \multicolumn{2}{c}{--} & \multicolumn{2}{c}{--} & \multicolumn{2}{c}{--} & \multicolumn{2}{c}{--} & 1&.38 & 58&.0${}^{\mathrm{e}}$ & 41&.6 & 41&.0 & \citenum{LIGOEtAl2019-SrGrvWvKPlTHrm20152017LD} \\
\hline
\hline
\end{longtable}
\paragraph{Notes to Table~\ref{tab:analysis_data}}\label{sec:analysis_data_notes}
\begin{itemize}
\item[a.] Abbreviations: ``G.C.'':~Galactic centre. Names of astronomical objects have also been abbreviated; see the reference for the full identifiers.
\item[b.] Abbreviations: ``$\mathcal{F}$-stat'': $\mathcal{F}$-statistic, ``2Spect'': TwoSpect, ``5-vec'': 5-vectors, ``Bayes'': Bayesian, ``BSHgh'': BinarySkyHough, ``coinc'': coincidence, ``FrHgh'': FrequencyHough, ``GCT'': Global Correlation Transform, ``PFlx'': PowerFlux (classic), ``PFlxLC'': PowerFlux (loosely coherent), ``pwr'': power, ``SkHgh'': SkyHough, ``SOAP'': Hidden Markov Model (SOAP), ``SSlide'': StackSlide, ``Vtrbi'': Hidden Markov Model (Viterbi), ``XCorr'': CrossCorr.
\item[c.] $\depth$ taken from~\cite{DreiEtAl2018-FAcSnsEsCntSr}.
\item[d.] $\depth$ derived from $h_0(f_{\depth})$ upper limit quoted in reference and $\Sh$ for O3~\cite{O3sens,O3Vsens} in a 1~Hz band around $f_{\depth}$.
\item[e.] $\depth$ derived from $h_0(f_{\depth})$ upper limit quoted in reference and $\Sh$ for O1~\cite{O1sens}, O2~\cite{O2sens} in a 1~Hz band around $f_{\depth}$.
\item[f.] $\depth$ derived from $h_0(f_{\depth})$ upper limit quoted in reference and $\Sh$ for VSR2~\cite{S6VSR23sens} in a 1~Hz band around $f_{\depth}$.
\item[g.] $\depth$ derived from $h_0(f_{\depth})$ upper limit quoted in reference and $\Sh$ for S6~\cite{S6VSR23sens}, VSR2~\cite{S6VSR23sens}, VSR4~\cite{VSR4sens} in a 1~Hz band around $f_{\depth}$.
\item[h.] $\depth$ derived from $h_0(f_{\depth})$ upper limit quoted in reference and $\Sh$ for O2~\cite{O2sens} in a 1~Hz band around $f_{\depth}$.
\item[i.] $\depth$ derived from $h_0(f_{\depth})$ upper limit quoted in reference and $\Sh$ for O2~\cite{O2sens}, O3~\cite{O3sens,O3Vsens} in a 1~Hz band around $f_{\depth}$.
\item[j.] $\depth$ derived from $h_0(f_{\depth})$ upper limit quoted in reference and $\Sh$ for O1~\cite{O1sens}, O2~\cite{O2sens}, O3a~\cite{O3sens,O3Vsens}, O3b~\cite{O3sens,O3Vsens} in a 1~Hz band around $f_{\depth}$.
\item[k.] $\depth$ derived from $h_0(f_{\depth})$ upper limit quoted in reference and $\Sh$ for O1~\cite{O1sens} in a 1~Hz band around $f_{\depth}$.
\item[l.] $\depth$ derived from $h_0(f_{\depth})$ upper limit quoted in reference and $\Sh$ for O3a~\cite{O3sens,O3Vsens}, O3b~\cite{O3sens,O3Vsens} in a 1~Hz band around $f_{\depth}$.
\item[m.] $\depth$ derived from $h_0(f_{\depth})$ upper limit quoted in reference and $\Sh$ for S5~\cite{S5sens} in a 1~Hz band around $f_{\depth}$.
\item[n.] Known pulsar search conducted after a glitch; see the reference for details.
\item[o.] Known pulsar search conducted prior to a glitch; see the reference for details.
\item[p.] $\rawMVfd$ computed from integral form of Eq.~\eqref{eq:rawMVfd} over non-rectangular $\fd$ parameter space.
\item[q.] $\rawMVfdd$ computed from integral form of Eq.~\eqref{eq:rawMVfdd} over non-rectangular $\fdd$ parameter space.
\item[r.] $\depth$ taken from reference.
\item[s.] $\depth$ derived from $h_0(f_{\depth})$ upper limit quoted in reference and $\Sh$ for O3a~\cite{O3sens,O3Vsens} in a 1~Hz band around $f_{\depth}$.
\item[t.] $\MVbin$ computed with range of $\tasc$ taken from~\cite{SammEtAl2014-ImpFrqSdSMGrWLMXBn}.
\item[u.] $\depth$ derived from $H_0(f_{\depth})$ upper limit quoted in reference, converted to $h_0$ using their Eq.~(5) and averaged over $\cos\iota \in [-1, 1]$, and $\Sh$ for O2~\cite{O2sens} in a 1~Hz band around $f_{\depth}$.
\item[v.] $\depth$ taken from~\cite{DreiEtAl2018-FAcSnsEsCntSr} and converted from PowerFlux circular/linear polarisation upper limits to population-averaged upper limits using conversion factors given in~\cite{Wett2012-EsSnWdpSrGrvP}.
\item[w.] $\depth$ derived from $h_0(f_{\depth})$ upper limit quoted in reference, converted from PowerFlux circular polarisation upper limits to population-averaged upper limits using the conversion factors given in~\cite{Wett2012-EsSnWdpSrGrvP}, and $\Sh$ for O1~\cite{O1sens} in a 1~Hz band around $f_{\depth}$.
\item[x.] $\MVbin$ computed using Eq.~\eqref{eq:rawMVbin-unknown-tasc-fixed-freq-mod-dpth} for fixed range of frequency modulation depth $\Delta f\uobs$.
\end{itemize}

\end{document}